\documentclass[aip,twocolumn,reprint,superscriptaddress]{revtex4-1}
\usepackage{amsmath}
\usepackage{amssymb}
\usepackage[latin9]{inputenc}
\usepackage{graphicx}
\usepackage{hyperref}
\usepackage{physics}

\usepackage{graphicx}
\usepackage[color= blue!20]{todonotes}
\usepackage[normalem]{ulem}
\usepackage{color}
\usepackage{ulem}
\usepackage[symbol]{footmisc}

\begin{document}

\onecolumngrid

\noindent\textbf{\textsf{\Large Cooling photon-pressure circuits into the quantum regime}}

\normalsize
\vspace{.3cm}
\noindent\textsf{I.~C.~Rodrigues$^{\dagger,1}$, D.~Bothner$^{\dagger,1,2}$ and G.~A.~Steele$^{1}$}

\vspace{.2cm}
\noindent\textit{$^1$Kavli Institute of Nanoscience, Delft University of Technology, PO Box 5046, 2600 GA Delft, The Netherlands\\$^2$Physikalisches Institut, Center for Quantum Science (CQ) and LISA$^+$, Universit\"at T\"ubingen, Auf der Morgenstelle 14, 72076 T\"ubingen, Germany\\$^\dagger$\normalfont{these authors contributed equally}}

\vspace{.5cm}

\date{\today}

{\addtolength{\leftskip}{10 mm}
\addtolength{\rightskip}{10 mm}

Quantum control of electromagnetic fields was initially established in the optical domain and has been advanced to lower frequencies in the gigahertz range during the past decades extending quantum photonics to broader frequency regimes. 
In standard cryogenic systems, however, thermal decoherence prevents access to the quantum regime for photon frequencies below the gigahertz domain. 
Here, we engineer two superconducting LC circuits coupled by a photon-pressure interaction and demonstrate sideband cooling of a hot radio frequency (RF) circuit using a microwave cavity. 
Because of a substantially increased coupling strength, we obtain a large single-photon quantum cooperativity $\mathcal{C_\mathrm{q0}}\sim1$ and reduce the thermal RF occupancy by $75\%$ with less than one pump photon. 
For larger pump powers, the coupling rate exceeds the RF thermal decoherence rate by a factor of 3, and the RF circuit is cooled into the quantum ground state. Our results lay the foundation for RF quantum photonics.

}
\vspace{.5cm}

\twocolumngrid
\section*{Introduction}

\begin{figure*}
	\centerline{\includegraphics[trim = {1cm, 0.3cm, 1cm, 0cm}, clip=True, width=0.99\textwidth]{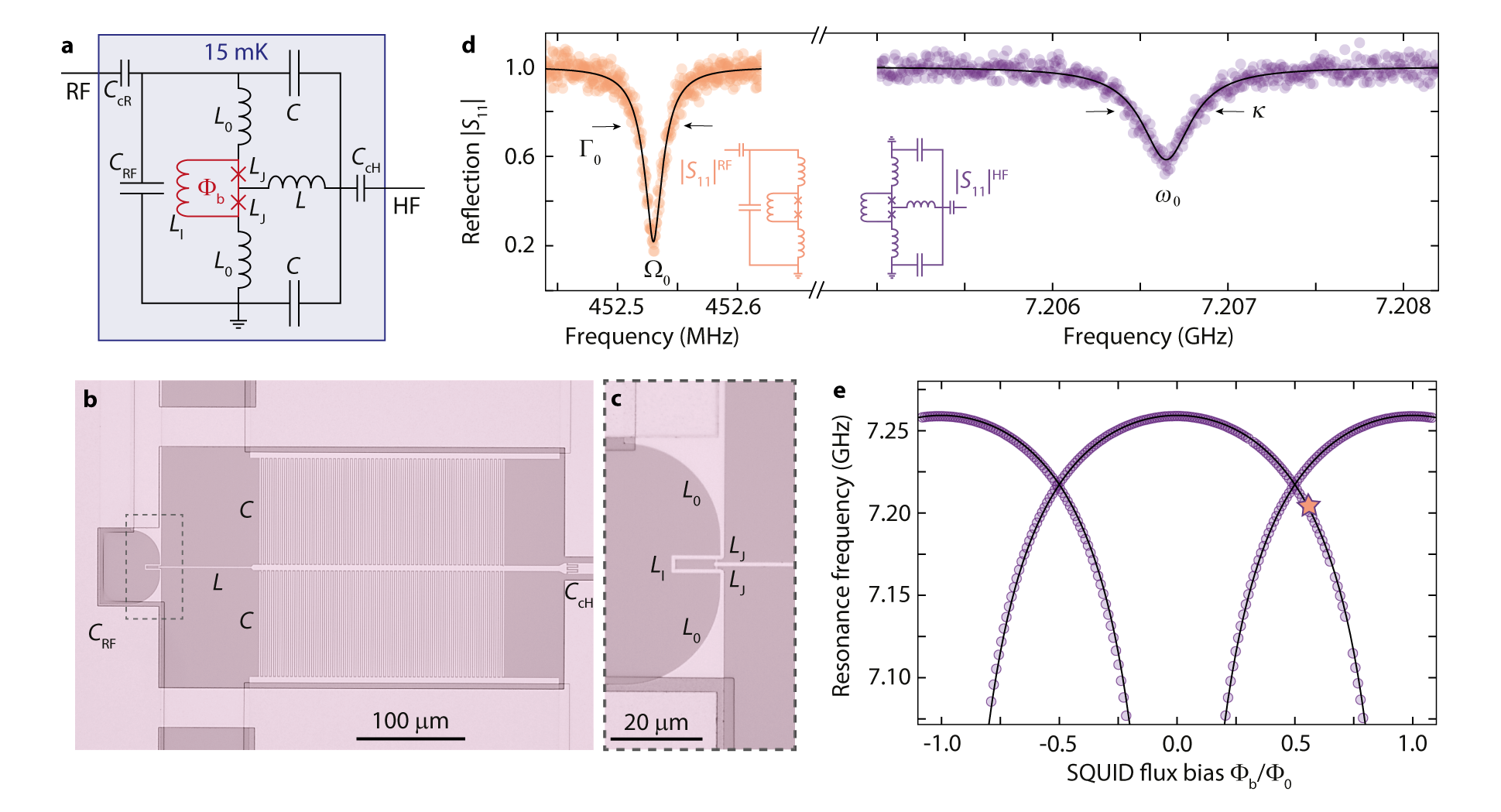}}
	\caption{\textsf{\textbf{A two-mode superconducting LC circuit with a tunable photon-pressure interaction.} \textbf{a} Circuit schematic. The full circuit has two modes, a low- and a high-frequency mode. The low, radio-frequency mode is formed by the capacitors and inductors $C_\mathrm{RF}$, $L_0$ and the parallel combination of $L_\mathrm{l}$ and $2L_\mathrm{J}$. The high-frequency microwave mode is formed by the combination of $L$, $C$ and $L_0$ and $L_\mathrm{J}$. The inductances $L_\mathrm{l}$ and $L_\mathrm{J}$ form a superconducting quantum interference device (SQUID). Both modes are capacitively coupled to individual feedlines for driving and readout. \textbf{b} Optical image of the device showing the circuit (full image is shown in Supplementary Fig.~2). The dashed box shows the zoom region for panel \textbf{c}. In \textbf{b}-\textbf{c}, brighter parts correspond to Aluminum, darker and transparent parts to Silicon. \textbf{d} shows the resonance curves of both modes vs excitation frequency in the reflection coefficient $|S_{11}|$, colored points are data and the black lines correspond to fits. The inset displays the reduced circuit equivalents for the two modes. \textbf{e} shows the resonance frequency of the high frequency mode vs magnetic flux bias through the SQUID loop, generated by an external magnetic coil. The dataset was obtained by combining data from a flux up-sweep with the data from a flux down-sweep. Due to non-negligible loop inductance $L_\mathrm{l}$, the flux-dependence is hysteretic and multi-valued for flux values around $\pm0.5\Phi_0\pm0.3\Phi_0$ \cite{LevensonFalk11, Kennedy19, Rodrigues19}. The flux operation point $\Phi_\mathrm{b}/\Phi_0 \sim 0.54$ used for the data shown in panel \textbf{d} and for the rest of this work is marked by a star.}}
	\label{fig:device}
\end{figure*}

In the recent decade, the parametric radiation-pressure coupling between two harmonic oscillators has been demonstrated to allow for groundbreaking experiments in the control and detection of harmonic oscillators from the kHz to the GHz frequency regime \cite{Aspelmeyer14, Teufel09, Wollman15, Xu16, Reed17, Riedinger18, OckeloenKorppi18, Xu19}.
The archetype of a radiation-pressure coupled system is an optomechanical cavity \cite{Aspelmeyer14}, where the interaction between a mechanical oscillator and light inside an electromagnetic resonator is used for displacement sensing and motion control of macroscopic objects with unprecedented precision. 
An outstanding feature of the radiation-pressure interaction is the possibility to cool a low-frequency mechanical oscillator orders of magnitude below its thermodynamic bath temperature using cavity red-sideband driving \cite{Cohadon99, Braginsky02, Metzger04, Gigan06, Arcizet06, Kleckner06, Schliesser06}.
The application of this technique to trapped ions and atoms \cite{Wineland78, Neuhauser78} or to mechanical oscillators has been used to place them in the phononic groundstate \cite{Diedrich89, Hamann98, WilsonRae07, Marquardt07, Teufel11, Chan11} and is the prerequisite for the preparation and investigation of quantum states of motion \cite{Meekhof96, Wollman15, Reed17}.
The implementation of photon-pressure coupling - radiation-pressure coupling between two photons of different frequency - using superconducting LC circuits has recently gained significant attention \cite{Johansson14, Hardal17, Weigand20, Eichler18, Bothner20}, as its realization promises to enable quantum control of electromagnetic fields in the radio-frequency domain.
This platform might also allow for the investigation of unexplored photon-pressure parameter regimes, as circuits provide an extremely high degree of design flexibility regarding resonance frequencies and linewidths.
Radiation-pressure coupled devices provide furthermore extensive possibilities for quantum signal processing, such as quantum-limited parametric amplification \cite{Massel11, Metelmann14, Nunnenkamp14, OckeloenKorppi16, Bothner20a}, nonreciprocal photon transport \cite{Malz18, Bernier17, Barzanjeh17, Xu19}, slow light \cite{SafaviNaeini11, Zhou13} and photonic reservoir engineering \cite{Fang17, Toth17}.
Implementing these technologies in a circuit-only platform would not only offer larger coupling rates and more architectural possibilities compared to mechanical oscillators, but would also be naturally compatible with superconducting quantum processors \cite{Arute19}.
Very recently, photon-pressure coupled circuits are also discussed as in the context of fault-tolerant quantum computing using bosonic codes \cite{Weigand20} and quantum-enhanced dark matter axion detection at low-energy scales \cite{Backes21, Chaudhuri19}.
To date, however, photon-pressure coupled superconducting circuits have only been realized in the classical regime and in presence of significant residual thermal fluctuations \cite{Eichler18, Bothner20}.
Here, we report photon-pressure coupling between a hot RF circuit and a high-frequency (HF) superconducting quantum interference cavity in the quantum regime.
By engineering galvanically connected circuits, we increase the single-photon coupling strength and single-photon cooperativity by about one order of magnitude compared to the best results reported to date \cite{Eichler18, Bothner20}.
This allows for sideband-cooling of the residual thermal occupation in the hot radio-frequency mode by a factor of $\sim 4$ with less than one pump photon and a single-photon quantum cooperativity $\mathcal{C}_\mathrm{q0} \sim 1$.
Due to the large single-photon coupling rate in our device, we reach the strong-coupling regime with only $0.7$ pump photons, where we observe the residual thermal fluctuations of the hybridized normal-modes and demonstrate groundstate cooling of the RF mode.
Simultaneuosly, the multi-photon coupling rate significantly exceeds the thermal decoherence rate of the RF mode and the decay rate of the HF cavity, which corresponds to the quantum-coherent strong-coupling regime, the basis for coherent quantum-state transfer between the two circuits \cite{Verhagen12}.
Our results pave the way towards quantum control of RF circuits and quantum-limited detection of photons in the radio-frequency regime.
\section*{Results}
Our device combines two integrated superconducting LC circuits, which are galvanically connected to each other at the heart of the circuit in a superconducting quantum interference device (SQUID).
A circuit schematic of the device and optical micrographs are shown in Fig.~\ref{fig:device}\textbf{a}-\textbf{c} and the multi-layer device fabrication is presented in detail in Supplementary Note~1.
The radio-frequency (RF) mode circuit consists of a large parallel plate capacitor using amorphous silicon as dielectric, and of a short inductor wire, which at the same time forms the loop of the SQUID.
The SQUID is completed by two constriction type Josephson junctions connecting the RF inductor wire to the high-frequency (HF) part of the circuit.
The remaining part of the HF mode consists of an additional linear inductor $L$ and two interdigitated capacitors $C$, cf. Fig.~\ref{fig:device}.
Both circuit modes are capacitively coupled to individual coplanar waveguide feedlines for driving and readout.
The chip is mounted into a printed circuit board, connected to microwave input/output cabling and packaged into a radiation tight copper (Cu) housing.
A small superconducting magnet is attached to the Cu housing below the chip, allowing for the application of an external out-of-plane magnetic field.
The experiments are carried out with the whole configuration placed inside a cryoperm magnetic shielding and attached to the mixing chamber of a dilution refrigerator with a base temperature of $T_\mathrm{b} \sim 15\,$mK.
More details on the device and the setup are given in Supplementary Notes~2 and 3 \cite{Vijay09, Vijay10, Igreja04}.
In Fig.~\ref{fig:device}\textbf{d}, the reflection response $|S_{11}|$ of the two modes is shown, measured through their individual feedlines.
The radio-frequency mode has a resonance frequency of $\Omega_0 = 2\pi\cdot452.5\,$MHz and a linewidth $\Gamma_0 = 2\pi\cdot 26\,$kHz.
For the high-frequency mode, the resonance frequency is $\omega_0 = 2\pi\cdot 7.207\,$GHz and the total linewidth $\kappa = 2\pi\cdot380\,$kHz.
The total linewidth of the HF mode is the sum of the internal contribution $\kappa_\mathrm{i} = 2\pi\cdot300\,$kHz and the external contribution due to the coupling to the feedline of $\kappa_\mathrm{e} = 2\pi\cdot80\,$kHz.
For the low-frequency circuit, we obtain $\Gamma_\mathrm{i} = 2\pi\cdot{10}\,$kHz and $\Gamma_\mathrm{e} = 2\pi\cdot16\,$kHz.
Details on the fitting function and routine are given in Supplementary Note~4.
When a magnetic flux $\Phi_\mathrm{b}$ is applied through the SQUID by the external coil, the resulting circulating current changes the inductance of the Josephson junctions and the HF resonance frequency is shifted accordingly.
In Fig.~1\textbf{e}, we show $\omega_0(\Phi_\mathrm{b})$ depending on the external bias flux $\Phi_\mathrm{b}$ through the SQUID loop, a theoretical description and modeling of the circuit and the flux dependence is detailed in Supplementary Note~3.
Any oscillating current flowing through the RF inductor induces additional flux through the SQUID loop and therefore modulates the resonance frequency of the HF mode.
As a result, the two modes interact via an effective photon-pressure coupling and the Hamiltonian of the device is given by \cite{Johansson14, Eichler18, Bothner20}
\begin{equation}
\hat{H}/\hbar = \omega_0\hat{a}^\dagger\hat{a} + \Omega_0\hat{b}^\dagger\hat{b} + g_0\hat{a}^\dagger\hat{a}\left(\hat{b}+\hat{b}^\dagger\right),
\end{equation}
where the creation (annihilation) operators for the HF and RF modes are given by $\hat{a}^\dagger$ ($\hat{a}$) and $\hat{b}^\dagger$ ($\hat{b}$), respectively.
The coupling constant is given by
\begin{equation}
g_0 = \frac{\partial\omega_0}{\partial\Phi_\mathrm{b}}\Phi_\mathrm{zpf}
\end{equation}
with the tunable flux responsivity of the HF cavity $\partial\omega_0/\partial\Phi_\mathrm{b}$ and the effective root-mean-square value of the RF vacuum flux fluctuations $\Phi_\mathrm{zpf} \approx 635\,\mu\Phi_0$.
Due to the low-power operation regime used for the experiments presented here, both circuits act in good approximation as harmonic oscillators and the Kerr nonlinearities arising from the Josephson junctions can be neglected.
With the small Josephson inductance $L_\mathrm{J0} = 40\,$pH of the constriction type junctions and due to the significant dilution by linear inductors, both Kerr nonlinearities $\chi_\mathrm{HF} = 2\pi\cdot2.4\,$kHz $\ll \kappa$ and $\chi_\mathrm{RF} = 2\pi\cdot1.3\,$Hz $\ll \Gamma_0$ \cite{Gely19a} are sufficiently low to justify this approximation.
From the resonance frequency fit curves shown in Fig.~\ref{fig:device}\textbf{e}, the flux responsivity at the operation point is found to be $\partial\omega_0/\partial\Phi \approx 2\pi\cdot 250\,$MHz$/\Phi_0$, and we get a coupling rate of $g_0 = 2\pi\cdot 160\,$kHz at the operation point.
At larger flux bias values $\Phi_\mathrm{b}/\Phi_0 \sim 0.75$ the single-photon coupling rate reaches values $g_0\sim 2\pi\cdot1\,$MHz$ \,\approx \kappa$, cf. Supplementary Note~5, a regime typically very difficult to reach in other photon-pressure coupled systems.
In the current setup, however, this operation point is related to considerable low-frequency flux noise, which leads to HF cavity fluctuations and slow frequency drifts.
Therefore, we chose to work at an operation point that simultaneously offers a large single-photon coupling rate and negligible HF cavity fluctuations.
\begin{figure}
	\centerline{\includegraphics[trim = {0.0cm, 3.0cm, 0.0cm, 0.0cm}, clip=True,scale=0.52]{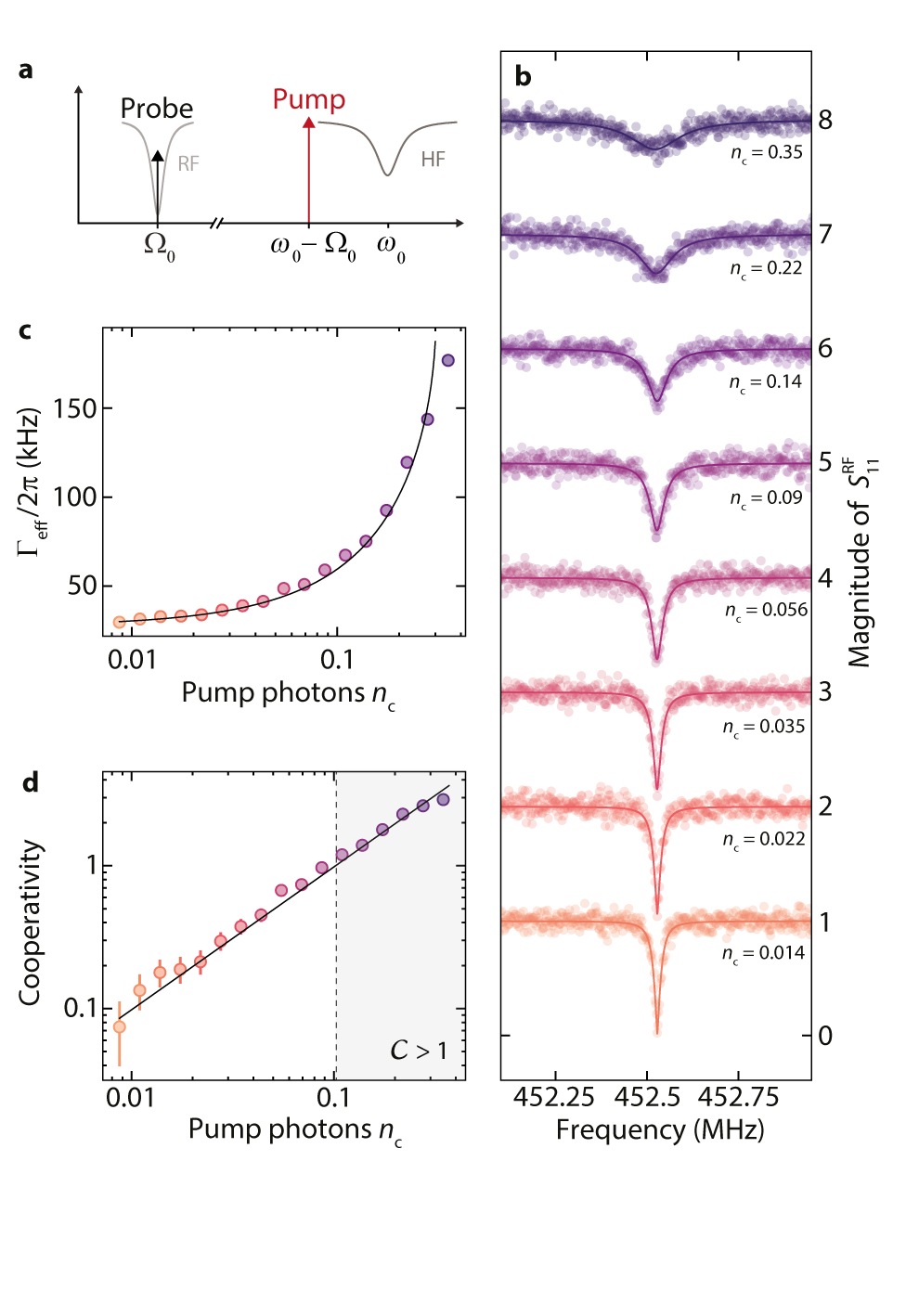}}
	\caption{\textsf{\textbf{Photon-pressure damping of the RF mode and large single-photon cooperativity.} \textbf{a} shows a schematic of the experiment. The high-frequency mode is driven by a pump tone on its red sideband $\omega_\mathrm{p} = \omega_0 - \Omega_0$ and the response of the radio-frequency mode is simultaneously measured with a weak probe tone around $\Omega \sim \Omega_0$. With increasing strength of the pump tone or intracavity pump photon number $n_\mathrm{c}$, respectively, the linewidth of the RF resonance broadens significantly as shown in panel \textbf{b}, indicating the regime of photon-pressure damping induced by the red-sideband pump field. Circles are data, lines are fits. Subsequent curves are shifted vertically by 1 for clarity. From the fits, we extract the effective RF mode linewidth $\Gamma_\mathrm{eff}$ depending on the number of intracavity pump photons. The extracted values are plotted in panel $\textbf{c}$. By fitting the data (circles) with Eq.~(6), fit curve is shown as line, we extract and quantify the multi-photon coupling strength $g$ and the cooperativity $\mathcal{C} = \frac{4g^2}{\kappa\Gamma_0}$ depending on the number of pump photons. The cooperativity extracted from the experimental data is shown as circles in \textbf{d}, the theoretical curve based on the fit in \textbf{c} is shown as line. The gray shaded area for $n_\mathrm{c} > 0.1$ indicates the regime of cooperativity $\mathcal{C}>1$. Error bars in $\textbf{d}$ correspond to a $1\,$kHz uncertainty in the bare RF linewidth $\Gamma_0$. Here,   the best agreement with the data was found with $\Gamma_0 = 2\pi\cdot27.5\,$kHz.}}
	\label{fig:coupling}
\end{figure}
Considering the parameter regime of our device with $g_0/\Omega_0 \sim 3\cdot 10^{-4}$, the photon-pressure nonlinearity \cite{Nunnenkamp11, Rabl11} induced in the HF cavity given by $2g_0^2/\Omega_0 \sim 2\pi\cdot 110\,$Hz is negligibly small.
Therefore, the interaction between the two modes with a coherently driven HF cavity can be linearized \cite{Aspelmeyer14} and the interaction part of the Hamiltonian with red-sideband driving is captured by a pump-tunable beam-splitter interaction
\begin{equation}
\hat{H}_\mathrm{int}/\hbar = g\left(\delta\hat{a}\hat{b}^\dagger +\delta\hat{a}^\dagger\hat{b} \right).
\end{equation}
Here, $g = \sqrt{n_\mathrm{c}}g_0$ is the multi-photon coupling strength and $\delta\hat{a}^\dagger$, $\delta\hat{a}$ describe the creation and annihilation of intracavity field fluctuations, respectively.
In this situation, photons from the pump will scatter mainly to the HF resonance frequency $\omega_0$, each event removing one photon from the RF circuit.
This process constitutes a cooling mechanism, which is exhibited by an additional damping term of the RF mode.
We characterize the total damping rate of the RF resonator by probing its response $S_{11}^\mathrm{RF}$ in reflection with a small probe tone while pumping the HF mode with a variable power microwave tone exactly on the red sideband $\omega_\mathrm{p} = \omega_0 - \Omega_0$.
The experimental scheme is shown in Fig.~\ref{fig:coupling}\textbf{a} and the result of the response measurement is plotted in \textbf{b} for varying HF sideband pump powers.
With increasing HF intracavity photon number $n_\mathrm{c}$, the total linewidth $\Gamma_\mathrm{eff}$ of the RF mode increases from about $2\pi\cdot30\,$kHz at low pump powers to $\sim2\pi\cdot180\,$kHz for pump powers that correspond to $n_\mathrm{c} \sim 0.4$ intracavity microwave photons.
From fits to the response data, the effective damping rate for each pump power is extracted, the result is shown in Fig.~\ref{fig:coupling}\textbf{c}.
The experimental data are fitted with the theoretical expression for the total damping
\begin{equation}
\Gamma_\mathrm{eff} = \frac{\kappa + \Gamma_0}{2} - \sqrt{\frac{\left(\kappa - \Gamma_0\right)^2}{4} - 4g^2}
\end{equation}
and as fit parameter we get the multi-photon coupling strength $g$ and subsequently the multi-photon cooperativity $\mathcal{C} = \frac{4g^2}{\kappa\Gamma_0}$.
The result is shown in Fig.~\ref{fig:coupling}\textbf{d} and demonstrates that we reach large values $\mathcal{C}>1$ for $0.1$ pump photon and a single-photon cooperativity $\mathcal{C}_0 = \frac{4g_0^2}{\kappa\Gamma_0} \approx 10$.
With the knowledge of $g$, $\Gamma_0$ and $\kappa$ for a given pump strength, the photon-pressure interaction is fully characterized.
\begin{figure*}
	\centerline{\includegraphics[trim = {0.0cm, 0.5cm, 0.0cm, 0.0cm}, clip=True, width=0.95\textwidth]{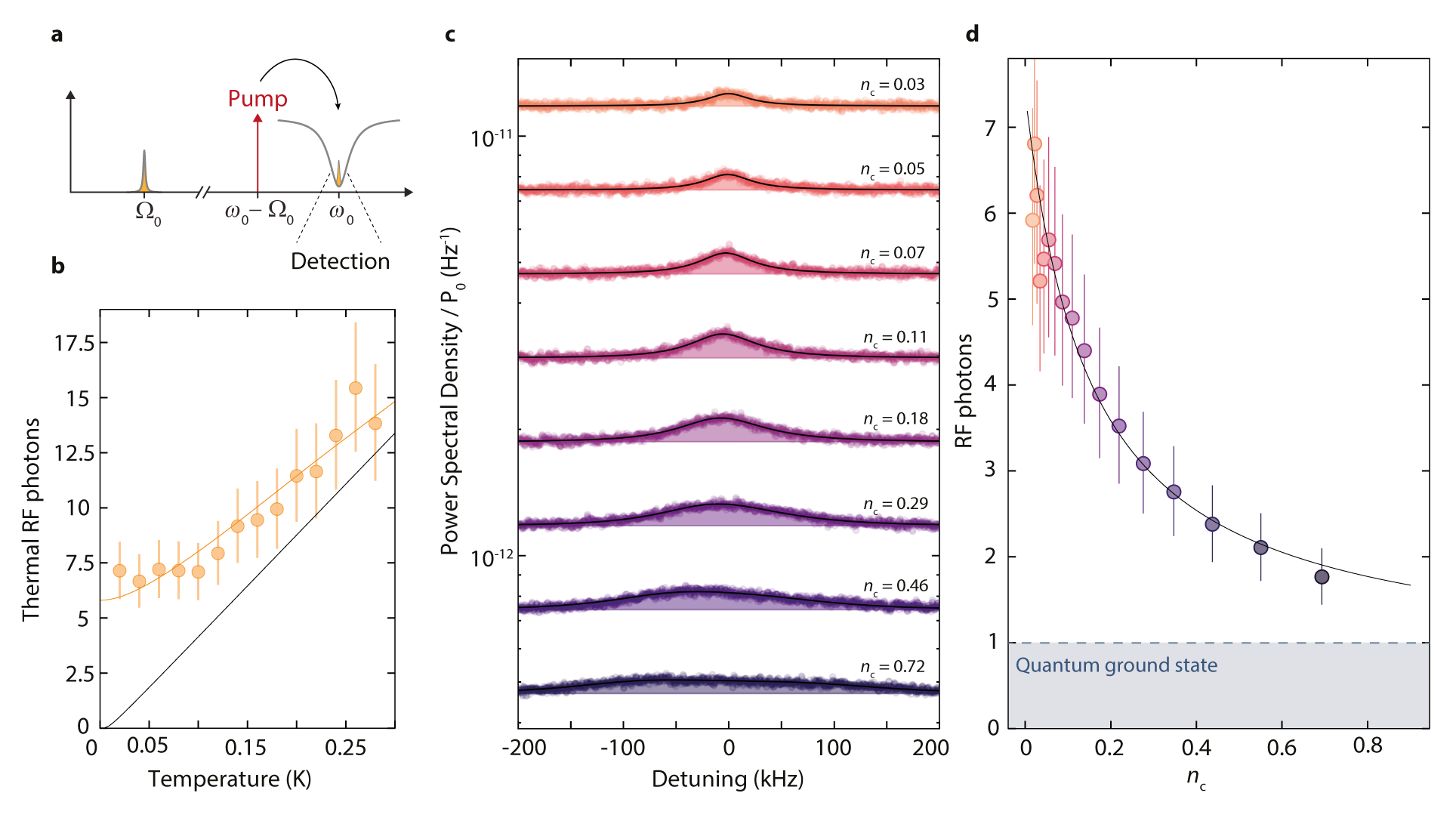}}
	\caption{\textsf{\textbf{Sideband-cooling of a hot radio-frequency resonator with less than a single pump photon.} \textbf{a} For the observation of upconverted thermal noise and cooling of the RF resonator, a pump tone is set to the red sideband of the high-frequency mode $\omega_\mathrm{p} = \omega_0-\Omega_0$ and the cavity output field around $\omega = \omega_0$ is detected with a signal analyzer. \textbf{b}, thermal RF photon number vs fridge temperature. Symbols are data, black line is the Bose factor, orange line is a more elaborate model taking into account thermal radiation on the feedline and imperfect chip thermalization, cf. Supplementary Note 7. From this thermal calibration, we determine the thermal occupation of the RF mode at fridge base temperature to be $n_\mathrm{th}^\mathrm{RF} \sim 7 \pm 1$. \textbf{c} shows the measured high-frequency output power spectral density for increasing red-sideband pump power, normalized to the on-chip pump power $P_0$. Frequency axis is given as detuning from the low-power noise center frequency. Circles are data, lines with shaded areas are fits. With increasing pump strength the RF resonance gets broadened by photon-pressure damping, and its total thermal noise power gets reduced, which corresponds to sideband-cooling of the RF mode. The slight asymmetry in the power spectral density for the largest pump powers originates from a small detuning $\delta \approx 2\pi\cdot 30\,$kHz of the pump from the red sideband, which is taken into account in our analysis. In \textbf{d}, the thermal mode occupation photon number is shown as symbols vs HF pump photon number. The initial thermal occupation is cooled by about a factor of $\sim 4$, theoretical expectation is shown as line. Error bars in \textbf{b} and \textbf{d} correspond to uncertainties of $\pm 2\,$ HF photons of added noise in the detection chain and $\pm2\,$kHz in bare RF linewidth $\Gamma_0$.}}
	\label{fig:cooling}
\end{figure*}

Without the radio-frequency probe tone applied in the previous experiment, the currents in the RF mode are given by residual thermal and quantum fluctuations.
These current fluctuations lead to resonance frequency fluctuations of the HF mode, mediated by the SQUID.
Therefore, when the HF mode is driven with a continuous frequency pump tone on the red sideband, the resonance frequency fluctuations induced by the LF mode lead to the generation of a sideband at $\omega_\mathrm{p} + \Omega_0$.
This sideband corresponds to upconverted thermal photons from the RF mode and its detection and analysis allows to determine the residual RF mode occupation.
The power spectral density at the detector (HF HEMT amplifier) input in units of quanta for a pump around the red sideband is in good approximation given by
\begin{equation}
\frac{S(\omega)}{\hbar\omega} = \frac{1}{2} + n_\mathrm{add}' + \frac{\kappa_e g^2 |\chi_0|^2 |\chi_c|^2 \Gamma_0 }{|1+g^2\chi_\mathrm{c}\chi_0|^2}n_\mathrm{th}^\mathrm{RF}
\label{eqn:PSD}
\end{equation}
with the RF mode occupation as weighted sum of internal and external bath occupations $n_\mathrm{th}^\mathrm{RF} = \frac{\Gamma_\mathrm{e}}{\Gamma_0}n_\mathrm{e}^\mathrm{RF} + \frac{\Gamma_\mathrm{i}}{\Gamma_0}n_\mathrm{i}^\mathrm{RF}$ and the HF and RF mode susceptibilities
\begin{eqnarray}
\chi_\mathrm{c}^{-1} & = & \frac{\kappa}{2} + i(\omega - \omega_0)\\
\chi_0^{-1} & = & \frac{\Gamma_0}{2} + i(\omega -\omega_\mathrm{p} - \Omega_0),
\end{eqnarray}
respectively.
For Eq.~(\ref{eqn:PSD}), we assumed that internal and external bath of the HF cavity are well equilibrated to the fridge temperature $T_\mathrm{i}^\mathrm{HF}, T_\mathrm{e}^\mathrm{HF} <100\,$mK which translates to $n_\mathrm{i}^\mathrm{HF}, n_\mathrm{e}^\mathrm{HF} \ll n_\mathrm{th}^\mathrm{RF}, n_\mathrm{add}', 1/2$,  for details cf. Supplementary Note 6 \cite{Weinstein14}.
From a thermal calibration of the RF mode occupation with varying fridge temperature, shown in Fig.~\ref{fig:cooling}\textbf{b}, we determine the residual occupation at base temperature to be $\sim 7\pm 1$ RF photons and the effective number of noise photons added by the detection chain $n_\mathrm{add}' \approx 11 \pm 2$, details are given in Supplementary Notes~7 and 8.
We note that we observe a dependence of the bare linewidth $\Gamma_0$ not only on the temperature of the mixing chamber, but also on the residual RF occupation at $T = T_\mathrm{b}$, which we attribute to two-level-system saturation in the amorphous silicon dielectric filling of the RF parallel plate capacitor.
For the fridge base temperature data in Fig.~\ref{fig:cooling}\textbf{b}, we find $\Gamma_0 \approx 2\pi\cdot40\,$kHz, an increased value compared to the values obtained from the reflection measurement.
This linewidth and $n_\mathrm{th}^\mathrm{RF} \sim 7$ correspond to a single-photon quantum cooperativity $\mathcal{C_\mathrm{q0}} = \mathcal{C}_0/n_\mathrm{th}^\mathrm{RF}\approx 1$.
The effective temperature of the RF mode occupied with 7 thermal photons is $T^\mathrm{RF} \approx 150\,$mK and thus considerably higher than the base temperature of the mixing chamber $T_\mathrm{f}\approx 15\,$mK.
We attribute this mostly to radiative noise heating through the RF input/output feedline, which is not strongly isolated from the cryogenic RF amplifier mounted in between the $800\,$mK and the $3.2\,$K plates, but we can also not completely exclude a small contibution from imperfect thermalization of the device to the mixing chamber.
This interpretation is supported by the significant increase of RF mode occupation when the cryogenic RF amplifier is switched on, in which case we find $n_\mathrm{th}^\mathrm{RF}\sim 21$.
For the thermal calibration and the cooling experiment presented in Fig.~\ref{fig:cooling}, however, the amplifier is switched off.
Additional thermal calibration and cooling data with the RF amplifier switched on can be found in Supplementary Note~8.
With the fridge temperature set back to its minimal value $T_\mathrm{b} = 15\,$mK, we measure the HF mode output spectra for varying red-sideband pump power, cf.  Fig.~\ref{fig:cooling}\textbf{c}.
For the smallest pump power shown, the upcoverted thermal noise spectrum displays a Lorentzian lineshape with an effective linewidth $\Gamma_\mathrm{eff} \approx 2\pi\cdot 65\,$kHz, broadened by dynamical backaction.
With increasing sideband pump power the thermal noise peak broadens further, until for the largest powers the lineshape deviates from a Lorentzian due to the onset of normal-mode splitting.
Additional spectra for a larger residual RF occupation $n_\mathrm{th}^\mathrm{RF} \sim 21$ are given in Supplementary Note~8. 
By fitting the spectra with Eq.~(\ref{eqn:PSD}), shown as lines and shaded areas in Fig.~\ref{fig:cooling}\textbf{c}, the equilibrium RF photon numbers are determined and converted to the sideband-cooled photon occupation
\begin{eqnarray}
n_\mathrm{cool}^\mathrm{RF} & = & n_\mathrm{th}^\mathrm{RF}\frac{\Gamma_0}{\kappa+\Gamma_0}\frac{4g^2 + \kappa(\kappa + \Gamma_0)}{4g^2 + \kappa\Gamma_0}\nonumber\\
& & +~ n_\mathrm{th}^\mathrm{HF}\frac{\kappa}{\kappa+\Gamma_0}\frac{4g^2}{4g^2 + \kappa\Gamma_0}.
\label{eqn:CooledPhotons}
\end{eqnarray}
Note that this relation differs from the result usually quoted in optomechanics \cite{Dobrindt08, Teufel11, Aspelmeyer14}, which is only valid for $\kappa \gg \Gamma_0$ and significantly underestimates the cooling rate in the unusual regime $\Gamma_0 \lesssim \kappa$.
The full equation taking into account also finite pump detunings $\delta$ can be found at the end of Supplementary Note~6.
\begin{figure*}
	\centerline{\includegraphics[trim = {0.0cm, 0.0cm, 0.0cm, 0.0cm}, clip=True,width=0.95\textwidth]{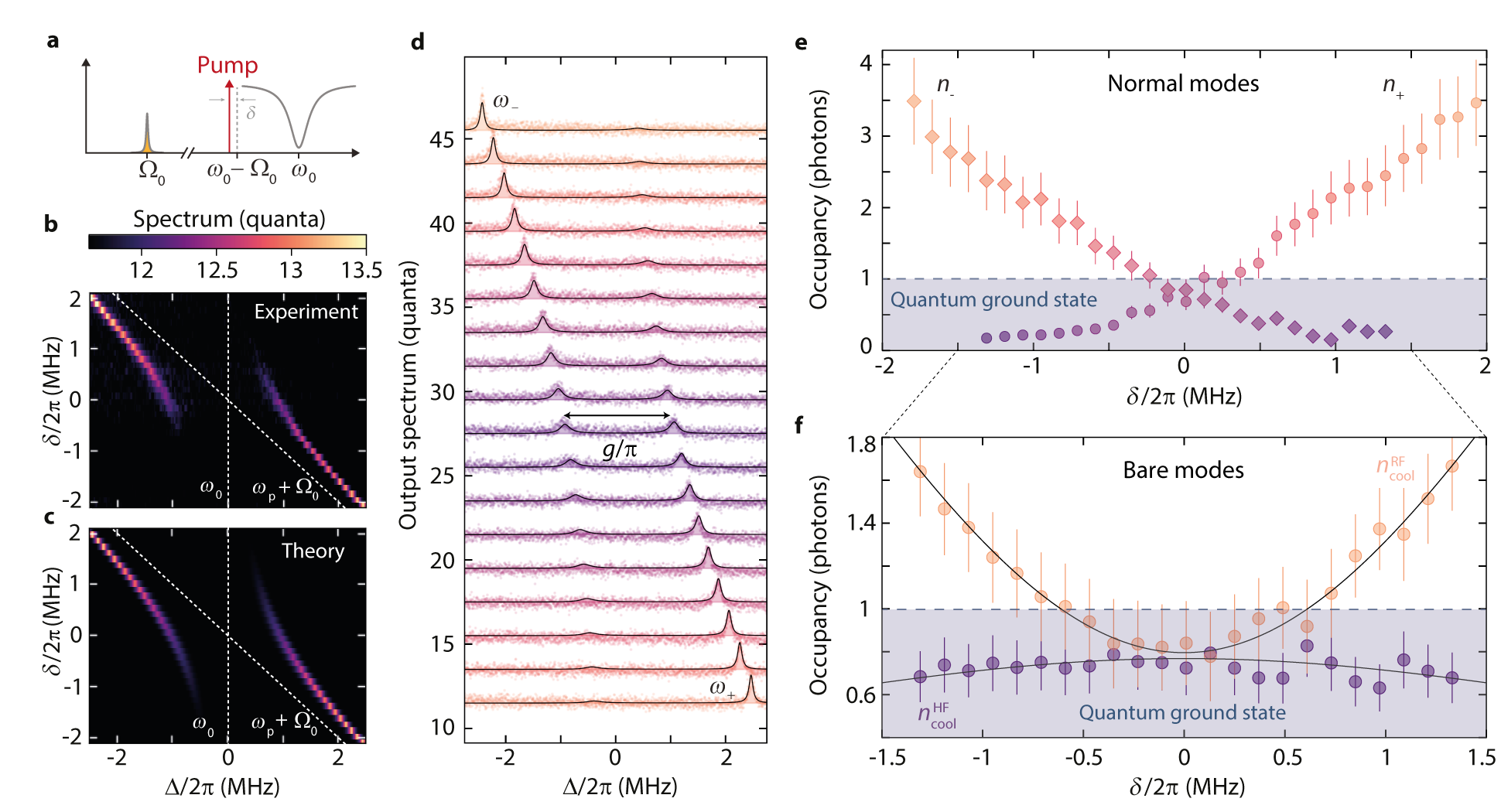}}
	\caption{\textsf{\textbf{Normal-mode thermometry and ground-state cooling in the quantum coherent strong-coupling regime.} \textbf{a} Experimental scheme. A strong pump tone is applied with detuning $\delta=\omega_\mathrm{p} - (\omega_0 - \Omega_0)$ from the red sideband of the HF mode. For each pump detuning $\delta$, the output spectrum around the bare HF mode resonance frequency $\omega_0 + \Delta$ is measured. The resulting power spectra are plotted color-coded in $\textbf{b}$. \textbf{c} shows the result of calculations based on the theoretical model. In \textbf{d}, we plot the corresponding linecuts from \textbf{b} and \textbf{c} on top of each other, displaying a high level of agreement between experiment and theory. The bottom data are displayed with an unmodified background level of $11.5$ and subsequent lines are offset by $+2$ quanta each. In the regime $\delta = 0$, the splitting between the modes is given by $g/\pi$ with $g = 2\pi\cdot{1}\,$MHz. We treat the two parametrically coupled normal-modes as individual HF modes with resonance frequencies $\omega_-$ and $\omega_+$ and extract the corresponding, effective thermal photon numbers $n_-$ and $n_+$ from the spectrum shown in \textbf{d}. The values are plotted in \textbf{e}. From the normal-mode occupations $n_\pm$, we determine the occupation of the bare HF and RF modes shown in \textbf{f}. Around $\delta = 0$, both bare modes are in the quantum groundstate with residual occupations $n_\mathrm{cool}^\mathrm{HF} \approx n_\mathrm{cool}^\mathrm{RF} \sim 0.8 \pm 0.2$. For the theoretical calculations, shown as lines, we assume $n_\mathrm{th}^\mathrm{HF} = 0.01$ and find as equilibrium thermal occupations $n_\mathrm{th}^\mathrm{RF} = 8.0$. Error bars in \textbf{e} and \textbf{f} correspond to uncertainties of $\pm 2\,$ HF photons of added noise in the detection chain and $\pm2\,$kHz in $\Gamma_0$.}}
	\label{fig:NMT}
\end{figure*}
The resulting sideband-cooled RF mode occupation is shown in Fig.~\ref{fig:cooling}\textbf{d}.
Therefore, with less than a single pump photon, the RF mode is cooled by about a factor of $\sim 4$ to an occupation of only $1.7$ RF quanta, demonstrating the applicability of sideband-cooling for photon-pressure coupled circuits and an extraordinarily large single-photon cooling rate.
At higher pump powers than the ones discussed so far, the two circuits hybridize in the parametric normal-mode splitting regime \cite{Dobrindt08, Teufel11, Teufel11a, Bothner20}.
In the strong-coupling regime, i.e, when the frequency-splitting of the normal modes exceeds the hybridized linewidths, the residual thermal occupation of the RF mode is distributed between the hybridized normal-modes.
Equation~(\ref{eqn:CooledPhotons}), however, remains valid and the onset of mode hybridization does not prevent the RF mode from being cooled further.
The theoretical limit for cooling in the regime $g \gg \kappa, \Gamma_0$ is given by $n_\mathrm{lim}^\mathrm{RF} = n_\mathrm{th}^\mathrm{RF}\Gamma_0/(\kappa + \Gamma_0)\approx 0.67$ photons, assuming a ground-state HF cavity.
The remaining thermal excitations of the system are then equally distributed between the RF and the HF modes, cf. also Supplementary Note~9.
To characterize the residual number of thermal RF photons in the strong-coupling regime, we detect the output noise of the normal-modes in the HF domain for varying detuning of the pump tone from the red sideband $\delta$ .
In Fig.\ref{fig:NMT}, the measured output spectra around $\Delta = \omega - \omega_0$ are shown color-coded in panel \textbf{b} and as individual linescans in panel \textbf{d}.
We find an excellent agreement between the data and theoretical calculations, shown color-coded in \textbf{c} and as lines in \textbf{d}.
For large detunings $|\delta|>2\pi\cdot2\,$MHz, a hot mode with a large noise amplitude is observed, whose frequency follows closely $\omega_\mathrm{p} + \Omega_0$ and corresponds to the normal-mode dominated by the RF circuit in this regime.
At the same time, no output noise field is detected around the HF-cavity-like normal mode close to $\Delta = 0$, indicating that this mode is cold and in thermal equilibrium with its bath.
For small detunings $\delta \sim 0$, we observe a pronounced avoided crossing of the RF mode with the driven HF cavity, centered at $\Delta = 0$.
The splitting between the two hybridized modes is given by $g/\pi \approx 2\,$MHz, indicating that we reach the so-called quantum-coherent coupling regime where $g > \kappa, \Gamma_0 n_\mathrm{th}^\mathrm{RF}$ \cite{Verhagen12} and a quantum cooperativity $\mathcal{C}_\mathrm{q} = \mathcal{C}/n_\mathrm{th}^\mathrm{RF} \approx 35$.
For a quantification of the effective normal-mode thermal occupation, we treat the modes as two independent HF modes, a detailed description is given in Supplementary Notes~9 and 10.
The lower frequency mode has the resonance frequency $\omega_-$, linewidth $\kappa_-$ and external linewidth $\kappa_{\mathrm{e}-}$, the higher frequency mode $\omega_+$, $\kappa_+$ and $\kappa_{\mathrm{e}+}$, respectively.
The complex resonances of the normal modes are given by \cite{Dobrindt08}
\begin{equation}
\tilde{\omega}_\pm = \omega_0 + \frac{\delta}{2} + i\frac{\kappa + \Gamma_0}{4} \pm \sqrt{g^2 - \left(\frac{\kappa - \Gamma_0 + 2i\delta}{4}\right)^2}
\end{equation}
and the resonance frequencies and linewidths are $\omega_\pm = \mathrm{Re}[\tilde{\omega}_\pm]$ and $\kappa_\pm = 2\mathrm{Im}[\tilde{\omega}_\pm]$, respectively.
The power spectral density of the HF output field in terms of these normal-mode parameters can be written as
\begin{equation}
\frac{S_\mathrm{nms}}{\hbar\omega} = \frac{1}{2} + n_\mathrm{add}' + 4\frac{\kappa_{\mathrm{e}-}\kappa_-}{\kappa_-^2 + 4\Delta_-^2}n_- + 4\frac{\kappa_{\mathrm{e}+}\kappa_+}{\kappa_+^2 + 4\Delta_+^2}n_+.
\label{eqn:Snms}
\end{equation}
where $n_\pm$ are the effective photon occupations of the two normal modes, $\Delta_\pm = \omega - \omega_\pm$ and the external linewidths $\kappa_{\mathrm{e}\pm} = \frac{\kappa_\mathrm{e}}{2}\left(1\pm\frac{\delta}{\sqrt{\delta^2 + 4g^2}}\right)$, cf. Supplementary Notes~9 and 10.
The effective normal-mode occupation $n_-$ and $n_+$ depending on the pump detuning is shown in Fig.~\ref{fig:NMT}\textbf{e}.
While for large detunings the RF-like normal mode is still hot and the HF-like mode is in the quantum ground-state, both normal modes appear to be in the quantum ground-state when they are close to the full mode hybridizaton at $\delta \sim 0$.
The minimum occupation that we observe at the symmetry point is $n_- = n_+ = 0.8$.
From comparison between the two versions of the HF power spectral densities Eq.~(\ref{eqn:Snms}) and Eq.~(\ref{eqn:PSD}) and the condition $S_\mathrm{nms} = S$, we obtain the HF mode occupation in the cooling regime $n_\mathrm{cool}^\mathrm{HF}$ by 
\begin{equation}
\kappa_\mathrm{e}n_\mathrm{cool}^\mathrm{HF} = \kappa_{\mathrm{e}-}n_- + \kappa_{\mathrm{e}+}n_+,
\end{equation}
For zero pump detuning, this simplifies to $n_\mathrm{cool}^\mathrm{HF} = n_- = n_+ = 0.8$.
Using the equation for the total thermal occupation $n_\mathrm{cool}^\mathrm{tot}$ of the system in the strong-coupling regime (cf. Supplementary Note~9), we calculate the residual occupation of the RF mode $n_\mathrm{cool}^\mathrm{RF} = n_\mathrm{cool}^\mathrm{tot} - n_\mathrm{cool}^\mathrm{HF} $.
The resulting occupation of both bare modes is shown in Fig.~\ref{fig:NMT}\textbf{f}, showing that $n_\mathrm{cool}^\mathrm{RF} \approx n_\mathrm{cool}^\mathrm{HF} = 0.8 \pm 0.2$, i.e., that both are in the quantum ground-state for $\delta \sim 0$.

\section*{Discussion}

In future devices, both the cooling factor and the thermal noise detection efficiency could be considerably improved by reducing the external linewidth of the HF cavity by one to two orders of magnitude.
A correspondingly optimized device might also be suitable to generate quantum-squeezed radio-frequency states or to entangle distinct radio-frequency circuits, similar to what has been reported for optomechanical devices \cite{Wollman15, OckeloenKorppi18}
If at the same time the HF cavity flux noise can be reduced by, for example, improved shielding and/or lower-noise current sources, an HF cavity operation point with larger $g_0$ can be chosen, while maintaining the high values for the single-photon cooperativity reported in the current work.
With the results presented in this work, we demonstrated photon-pressure coupling of a hot radio-frequency circuit to a superconducting microwave cavity in the quantum regime.
By a galvanically connected circuit design, we dramatically increased the single-photon coupling strength and achieved a single-photon quantum cooperativity of unity.
Based on the large single-photon coupling rate, we were able to demonstrate both, sideband-cooling of the RF mode by a factor of $4$ and the strong-coupling regime, with less than a single pump photon.
For stronger pump powers, we enter the quantum-coherent coupling regime and demonstrate photon-pressure groundstate cooling of the originally hot RF mode.
Compared to other recently developed radiative cooling techniques of circuits and other systems \cite{Xu20a, Albanese20}, sideband-cooling can reduce the effective mode temperature far below the physical temperature of any bath \cite{Yuan15}.
Furthermore, in contrast to previous reports of sideband-cooling techniques with circuits using highly nonlinear systems such as superconducting qubits \cite{Valenzuela06, Gely19}, our approach allows for both participating circuits to have a very high degree of linearity which is highly desirably for many signal processing applications.
This work lays the foundation for radio-frequency quantum photonics, for quantum-limited RF sensing and has potential applications in quantum-limited microwave signal and bosonic code quantum information processing based on photon-pressure coupled circuits.

\section*{Materials and Methods}
The photon-pressure system used in this experiment consists of two galvanically connected lumped-element LC circuits and it was engineered via a multi-layer nanofabrication process. The HF mode of the circuit comprises two interdigitated capacitors, a SQUID containing two Josephson junctions and two linear inductors. Aside from the firstly patterned $50$~nm wide, $100$~nm long and $15$~nm thick aluminum nanobridge junctions and their respective $500 \times 500\, \textrm{nm}^2$ contact pads, the circuit is made of a $\sim 70$ nm thick aluminum layer on a silicon substrate. Both layers were patterned via a sputtering deposition in combination with electron beam lithography and liftoff. In addition, an argon milling process ($\sim 2$ minutes) was performed in-situ prior to the second deposition step to provide good electrical contact between the two layers. The RF mode is formed by a parallel plate capacitor with a $\sim 130$ nm thick amorphous silicon layer as dielectric and a short inductor wire, that simultaneously acts as the loop of the SQUID. The inductor wire was patterned together with the bottom capacitor plate and the HF circuit mode components. Subsequent to this step the dielectric deposition takes place, i.e. a PECVD (Plasma-Enhanced Chemical Vapor Deposition) process followed by a reactive ion etching step and $\textrm{O}_2$ plasma ashing. Lastly, we patterned the top capacitor plate. This one is made of a $250$ nm aluminum layer and it was fabricated via another sputtering-liftoff procedure, which once again included an in-situ argon milling in order to guarantee good contact between the plates. A step-by-step description of the device fabrication is given in Supplementary Note 1.

\subsection*{Acknowledgements}
\vspace{-2mm}

This research was supported by the Netherlands Organisation for Scientific Research (NWO) in the Innovational Research Incentives Scheme -- VIDI, project 680-47-526.
This project has received funding from the European Research Council (ERC) under the European Union's Horizon 2020 research and innovation programme (grant agreement No 681476 - QOMD) and from the European Union's Horizon 2020 research and innovation programme under grant agreement No 732894 - HOT.

\subsection*{Author contributions}
\vspace{-2mm}

ICR and DB designed and fabricated the device, performed the measurements, analysed the data and developed the theoretical treatment.
GAS conceived the experiment and supervised the project.
ICR and DB edited the manuscript with input from GAS.
All authors discussed the results and the manuscript.

\subsection*{Competing interest}
\vspace{-2mm}
The authors declare no competing interests.

\subsection*{Data availability}
\vspace{-2mm}
All data and processing scripts of the results presented in this paper, including those in the Supplementary Information, are available on Zenodo with the identifier https://doi.org/10.5281/zenodo.4733659.

\clearpage

\widetext

\noindent\textbf{\textsf{\Large  Supplementary Material for: Cooling photon-pressure circuits into the quantum regime}}

\normalsize
\vspace{.3cm}

\noindent\textsf{I.C.~Rodrigues$^\dagger$, D.~Bothner$^\dagger$, and G.~A.~Steele}

\vspace{.2cm}
\noindent{$^\dagger$these authors contributed equally}

\renewcommand{\thefigure}{S\arabic{figure}}
\renewcommand{\theequation}{S\arabic{equation}}

\renewcommand{\thesection}{S\arabic{section}}
\renewcommand{\bibnumfmt}[1]{[S#1]}
\renewcommand{\citenumfont}[1]{S#1}

\newpage

\setcounter{figure}{0}
\setcounter{equation}{0}

\section*{Supplementary Note 1: Device fabrication}
\label{Section:Fab}
\begin{figure}[h]
	\centerline {\includegraphics[trim={0.5cm 6cm 0.5cm 0cm},clip=True,scale=0.75]{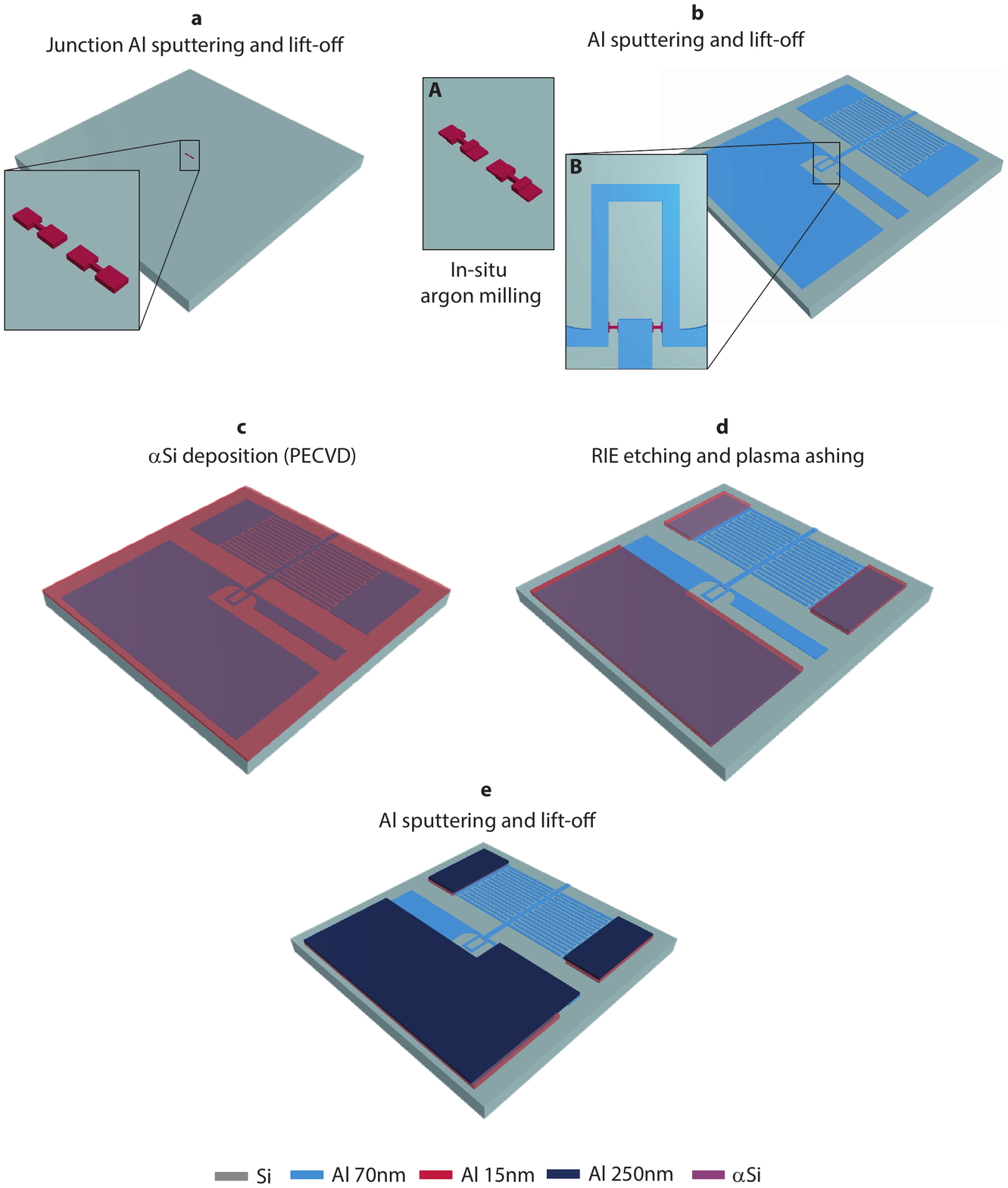}}
	\caption{\textsf{\textbf{Schematic device fabrication.} \textbf{a} shows the weak-link Josephson junctions with contact pads, patterned in the first fabrication step. \textbf{b} shows the patterned second Aluminum layer, forming the bottom of the RF parallel plate capacitor, the SQUID loop and the HF cavity. Inset \textbf{A} showing the in-situ argon milled Josephson junctions prior to the deposition (the existing resist is not shown for better visibility of the milled structures). Inset \textbf{B}  shows a zoom-in of the 3D SQUID. \textbf{c} shows the sample after the deposition of $\alpha$Si. \textbf{d} shows the device after the subsequent SF$_6/$He reactive ion etching step, finished by an in-situ $\textrm{O}_{2}$ plasma ashing. \textbf{e} shows the final device after the deposition of the last Aluminum layer. }}
	\label{fig:Fab}
\end{figure}

\begin{itemize}
	\item \textbf{Step 0: Marker patterning.} Prior to the device fabrication, we perform the patterning of alignment markers on a full $4\,$inch Silicon wafer, required for the electron-beam lithography (EBL) alignment of the following fabrication steps.
	The structures were patterned using a CSAR62.13 resist mask and sputter deposition of $50\,$nm Molybdenum-Rhenium alloy.
	After undergoing a lift-off process, the only remaining structures on the wafer were the markers.
	The complete wafer was diced into $14\times14\,$mm$^2$ chips, which were used individually for the subsequent fabrication steps.
	The step was finalized by a series of several acetone and IPA rinses.
	\item \textbf{Step 1: Junctions patterning.} As first step in the fabrication, we pattern weak links which afterwards result in constriction type Josephson junctions between the arms of the SQUID.
	The weak link nanowires were patterned together with larger pads, cf. Supplementary Fig.~\ref{fig:Fab}\textbf{a}, which were used to achieve good electrical contact with the rest of the circuit, cf. Step~3.
	The nanowires are designed to be $\sim 50\,$nm wide and $\sim 100\,$nm long at this point of the fabrication, and each pad is $500\times500\,$nm$^2$ large.
	For this fabrication step, a CSAR62.09 was used as EBL resist and the development was done by dipping the exposed sample into Pentylacetate for $60\,$seconds, followed by a solution of MIBK:IPA (1:1) for $60\,$seconds, and finally rinsed in IPA, where MIBK is short for methyl isobutyl ketone and IPA for isopropyl alcohol. 
	The sample was subsequently loaded into a sputtering machine where a $15\,$nm layer of Aluminum was deposited.
	Finally, the chip was placed at the bottom of a beaker containing a small amount of Anisole and inserted into an ultrasonic bath for a few minutes where the sample underwent a lift-off process.
	The step was finalized by a series of several acetone and IPA rinses.
	\item \textbf{Step 2: Bottom RF capacitor plate and HF resonator patterning.} As second step in the fabrication, we pattern the bottom plate of the parallel plate capacitor, the inductor wire of the radio-frequency cavity, which also forms part of the  SQUID loop, the remaining part of the SQUID cavity (cf. Supplementary Fig.~\ref{fig:Fab}\textbf{b}.) and the center conductor of the SQUID cavity feedline by means of EBL using CSAR62.13 as resist. 
	After the exposure, the sample was developed in the same way as in the first fabrication step and loaded into a sputtering machine.
	In the sputter system, we performed an argon milling step for two minutes and afterwards deposited $70\,$nm of Aluminum.
	The milling step, performed in-situ and prior to the deposition, very efficiently removes the oxide layer which was formed on top of the previously sputtered weak link pads, and therefore allows for good electrical contact between the two layers.
	After the deposition, the unpatterned area was lifted-off by means of an ultrasonic bath in room-temperature Anisole for a few minutes. 
	The step was finalized by a series of several acetone and IPA rinses.
	\item \textbf{Step 3: Amorphous silicon deposition.} The deposition of the dielectric layer of the parallel plate capacitor was done using a plasma-enhanced chemical vapor deposition (PECVD) process.
	To guarantee low dielectric losses in the material, the chamber underwent an RF cleaning process overnight and only afterwards the deposition of $\sim130\,$nm of amorphous silicon was performed.
	At this point of the fabrication, the whole sample is covered with dielectric, cf. Supplementary Fig.~\ref{fig:Fab}\textbf{c}.
	\item \textbf{Step 4: Reactive ion etch patterning of $\alpha$Si.} We spin-coat a double layer of resist (PMMA 950K A4 and ARN-7700-18) on top of the $\alpha$Si-covered sample, and expose the next pattern with EBL. 
	Prior to the development of the pattern, a post-bake of 2 minutes at $\sim115\,^{\circ}{\rm C}$ was required.
	Directly after, the sample was dipped into MF-321 developer for 2 minutes and 30 seconds, followed by H$_2$O for 30 seconds and lastly rinsed in IPA.
	To finish the third step of the fabrication, the developed sample underwent a SF$_6$/He reactive ion etching (RIE) to remove the amorphous Silicon.
	To conclude the etching step, we performed a O$_2$ plasma ashing in-situ with the RIE process to remove resist residues, the result is shown schematically in Supplementary Fig.~\ref{fig:Fab}\textbf{d}.
	\item \textbf{Step 5: Top capacitor plate and ground-plane patterning.} As final step, the sample was again coated in CSAR62.13 and the top plate of the RF capacitor as well as all ground plane and the low-frequency feedline was patterned with EBL.
	The resist development was done identical to the ones in the second and third steps.
	Afterwards, the sample was loaded into a sputtering machine where an argon milling process was performed in-situ for 2 minutes, in order to have good electrical contact between the top and bottom plates of the low-frequency capacitor, similar to what was done between the second and third fabrication steps.
	After the milling, a $250\,$nm layer of Aluminum was deposited and finally an ultrasonic lift-off procedure was performed.
	The step was finalized by a series of several acetone and IPA rinses.
	With this, the sample fabrication process was essentially completed, cf. Supplementary Fig.~\ref{fig:Fab}\textbf{e}.
	\item \textbf{Step 6: Dicing and mounting.} At the end of the fabrication, the sample was diced to a $10\times10\,$mm$^2$ size and mounted to a printed circuit board (PCB), wire-bonded to microwave feedlines and ground and packaged into a radiation tight copper housing.
\end{itemize}

A schematic representation of this fabrication process can be seen in Supplementary Fig.~\ref{fig:Fab}, omitting the initial patterning of the electron beam markers and the sample mounting.
In addition, an optical image of the full device is shown in Supplementary Fig.~\ref{fig:DevicePhoto}.
\clearpage
\begin{figure}[h]
	\centerline {\includegraphics[trim={0cm 0cm 0cm 0cm},clip=True,scale=0.75]{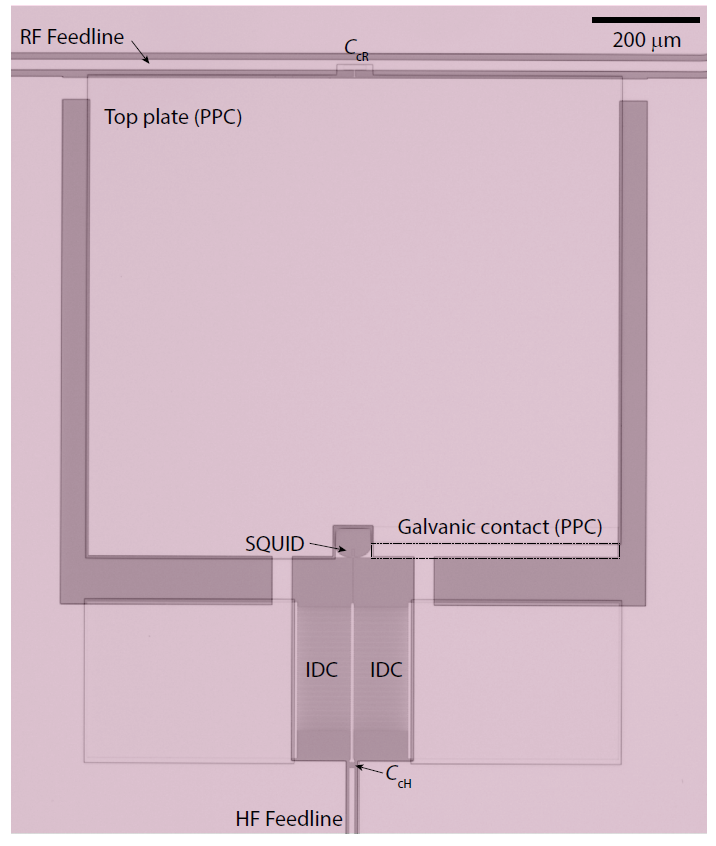}}
	\caption{\textsf{\textbf{Optical image of the full device.} Visible are both, the radio-frequency (top) and microwave (bottom) resonators including their corresponding RF and HF feedlines. The coupling capacitors are labelled with $C_\mathrm{cR}$ for the RF mode and $C_\mathrm{cH}$ for the HF mode, respectively. Also labelled are the RF parallel plate capacitor (PPC), the HF interdigitated capacitors (IDC), and the SQUID for orientation. Zoom-ins to the HF mode circuit including labels for the inductors are shown in main paper Fig.~1. The galvanic contact area of the PPC top and bottom plates is marked with a dashed rectangle.}}
	\label{fig:DevicePhoto}
\end{figure}
\section*{Supplementary Note 2: Measurement setup}

\begin{figure}[h]
	\centerline {\includegraphics[trim={0cm 6cm 0cm 0.5cm},clip=True,scale=0.8]{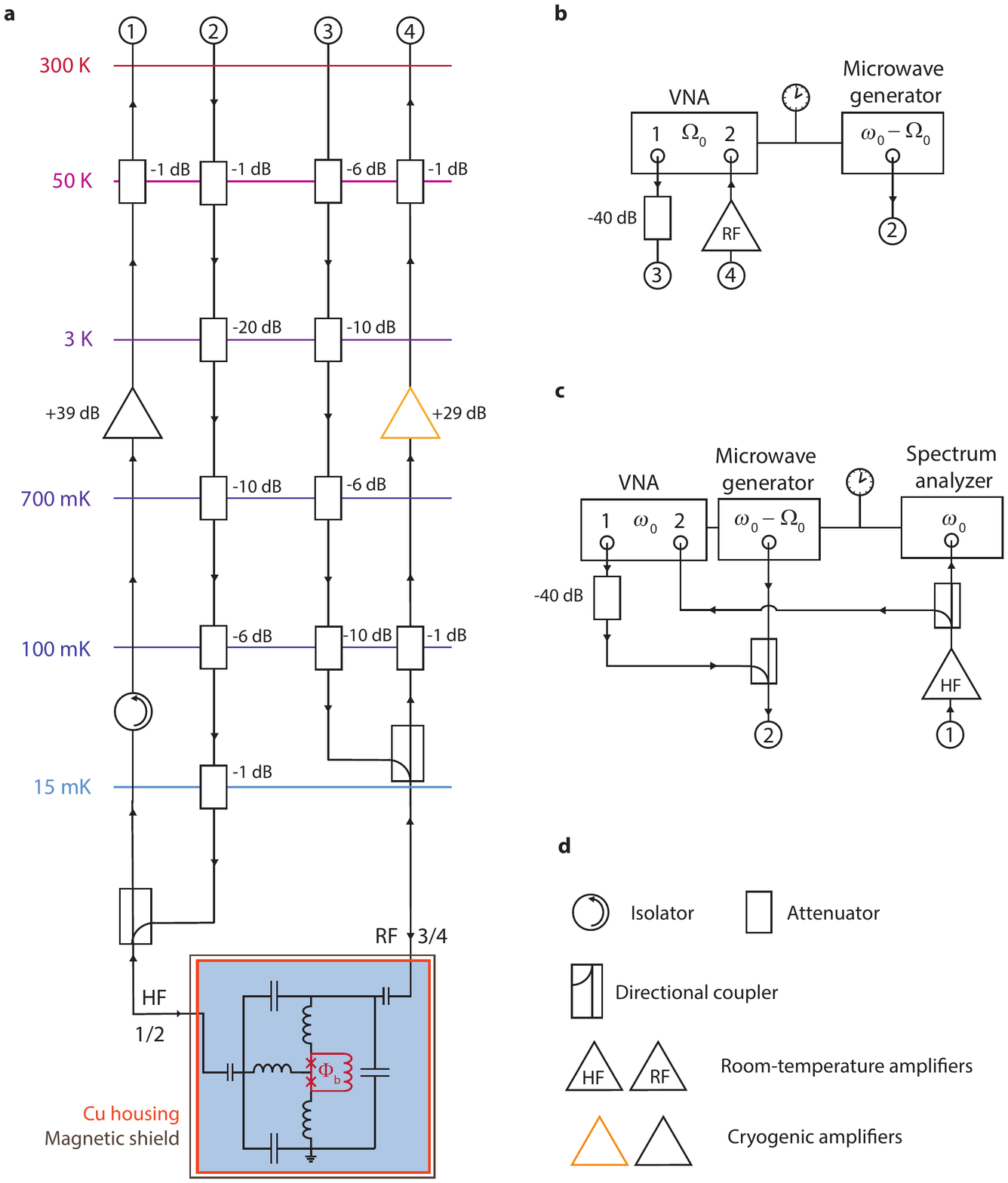}}
	\caption{\textsf{\textbf{Schematic of the measurement setup.} Detailed information is provided in text.}}
	\label{fig:Setup}
\end{figure}
All the experiments reported in this paper were performed in a dilution refrigerator operating at a base temperature close to $T_\mathrm{b} = 15\,$mK.
A schematic of the experimental setup and of the external configurations used in the different performed experiments can be seen in Supplementary Fig.~\ref{fig:Setup}.

The printed circuit board (PCB), onto which the fabricated sample was glued and wire-bonded, was placed in a radiation tight copper housing and connected to two coaxial lines. 
One of the lines was used as input/output port for the high-frequency (HF) SQUID cavity and the second line was set in a similar way for the radio-frequency resonator.
Both of the cavities were measured in a reflection geometry, and therefore the input and output signals were split via a directional coupler.
For the HF line, the directional coupler was positioned on the $15\,$mK stage, while for the RF line it was mounted in between the $15\,$mK plate and the $100\,$mK plate. 
Both output signals went into a cryogenic amplifier for their particular frequency range.
Furthermore, in order to generate an out-of-plane magnetic field, required to flux bias the SQUID cavity, an external magnet (not shown in the figure) was put in very close proximity below the device and the two were placed inside a cryoperm magnetic shield.
The magnet was connected with DC wires, allowing for the field to be tuned by means of a DC current (not shown).
Both input lines were heavily attenuated in order to balance the thermal radiation from the line to the base temperature of the fridge. 
The low-frequency line, however, is not fully equilibrated to the fridge base temperature due to the lack of cryogenic circulators/isolators for the particular frequency range.
Outside of the refrigerator, we used different configurations of microwave signal sources and high-frequency electronics for the different experiments.
In \textbf{b} we show the configuration used to measure the photon-pressure damping of the radio-frequency mode (main paper Fig.~2).
A microwave generator sends a continuous wave signal to the SQUID cavity around its red sideband, while the RF resonator is probed in reflection with a vector network analyzer (VNA). 

In \textbf{c} we show the setup for photon-pressure sideband-cooling experiment and the normal-mode thermometry (main paper Figs.~3 and 4), where a continuous wave tone is send to the red sideband of the SQUID cavity.
In addition, in order to observe the cavity response and adjust the pump tone frequency with respect to the power-dependent cavity resonance, a weak VNA signal is combined with the pump tone via a directional coupler.
The output signal is analyzed individually by a spectrum analyzer and a VNA after being amplified.
During the detection of thermal noise with the signal analyzer, the VNA scan was stopped and the VNA output power was completely switched off.
For the normal-mode thermometry experiment, we replaced the $20\,$dB attenuator on the $3\,$K plate of HF input line 2 by a $10\,$dB one.
This allowed for stronger red-sideband pumping and therefore for reaching deeper into the strong-coupling regime.
For all experiments, the microwave sources and vector network analyzers (VNA) as well as the spectrum analyzer used a single reference clock of one of the devices.

\subsection*{Power Calibration}

In order to estimate the input power on the on-chip high-frequency feedline of the device, we use the thermal noise of the HF HEMT (High-Electron-Mobility Transistor) amplifier as calibration method.
The cryogenic HEMT amplifier thermal noise power is given by
\begin{equation}
P_\mathrm{HEMT} = 10\,\log\left(\frac{k_\mathrm{B} T_{\mathrm{HEMT}}  \Delta f}{1\,\textrm{mW}}\right)
\label{eq:HEMT}
\end{equation}
where $k_\mathrm{B}$ is the Boltzmann constant, $T_{\mathrm{HEMT}}$ is the noise temperature of the amplifier, which, according to the specification datasheet, is approximately $2\,$K, and $\Delta f = 2000\,$Hz is the measurement IF bandwidth.
The calculated noise power is $P_\mathrm{HEMT} = -162.58\,$dBm, or as noise RMS voltage $\Delta V = 1.66\,$nV.
\begin{figure}[h!]
	\centerline {\includegraphics[trim={2cm 2.5cm 0cm 2.5cm},clip=True,scale=0.7]{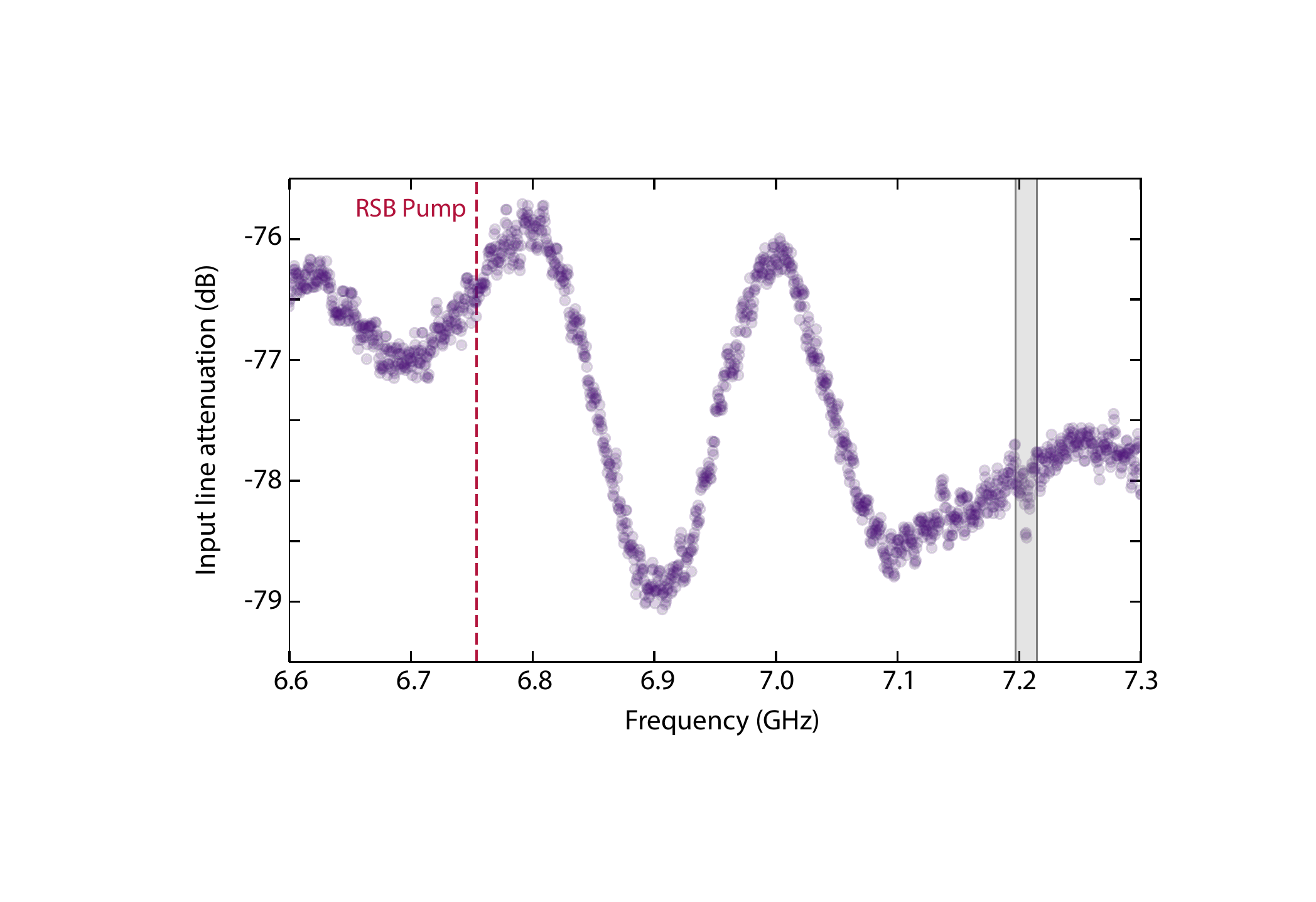}}
	\caption{\textsf{\textbf{Estimation of the frequency-dependent input line attenuation for the pump tone.} The data are obtained by measuring $501$ traces in the shown frequency range using the vector network analyzer, cf. Supplementary Fig.~\ref{fig:Setup}. For each frequency point, we determine from the $501$ traces the signal-to-noise ratio and with the assumption of a frequency-independent HEMT noise temperature and $2.7\,$dB losses between the sample and the HEMT, we get the input line attenuation as plotted. The gray area shows where the cavity was during the calibration. The frequency of the red sideband (RSB) pump tone utilized for the experiments reported in this work is indicated by the red dashed line. The input line attenuation estimated for that point is $\sim 76.5\,$dB.}}
	\label{fig:NoiseCal}
\end{figure}

Taking into account the room temperature attenuators of $40\,$dB as well as the directional coupler ($20\,$dB loss on the coupled port) for the VNA tone and assuming an attenuation between the sample and the HEMT of $2.7\,$dB (based on the effective added noise obtained from the thermal calibration, see Supplementary Note 7), we extract a frequency-dependent input line attenuation as shown in Supplementary Fig.~\ref{fig:NoiseCal}.
The input line attenuation for the frequency of our red sideband pump is $\sim 76.5\,$dB.
The deviation between this value and one extracted from the effects of photon-pressure dynamical backaction (cf. Fig.~2 of the main paper) is as small as $0.3\,$dB.

\section*{Supplementary Note 3: The circuit model and flux dependence}

\subsection*{The circuit model}

The diagram shown in Supplementary Fig.~\ref{fig:CM} represents the full circuit model of the device.
Similar to the simplified version shown in Fig.~1\textbf{a} of the main paper, it contains the high-frequency (drawn in purple) and a radio-frequency (drawn in orange) mode, which share the center part of the circuit (drawn in gray).
The shared part contains a non-linear, flux-tunable SQUID inductance.
The Josephson junctions which form part of the tunable SQUID are constriction type Josephson junctions, which are known to have a current-phase relation (CPR) that can differ significantly from the typical sinusoidal CPR [49,~50].
To include this effect in the flux dependence of the modes, we model each weak-link inductance to be a series combination of a non-linear element $L_\mathrm{j}$ with a sinusoidal current-phase relation and a linear inductor $L_\mathrm{a}$. 
\begin{figure}[h]
	\centerline {\includegraphics[trim={0cm 2cm 0cm 1cm},clip=True,scale=0.8]{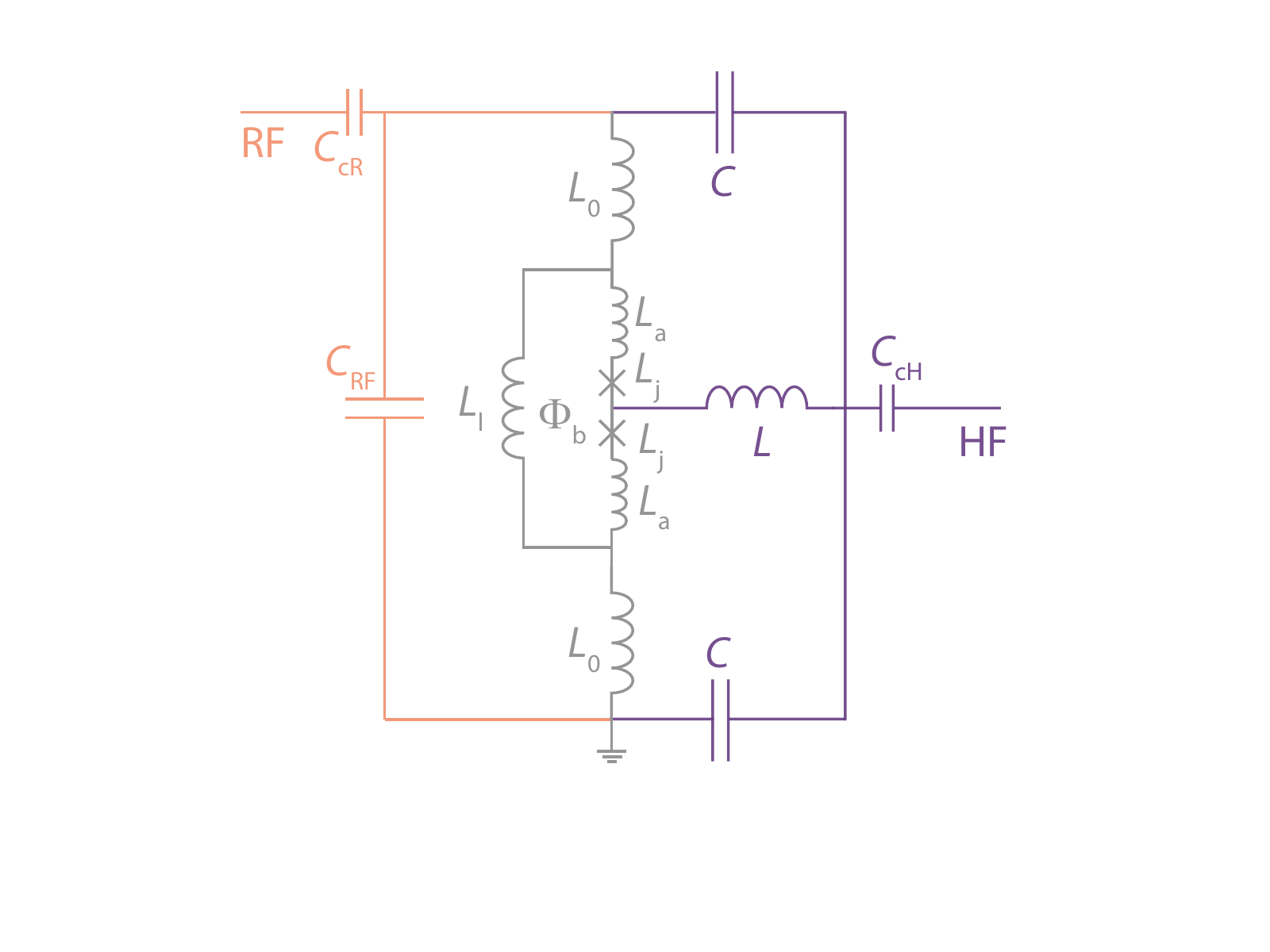}}
	\caption{\textsf{\textbf{Full circuit diagram of the device.} The orange part of the circuit represents the components belonging only to the radio-frequency mode. The circuit parts drawn in purple correspond to the high-frequency mode. Both modes share the part of the circuit drawn in gray.}}
	\label{fig:CM}
\end{figure}

\subsection*{Flux dependence of the HF mode}

The resonance frequency of a SQUID cavity with a symmetric SQUID can be described by
\begin{equation}
\omega_0(\Phi_\mathrm{b}) = \frac{\omega_0(0)}{\sqrt{\Lambda + \frac{1-\Lambda}{\cos{\left(\pi\frac{\Phi}{\Phi_0}\right)}}}}
\label{eqn:FluxDep}
\end{equation}
where $\Phi$ corresponds to the total flux threading the SQUID loop and $\omega_0(0)$ is the resonance frequency without external flux bias (sweetspot frequency).
The parameter $\Lambda = (L_\mathrm{HF}-\frac{1}{2}L_\mathrm{j0})/L_\mathrm{HF}$ with the total high-frequency inductance $L_\mathrm{HF}$ and the single junction Josephson inductance $L_{\mathrm{j}0}$ is a measure for the contribution of the Josephson inductance to the total inductance.
For zero bias current and the (magnetic plus kinetic) loop inductance $L_\mathrm{loop} = L_\mathrm{l} + 2L_\mathrm{a}$ the total flux threading the SQUID is given by
\begin{eqnarray}
\frac{\Phi}{\Phi_0} = \frac{\Phi_\mathrm{b}}{\Phi_0} + \frac{L_\mathrm{loop}J}{\Phi_0}
\end{eqnarray}
with the circulating current $J$. 
In the absence of a bias current and symmetric junctions, the circulating current is given by 
\begin{equation}
J = -I_\mathrm{c}\sin{\left(\pi\frac{\Phi}{\Phi_0}\right)}
\end{equation}
with the zero bias critical current of a single junction $I_\mathrm{c} = \frac{\Phi_0}{2\pi L_\mathrm{j0}}$.
Using the screening parameter $\beta_L = \frac{2L_\mathrm{loop}I_\mathrm{c}}{\Phi_0} = \frac{L_\mathrm{loop}}{\pi L_\mathrm{j0}}$ the relation for the total flux can be written as
\begin{eqnarray}
\frac{\Phi}{\Phi_0} = \frac{\Phi_\mathrm{b}}{\Phi_0} - \frac{\beta_L}{2}\sin{\left(\pi\frac{\Phi}{\Phi_0}\right)}.
\label{eqn:TotFlux}
\end{eqnarray}
Figure 1\textbf{e} of the main paper shows the experimentally determined SQUID cavity resonance frequency modulating with external magnetic flux $\Phi_\mathrm{b}$ and a fit curve using Eq.~\ref{eqn:FluxDep}, where the relation between the applied external flux $\Phi_\mathrm{b}$ and the total flux in the SQUID $\Phi$ is given by Eq.~(\ref{eqn:TotFlux}).
As fit parameters we obtain $\beta_L =1.07$ and $\Lambda = 0.946$, i.e., the SQUID Josephson inductance contributes about $5.4\%$ to the total HF inductance.
Furthermore, we estimate the capacitance of the SQUID cavity $C_{\textrm{HF}} = 2C + C_{\mathrm{cH}}$ with the expressions given in Ref.~[51] to be $\sim 1.3\,$pF.
Based on the sweetspot frequency of the SQUID cavity 
\begin{equation}
\omega_0 = \frac{1}{\sqrt{L_\mathrm{HF}({2C + C_{\mathrm{cH}}})}},
\end{equation} 
we extract the total inductance of the high frequency mode to be $L_{\textrm{HF}} = 370\,$pH and with $\Lambda$ we get the inductance of a single junction $L_\mathrm{j0} = 40\,$pH.
This inductance corresponds to a critical junction current $I_\mathrm{c} = \frac{\Phi_0}{2\pi L_\mathrm{jc}} \approx 8.3\,\mu$A.
From the screening parameter $\beta_L = 1.07$ and the single-junction inductance $L_\mathrm{j0} = 40\,$pH, we get a loop inductance $L_\mathrm{loop} = 2L_\mathrm{a} + L_\mathrm{l} = \pi\beta_L L_\mathrm{j0} \approx 134\,$pH.

\subsection*{Flux dependence of the RF mode}

Based on the circuit diagram shown in Supplementary Fig.~\ref{fig:CM}, we find the total inductance of the radio-frequency mode $L_{\mathrm{RF}}$ as
\begin{equation}
L_\mathrm{RF}  = 2L_\mathrm{0} + \frac{2(L_\mathrm{j} + L_\mathrm{a}) L_\mathrm{l}}{2(L_\mathrm{j} + L_\mathrm{a}) + L_\mathrm{l}},
\label{eq:L_RF}
\end{equation} 
with the Josephson inductance of a single junction $L_\textrm{j} = \frac{L_\mathrm{j0}}{\cos\left(\pi\frac{\Phi}{\Phi_0}\right)}$ and the SQUID loop inductance $L_\mathrm{loop} = 2L_\mathrm{a} + L_\mathrm{l}  \approx 134\,$pH as boundary condition for $L_\mathrm{l}$ and $L_\mathrm{a}$.
In addition, we independently estimate the parallel plate capacitance $C_{\textrm{RF}} = 659.7\,$pF and the parallel plate coupling capacitance $C_\mathrm{cR} = 0.3\,$pF.
From the total capacitance $C_{\textrm{tot}} = C_{\textrm{RF}}+C_\mathrm{cR}$ and the resonance frequency $\Omega_0$, we determine the total inductance of the RF resonator as $L_\mathrm{RF} \sim 188\,$pH. 
\begin{figure}[h]
	\centerline {\includegraphics[trim={0cm 0cm 0cm 1cm},clip=True,scale=0.8]{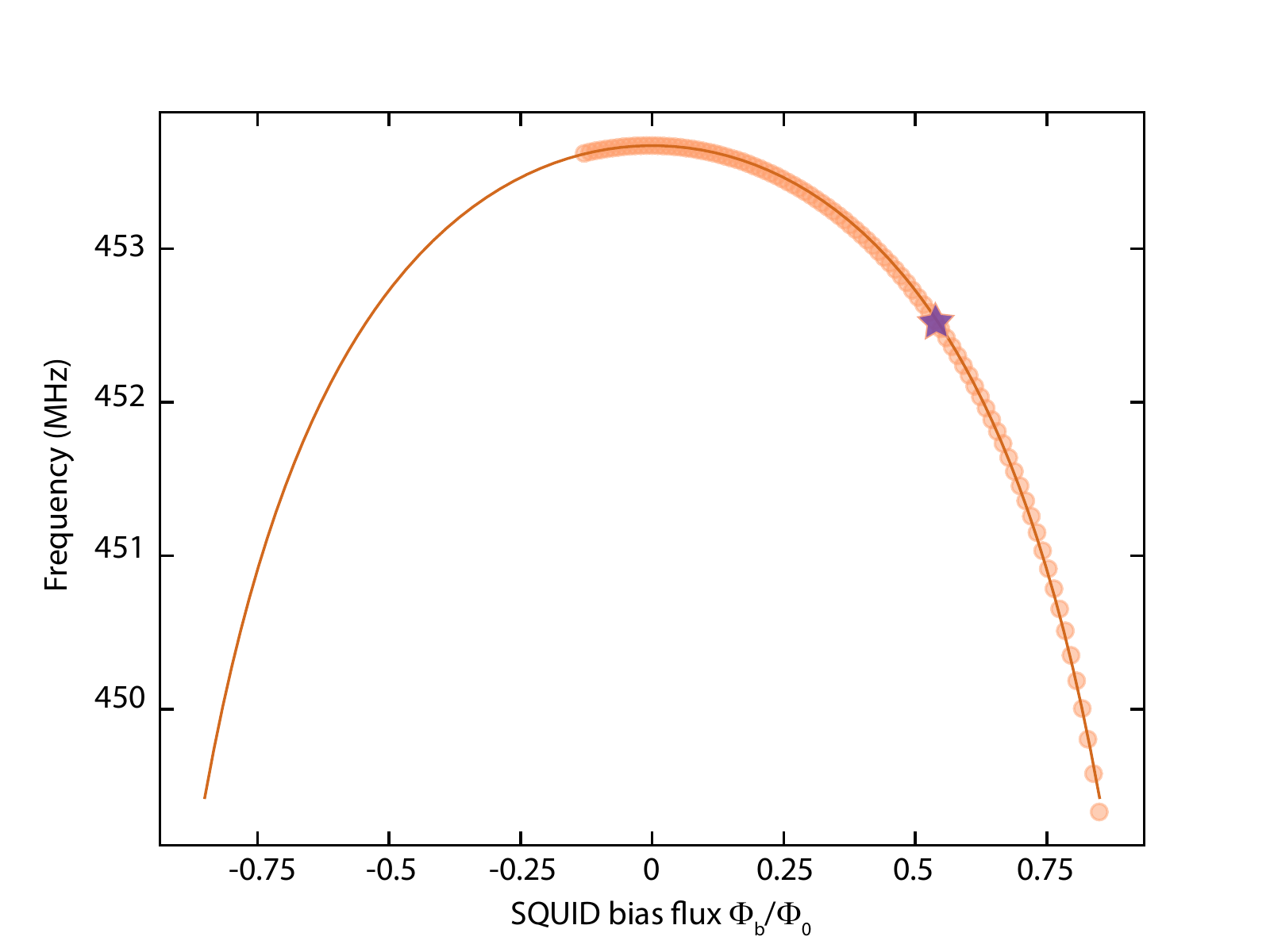}}
	\caption{\textsf{\textbf{RF mode resonance frequency depending on external magnetic flux.} Points are data and line is a fit curve using Eq.~(\ref{eq:RFflux}). The operation point in this work $\Phi_\mathrm{b}/\Phi_0 = 0.54$ is marked by a star.}}
	\label{fig:RFfluxarch}
\end{figure}
As the total inductance of the radio-frequency mode is partly composed by the field-dependent Josephson inductance $L_\mathrm{j}$, the resonator resonance frequency will as well modulate with applied magnetic flux $\Phi_{\textrm{b}}$ as 
\begin{equation}
\Omega_0 = \frac{1}{\sqrt{C_{\textrm{tot}} \left( 2L_0 + L_\mathrm{l} \left(1 + \frac{L_\mathrm{l}}{2} \frac{\cos{\left( \pi \frac{\Phi}{\Phi_0} \right)}}{L_\mathrm{j0} + L_\mathrm{a}\cos{\left( \pi \frac{\Phi}{\Phi_0} \right)}} \right)^{-1}\right) }}.
\label{eq:RFflux}
\end{equation}
where the relation between $\Phi$ and $\Phi_\mathrm{b}$ is again given by Eq.~(\ref{eqn:TotFlux}).
Supplementary Fig.~\ref{fig:RFfluxarch} shows the RF resonance frequency depending on the applied magnetic flux together with a fit curve using Eq.~(\ref{eq:RFflux}), where the parameter $\beta_L = 1.07$ was kept constant as determined from the HF mode flux dependence.
From the fit, we extract the parameters $L_\textrm{a} = 43.5\,$pH and $L_\textrm{l} = 47\,$pH.
Based on the returned fit parameters and on Eq.~(\ref{eq:RFflux}), we obtain $L_\mathrm{0} = 75\,$pH.
We note here, that without the linear junction inductances $L_\mathrm{a}$, it is not possible to fit both flux dependences with a single set of reasonable parameters.

\subsection*{Flux dependence of the decay rates $\kappa$ and $\Gamma_0$}

As the external magnetic field is kept at a non-zero value during the experiment, it is of interest to analyze how the applied magnetic flux $\Phi_\textrm{b}$ affects the linewidths of the circuit.
For that, we extract the decay rates of both modes $\kappa$ and $\Gamma_0$ while changing the flux bias point.
The result is shown in Supplementary Fig.~\ref{fig:decays_flux} for the positive tuning range.
Both linewidths clearly show a strong dependence for values larger than $\sim 0.7 \Phi_0$ to $\sim 0.8 \Phi_0$.
For the operating point used here of $\Phi_\mathrm{b}/\Phi_\mathrm{0} = 0.54$, however, they are nearly unmodified compared to the sweetspot values. 
\begin{figure}[h]
	\centerline {\includegraphics[trim={1cm 13.5cm 1cm 6.5cm},clip=True,scale=0.8]{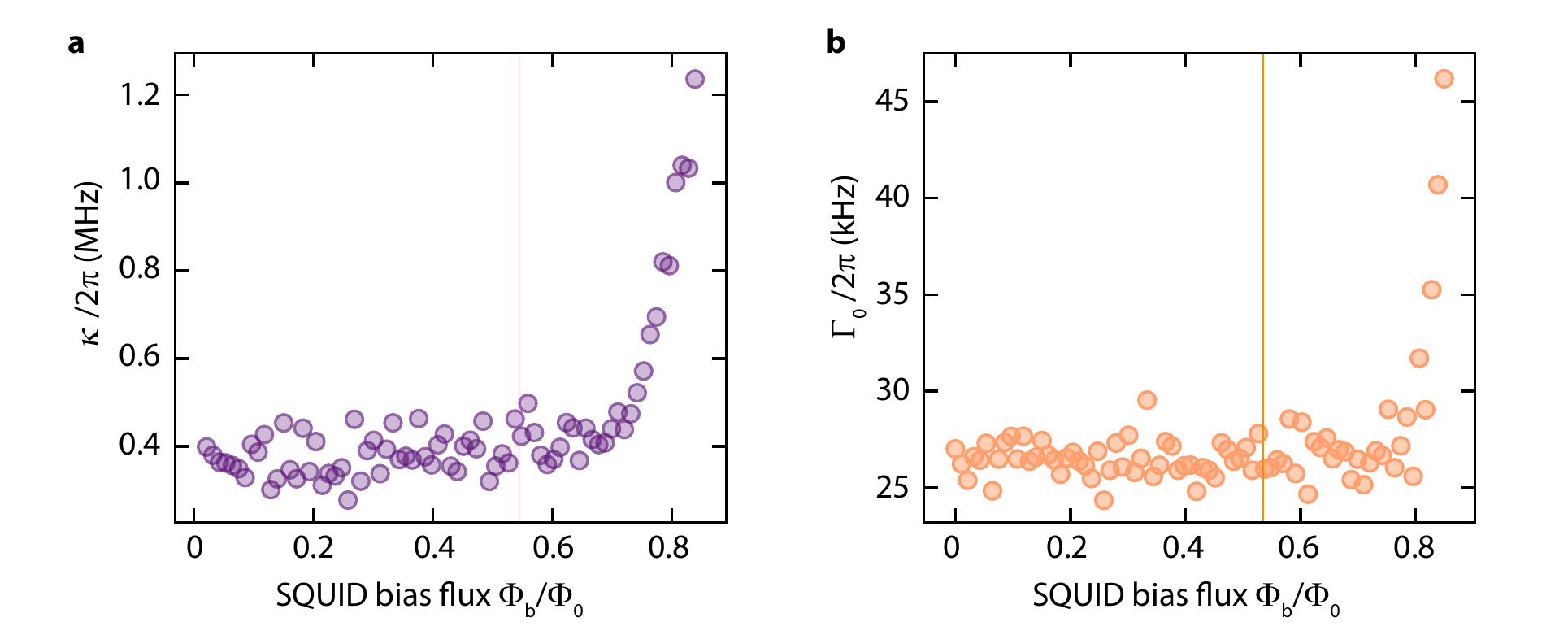}}
	\caption{\textsf{\textbf{Linewidths for varying SQUID bias flux.} Decay rate of the microwave mode $\kappa$ (\textbf{a}) and of the radio-frequency mode $\Gamma_0$ (\textbf{b}) depending on magnetic bias flux in units of flux quanta $\Phi_0$. The operation point $\Phi_\mathrm{b}/\Phi_0 \approx 0.54$ for the experiments reported here is marked by vertical lines.}}
	\label{fig:decays_flux}
\end{figure}

\section*{Supplementary Note 4: Response functions and fitting routine}

\subsection*{Ideal HF and RF resonators response functions}
Both, our HF SQUID cavity and the RF resonator, can be modeled as a parallel LC circuit capacitively coupled to a transmission line in a reflection geometry.
The $S_{11}$ response function of such a circuit (here for the HF mode) is given by
\begin{equation}
S_{11}^\mathrm{HF} = 1 - \frac{2\kappa_\mathrm{e}}{\kappa_\mathrm{i}+\kappa_\mathrm{e}+2i\Delta}
\label{eq:ResponsefuncHF}
\end{equation}
with detuning from the resonance frequency
\begin{eqnarray}
\Delta = \omega - \omega_0.
\end{eqnarray}
For the RF resonator, we get fully equivalently
\begin{equation}
S_{11}^\mathrm{RF} = 1- \frac{2\Gamma_\mathrm{e}}{\Gamma_\mathrm{i}+\Gamma_\mathrm{e}+2i\Delta_0}
\label{eq:ResponsefuncRF}
\end{equation}
with $\Delta_0 = \Omega - \Omega_0$.

\subsection*{Real response function and fitting routine}

When analyzing the measured cavity response, we consider a frequency-dependent complex-valued reflection background with amplitude and phase modulations originating from a variety of microwave components in our input and output lines and possible interfering signal paths.
Under this assumption, we model the modified cavity response with
\begin{eqnarray}
S_{11} = (\alpha_0 + \alpha_1\omega)\left(1-\frac{2\kappa_\mathrm{e} e^{i\theta}}{\kappa_\mathrm{i}+\kappa_\mathrm{e}+2i\Delta}\right)e^{i(\beta_1\omega + \beta_0)}
\label{eq:fitS11}
\end{eqnarray}
where we consider a frequency dependent complex background
\begin{eqnarray}
S_{11} = (\alpha_0 + \alpha_1\omega)e^{i(\beta_1\omega + \beta_0)}
\label{eqn:Fit_BG}
\end{eqnarray}
and an additional rotation of the resonance circle with the phase factor $e^{i\theta}$.
The first step in the fitting routine removes the cavity resonance part from the data curve and fits the remaining background with Eq.~(\ref{eqn:Fit_BG}).
After removing the background contribution from the full dataset by complex division, the resonator response is fitted using the ideal response function.
In the final step, the full function is re-fitted to the bare data using as starting parameters the individually obtained fit numbers from the first two steps.
From this final fit, we extract the final background fit parameters and remove the background of the full dataset by complex division.
Also, we correct for the additional rotation factor $e^{i\theta}$.
As result we obtain clean resonance curves as shown in Fig.~1\textbf{d} of the main paper.

\subsection*{Complex fits and mode parameters at the operation point}

\begin{figure}[h]
	\centerline {\includegraphics[trim={0cm 0.0cm 0cm 0cm},clip=True,scale=0.47]{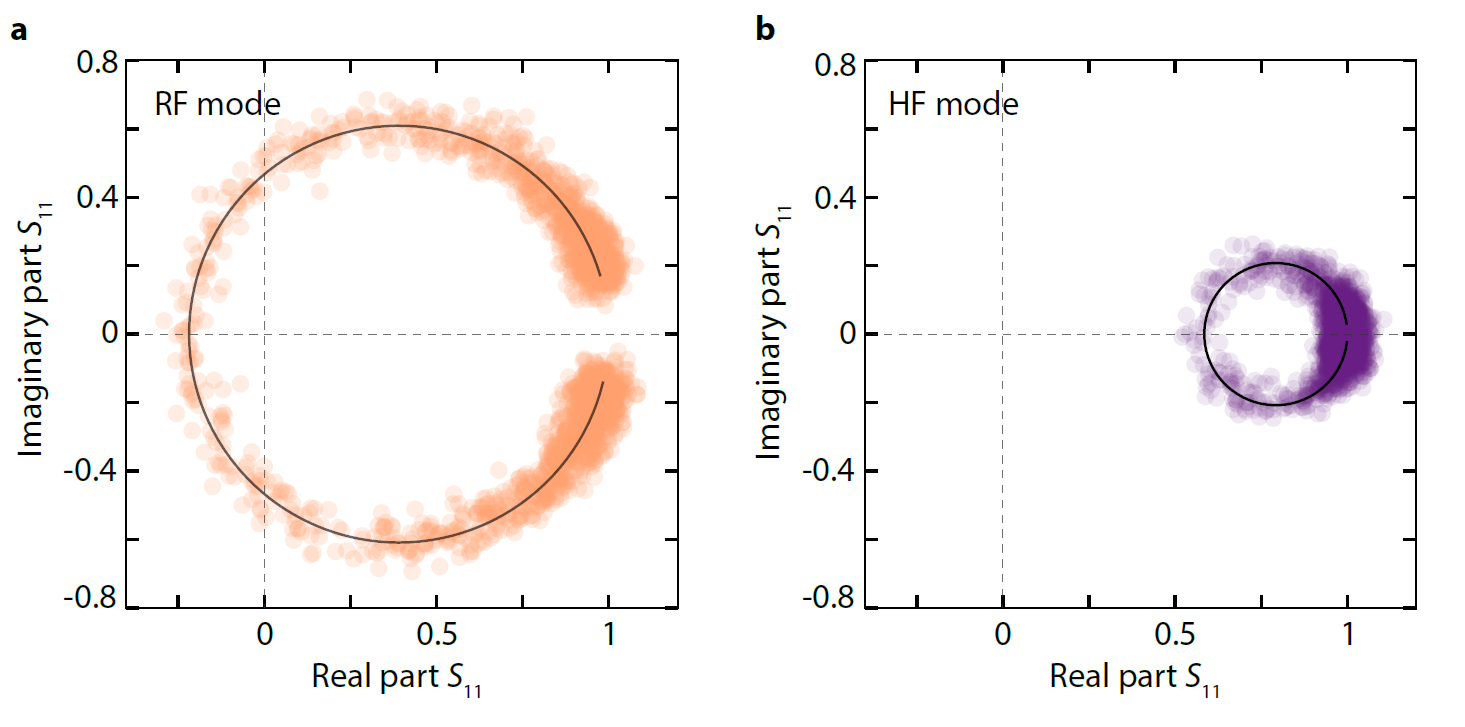}}
	\caption{\textsf{\textbf{RF and HF complex resonance fits.} In \textbf{a} the resonance of the radio-frequency mode is shown, in \textbf{b} the resonance of the high-frequency mode. Circles are data, line is a fit. Extracted fit parameters are given in the text.}}
	\label{fig:ComplexFits}
\end{figure}
In Supplementary Fig.~\ref{fig:ComplexFits} we show the complex reflection signal obtained from the measurements for the RF mode in \textbf{a} and the HF mode in \textbf{b}.
The background has been removed by complex division and the corresponding fit curves are added as lines.
From the fits, we extract for the radio-frequency mode data shown in \textbf{a} the resonance frequency $\Omega_0 = 2\pi\cdot 452.5\,$MHz, and the linewidths $\Gamma_\mathrm{i} = 2\pi\cdot 10\,$kHz and $\Gamma_\mathrm{e} = 2\pi\cdot 16\,$kHz.
The fit parameters of the high-frequency mode data shown in \textbf{b} are $\omega_0 = 2\pi\cdot 7.207\,$GHz, $\kappa_\mathrm{i} = 2\pi\cdot 310\,$kHz and $\kappa_\mathrm{e} = 2\pi\cdot 80\,$kHz 
The absolute values of these two resonance are shown in main paper Fig.~1\textbf{d}.

\section*{Supplementary Note 5: Zero-point fluctuations $\Phi_\mathrm{zpf}$ and coupling rate $g_0$}

The zero-point current fluctuations of the radio-frequency mode at the operation point are given by

\begin{equation}
I_\textrm{zpf} = \sqrt{\frac{\hbar \Omega_0}{2 L_{\textrm{RF}}}} \approx 28\,\mathrm{nA},
\end{equation}

where $\Omega_0 = 2\pi \cdot 452.5\,$MHz and $L_{\textrm{RF}} = 188\,$pH.
In presence of this zero-point current, which flows asymmetrically through the loop wire arm and the junction arm of the SQUID, the total flux in the SQUID is given by
\begin{equation}
\Phi = \Phi_\mathrm{b} + L_\mathrm{loop}J - \alpha L_\mathrm{loop}\frac{I_\mathrm{zpf}}{2}
\end{equation}
where $\alpha = \frac{2L_\mathrm{a} - L_\mathrm{l}}{L_\mathrm{loop}} \approx 0.3$ describes the inductance asymmetry of the SQUID from the perspective of the RF currents.
With the current-phase relation of the JJ, this can also be written as
\begin{equation}
\Phi = \Phi_\mathrm{b} + L_\mathrm{loop}(1-\alpha)\frac{I_\mathrm{zpf}}{2} - L_\mathrm{loop}I_\mathrm{c}\sin{\left(\pi\frac{\Phi}{\Phi_0}\right)}.
\end{equation}
Therefore, the zero-point fluctuation current is formally equivalent to a fluctuating external flux with
\begin{eqnarray}
\Phi_\mathrm{zpf} & = & L_\mathrm{loop}(1-\alpha)\frac{I_\mathrm{zpf}}{2}\\
& = & L_\mathrm{l}I_\mathrm{zpf}\\
& \approx & 635\,\mu\Phi_0.
\end{eqnarray}
Using the derivative  $\frac{\partial\omega_0}{\partial\Phi_\mathrm{b}}$ of the flux-dependence fit curve $\omega_0(\Phi_\mathrm{b})$, we can therefore calculate the single-photon coupling strength
\begin{equation}
g_0 = \frac{\partial\omega_0}{\partial\Phi_\mathrm{b}}\Phi_\mathrm{zpf},
\end{equation}
the result is shown in Supplementary Fig.~\ref{fig:g0}.

\begin{figure}[h]
	\centerline {\includegraphics[trim={1cm 13.5cm 1cm 6.5cm},clip=True,scale=0.8]{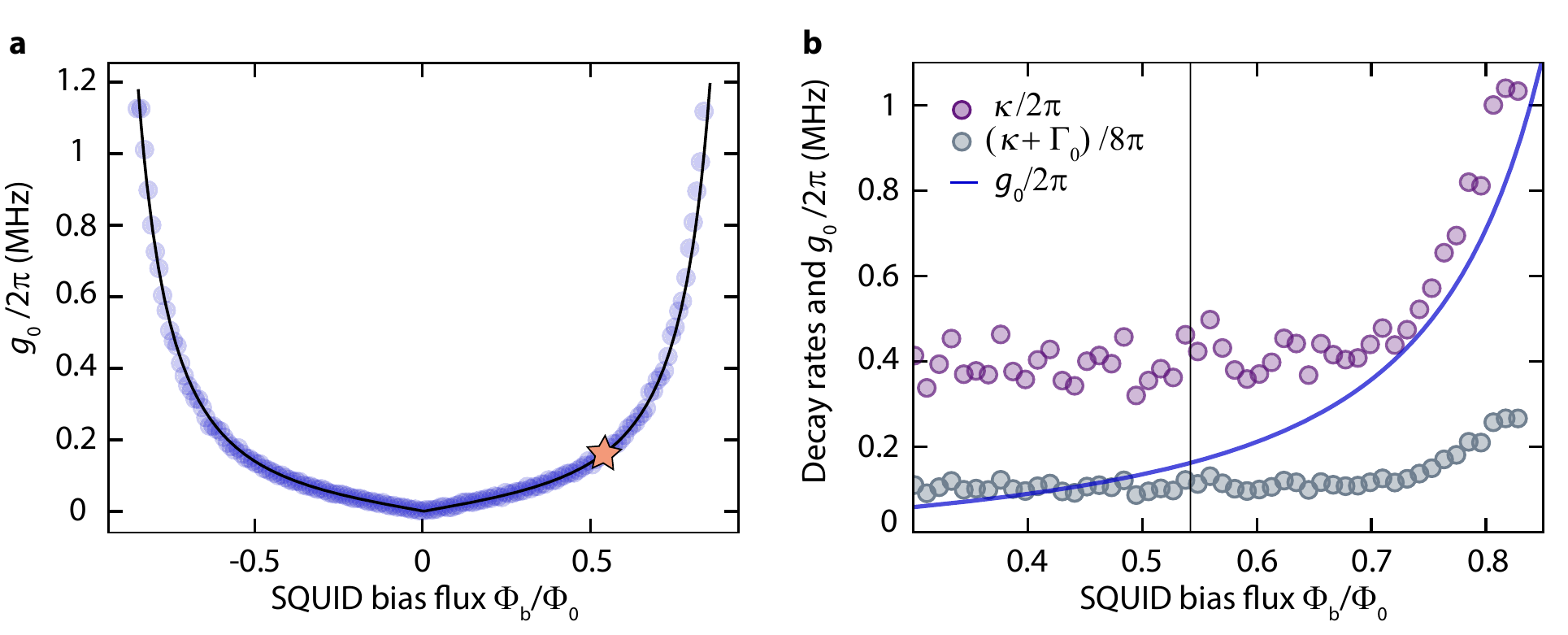}}
	\caption{\textsf{\textbf{Single photon-coupling rate $g_0$ and decay rates vs SQUID bias flux.} \textbf{a} The points are calculated from the experimentally determined flux arch, the line is based on the arch fit, cf. main paper Fig. 1. For both, we calculate the derivative and multiply with the theoretical value for $\Phi_\mathrm{zpf} = 635\,\mu\Phi_0$. \textbf{b} shows the theoretically obtained $g_0$, the extracted cavity decay rate $\kappa$ and $(\kappa + \Gamma_0)/4$ versus SQUID bias flux. The last quantity describes the $g_0$ limit for where a single sideband photon will induce parametric normal mode splitting. The operation point for the experiments reported here $\Phi/\Phi_0 = 0.54$ is marked by a star in \textbf{a} and as vertical gray line in \textbf{b}.}}
	\label{fig:g0}
\end{figure}

At the bias point used in the reported experiments, we get $g_0 \approx 2\pi\cdot 160\,$kHz.
Around $\Phi_\mathrm{b}/\Phi_0 \sim 0.7$ though we get $g_0 \sim \kappa$, cf. also Supplementary Fig.~\ref{fig:decays_flux}.
Also, for the regime $ \Phi_\mathrm{b}/\Phi_0 > 0.5$, we get $g_0 \gtrsim \frac{\kappa + \Gamma_0}{4}$, i.e, the regime where a single red-sideband photon will induce well-resolved, parametric normal-mode splitting [26,~57].
From a comparison of the power calibration presented in Supplementary Note 2 with the intracavity photons required to obtain the observed total coupling rates $g = \sqrt{n_\mathrm{c}}g_0$ in Figs.~2-4 of the main paper we find only a very small disagreement in input line attenuation/intracavity pump photon number of $\sim0.3\,$dB.
If we assume this disagreement to originate from uncertainties in the calculation of $g_0$ and $\Phi_\mathrm{zpf}$, respectively, we find the magnitude of the relative error in $\Phi_\mathrm{zpf}$ to be $0.04$.

\section*{Supplementary Note 6: Photon-pressure sideband cooling}

\subsection*{Equations of motion}

We model the (approximately red-sideband) driven system with the linearized equations of motion for photon-pressure interacting harmonic oscillators [1,~20]
\begin{eqnarray}
\delta\dot{\hat{a}} & = & \left(-i\Delta - \frac{\kappa}{2}\right)\delta\hat{a} + ig(\hat{b} + \hat{b}^\dagger) + \sqrt{\kappa_\mathrm{i}}\hat{S}_\mathrm{i}^\mathrm{HF}+ i\sqrt{\kappa_\mathrm{e}}\hat{S}_\mathrm{e}^\mathrm{HF}\\
\dot{\hat{b}} & = & \left(i\Omega_0 - \frac{\Gamma_0}{2}\right)\hat{b} + ig(\delta\hat{a} + \delta\hat{a}^\dagger) + \sqrt{\Gamma_\mathrm{i}}\hat{S}_\mathrm{i}^\mathrm{RF}+ i\sqrt{\Gamma_\mathrm{e}}\hat{S}_\mathrm{e}^\mathrm{RF}.
\end{eqnarray}
Here, $\delta\hat{a}$ and $\delta\hat{a}^\dagger$ describe the annihilation and creation operator for HF cavity field fluctuations, respectively, $\Delta = \omega_\mathrm{p} - \omega_0$, $g = \sqrt{n_\mathrm{c}}g_0$, and $\hat{S}_\mathrm{i}^\mathrm{HF}$ and $\hat{S}_\mathrm{e}^\mathrm{HF}$ corresponds to internal and external HF noise input fields.
The RF mode annihilation and creation operators are given by $\hat{b}$ and $\hat{b}^\dagger$ and the internal and external RF noise input fields are taken into account by $\hat{S}_\mathrm{i}^\mathrm{RF}$ and $\hat{S}_\mathrm{e}^\mathrm{RF}$.
The input noise operators $\hat{S}$ follow $\expval{ \hat{S}^\dagger\hat{S}} = n$ and $\expval{\hat{S}\hat{S}^\dagger } = n + 1$.
These equations can be solved by Fourier transform and the solutions read in frequency space
\begin{eqnarray}
\delta\hat{a}(\Omega) & = & ig\chi_\mathrm{c}\left[\hat{b}(\Omega) + \hat{b}^\dagger(-\Omega)\right] + \chi_\mathrm{c}\left[\sqrt{\kappa_\mathrm{i}}\hat{S}_\mathrm{i}^\mathrm{HF}(\Omega)+i\sqrt{\kappa_\mathrm{e}}\hat{S}_\mathrm{i}^\mathrm{HF}(\Omega)\right]
\label{eqn:EOM_FS_HF}\\
\hat{b}(\Omega) & = & ig\chi_0\left[\delta\hat{a}(\Omega) + \delta\hat{a}^\dagger(-\Omega)\right] + \chi_0\left[\sqrt{\Gamma_\mathrm{i}}\hat{S}_\mathrm{i}^\mathrm{RF}(\Omega) + i\sqrt{\Gamma_\mathrm{e}}\hat{S}_\mathrm{e}^\mathrm{RF}(\Omega)\right]
\label{eqn:EOM_FS_RF}
\end{eqnarray}
with the susceptibilities
\begin{eqnarray}
\chi_\mathrm{c} & = & \frac{1}{\frac{\kappa}{2}+i(\Delta + \Omega)}\\
\chi_0 & = & \frac{1}{\frac{\Gamma_0}{2}+i\Delta_0},
\end{eqnarray}
$\Omega$ being the frequency relative to the pump tone and $\Delta_0 = \Omega - \Omega_0$.
\subsection*{Simplified equations of motion in the sideband-resolved regime with red-sideband pumping}
Under red-sideband pumping $\Delta\approx -\Omega_0$ and in the sideband-resolved regime $\Omega_0 \gg \kappa$ the equations of motion can be simplified as
\begin{eqnarray}
\delta\hat{a}(\Omega) & = & ig\chi_\mathrm{c}\hat{b}(\Omega) + \chi_\mathrm{c}\left[\sqrt{\kappa_\mathrm{i}}\hat{S}_\mathrm{i}^\mathrm{HF}(\Omega)+i\sqrt{\kappa_\mathrm{e}}\hat{S}_\mathrm{i}^\mathrm{HF}(\Omega)\right]
\label{eqn:EOM_FS_HF}\\
\hat{b}(\Omega) & = & ig\chi_0\delta\hat{a}(\Omega) + \chi_0\left[\sqrt{\Gamma_\mathrm{i}}\hat{S}_\mathrm{i}^\mathrm{RF}(\Omega) + i\sqrt{\Gamma_\mathrm{e}}\hat{S}_\mathrm{e}^\mathrm{RF}(\Omega)\right]
\label{eqn:EOM_FS_RF}
\end{eqnarray}

\subsection*{Solution for the RF mode response function}

To calculate the response to a radio-frequency probe tone, we replace the noise input by a probe tone input $\hat{S}_0$ and get 
\begin{eqnarray}
\delta\hat{a}(\Omega) & = & ig\chi_\mathrm{c}\hat{b}(\Omega)
\\
\hat{b}(\Omega) & = & ig\chi_0\delta\hat{a}(\Omega) + i\chi_0\sqrt{\Gamma_\mathrm{e}}\hat{S}_0(\Omega)
\end{eqnarray}
For the response function and the input-output relations [1] we find from this the result
\begin{equation}
S_{11}^\mathrm{RF} = 1-\Gamma_\mathrm{e}\frac{\chi_0}{1+g^2\chi_\mathrm{c}\chi_0}.
\end{equation}
The resonance condition $\left(\chi_0^\mathrm{eff}\right)^{-1} = 0$ for effective RF susceptibility
\begin{equation}
\chi_0^\mathrm{eff} = \frac{\chi_0}{1+g^2\chi_\mathrm{c}\chi_0}
\end{equation}
delivers the complex solutions
\begin{equation}
\tilde{\Omega}_\pm = \Omega_0 - \frac{\delta}{2} + i\frac{\kappa + \Gamma_0}{4} \pm \sqrt{g^2 - \left(\frac{\kappa - \Gamma_0 + 2i\delta}{4}\right)^2}
\end{equation}
where $\delta$ is the pump detuning from the red sideband defined by $\Delta = -\Omega_0 + \delta$.

\subsection*{Solution for the HF mode response function}

In full analogy to the RF mode, we get as probe tone response function for the HF mode
\begin{equation}
S_{11}^\mathrm{HF} = 1-\kappa_\mathrm{e}\frac{\chi_\mathrm{c}}{1+g^2\chi_\mathrm{c}\chi_0}.
\end{equation}
with the complex solutions
\begin{equation}
\tilde{\omega}_\pm = \omega_0 + \frac{\delta}{2} + i\frac{\kappa + \Gamma_0}{4} \pm \sqrt{g^2 - \left(\frac{\kappa - \Gamma_0 + 2i\delta}{4}\right)^2}
\end{equation}

\subsection*{Solution for the HF mode thermal noise power spectral density}
We solve Eqs.~(\ref{eqn:EOM_FS_HF}), (\ref{eqn:EOM_FS_RF}) with noise input now and get
\begin{equation}
\delta\hat{a} = \frac{ig\chi_\mathrm{c}\chi_0\left[\sqrt{\Gamma_\mathrm{i}}\hat{S}_\mathrm{i}^\mathrm{RF} + i\sqrt{\Gamma_\mathrm{e}}\hat{S}_\mathrm{e}^\mathrm{RF}\right] + \chi_\mathrm{c}\left[\sqrt{\kappa_\mathrm{i}}\hat{S}_\mathrm{i}^\mathrm{HF} + i\sqrt{\kappa_\mathrm{e}}\hat{S}_\mathrm{e}^\mathrm{HF} \right]}{1+g^2\chi_\mathrm{c}\chi_0}.
\end{equation}
which leads to the symmetrized output field power spectral density in units of photons [55]
\begin{equation}
\frac{S(\omega)}{\hbar\omega} = \frac{1}{2} + n_\mathrm{e}^\mathrm{HF} + \kappa_\mathrm{e}\kappa_\mathrm{i} \frac{|\chi_\mathrm{c}(\omega)|^2}{|1+g^2\chi_0(\Omega)\chi_\mathrm{c}(\omega)|^2}\left(n_\mathrm{i}^\mathrm{HF} - n_\mathrm{e}^\mathrm{HF}\right) + \kappa_\mathrm{e}\Gamma_0\frac{g^2|\chi_\mathrm{c}(\omega)|^2|\chi_0(\Omega)|^2}{|1+g^2\chi_0(\Omega)\chi_\mathrm{c}(\omega)|^2}\left(n_\mathrm{th}^\mathrm{RF} - n_\mathrm{e}^\mathrm{HF}\right)
\end{equation}
where the effective thermal photon occupation of the RF mode is given by the weighted sum
\begin{eqnarray}
n_\mathrm{th}^\mathrm{RF} & = & \frac{\Gamma_\mathrm{i}}{\Gamma_0}n_\mathrm{i}^\mathrm{RF} + \frac{\Gamma_\mathrm{e}}{\Gamma_0}n_\mathrm{e}^\mathrm{RF}.
\end{eqnarray}
%
%
%
%
%
%

\subsection*{Added noise}

The effective number of added noise photons by the amplifier chain is given by [20]
\begin{equation}
n_\mathrm{add}' = \frac{n_\mathrm{add}}{\eta}+\left(\frac{1-\eta}{\eta}\right)\frac{1}{2}
\end{equation}
where $n_\mathrm{add}$ is the actual number of photons added by the HEMT amplifier noise in our case, and $\eta \sim 0.5$ accounts for losses of the cavity output field on its way to the HEMT. 
We will estimate the number of added noise photons based on a temperature sweep calibration presented below.
As a rough first estimate, we can use the datasheet noise temperature of the amplifier of $\sim 2\,$K to find $n_\mathrm{add} \approx 5.3$ and $n_\mathrm{add}' \approx 11.1$.

\subsection*{The total power spectral density}

For the power spectral density in units of photons at frequency $\omega = \omega_\mathrm{p} + \Omega$ of the SQUID cavity with a drive around the red sideband, we get for $n_\mathrm{add}', n_\mathrm{th}^\mathrm{RF}\gg n_\mathrm{e}^\mathrm{HF}, n_\mathrm{i}^\mathrm{HF}\ll 1/2$ (where the latter corresponds to the reasonable assumption $T_\mathrm{e}^\mathrm{HF}, T_\mathrm{i}^\mathrm{HF} \lesssim 100\,$mK)
\begin{equation}
\frac{S(\omega)}{\hbar\omega} = \frac{1}{2} + n_\mathrm{add}' + \kappa_\mathrm{e}\Gamma_0\frac{g^2|\chi_\mathrm{c}(\omega)|^2|\chi_0(\Omega)|^2}{|1+g^2\chi_0(\Omega)\chi_\mathrm{c}(\omega)|^2}n_\mathrm{th}^\mathrm{RF}
\end{equation}
This can be also written as
\begin{equation}
\frac{S(\omega)}{\hbar\omega} = \frac{1}{2} + n_\mathrm{add}' + \frac{16 \kappa_\mathrm{e} g^2\Gamma_0 n_\mathrm{th}^\mathrm{RF}}{\left|4g^2 + \left[\kappa + 2i(\delta + \Delta_0)\right]\left[\Gamma_0 + 2i\Delta_0\right]\right|^2}
\label{eq:PSD}
\end{equation}
where $\Delta_0 = \Omega - \Omega_0$ takes into account the detuning from the RF resonance frequency and $\delta =  \omega_\mathrm{p} -(\omega_0 - \Omega_0)$ takes into account the detuning of the pump from the red sideband of the cavity.
By fitting the measured power spectral density with Eq.~(\ref{eq:PSD}), as shown as fit curves in Fig.~3\textbf{c} of the main paper, we obtain for each curve the thermal photon number occupancy of the RF mode as detailed below. 

\subsection*{Cooled RF photons}
For the RF mode we get from the equations of motion
\begin{equation}
\hat{b} = \frac{ig\chi_\mathrm{c}\chi_0\left[\sqrt{\kappa_\mathrm{i}}\hat{S}_\mathrm{i}^\mathrm{HF} + i\sqrt{\kappa_\mathrm{e}}\hat{S}_\mathrm{e}^\mathrm{HF} \right] + \chi_0\left[\sqrt{\Gamma_\mathrm{i}}\hat{S}_\mathrm{i}^\mathrm{RF} + i\sqrt{\Gamma_\mathrm{e}}\hat{S}_\mathrm{e}^\mathrm{RF}\right]}{1+g^2\chi_\mathrm{c}\chi_0}.
\end{equation}
We can use this to calculate the RF photon population with a sideband drive exactly on the red sideband and get
\begin{equation}
n_\mathrm{cool}^\mathrm{RF} = \frac{\Gamma_0}{\kappa + \Gamma_0}\frac{4g^2 + \kappa(\kappa + \Gamma_0)}{4g^2 + \kappa\Gamma_0}n_\mathrm{th}^\mathrm{RF} + \frac{\kappa}{\kappa + \Gamma_0}\frac{4g^2}{4g^2 + \kappa\Gamma_0}n_\mathrm{th}^\mathrm{HF}.
\end{equation}
Compared to the usually quoted result [1,~20,~56], we find some corrections in the cooled RF occupation, in particular the appearance of the factor $\kappa + \Gamma_0$ instead of $\kappa$.
These corrections are negligble for $\kappa \gg \Gamma_0$, which in our case, however is not strictly true anymore.
For non-vanishing detuning we get
\begin{equation}
n_\mathrm{cool}^\mathrm{RF} = \frac{\Gamma_0}{\kappa + \Gamma_0}\frac{4g^2 + \kappa(\kappa + \Gamma_0)\left[1+\frac{4\delta^2}{(\kappa + \Gamma_0)^2}\right]}{4g^2 + \kappa\Gamma_0\left[1+\frac{4\delta^2}{(\kappa + \Gamma_0)^2}\right]}n_\mathrm{th}^\mathrm{RF} + \frac{\kappa}{\kappa + \Gamma_0}\frac{4g^2}{4g^2 + \kappa\Gamma_0\left[1+\frac{4\delta^2}{(\kappa + \Gamma_0)^2}\right]}n_\mathrm{th}^\mathrm{HF}.
\end{equation}

\section*{Supplementary Note 7: Temperature calibration}

To perform a calibration of the RF resonator thermal occupation, we vary the fridge temperature $T_\mathrm{f}$ in steps of $20\,$mK and take a series of measurements for each $T_\mathrm{f}$.
During this procedure, we keep the flux bias constant at $\Phi_\mathrm{b}/\Phi_0 = 0.54$.
First, we take bare response measurements of the two modes $S_{11}^\mathrm{RF}$ and $S_{11}^\mathrm{HF}$.
From the response curves, we extract the resonance frequencies $\omega_0$ and $\Omega_0$ as well as the linewidths $\kappa$, $\kappa_\mathrm{e}$ and $\Gamma_{i}$, $\Gamma_\mathrm{e}$ by fitting the responses using Eq.~(\ref{eq:ResponsefuncHF}).
The linewidths are shown in Supplementary Fig.~\ref{fig:decays_Ts}.

\begin{figure}[h]
	\centerline {\includegraphics[trim={1.5cm 13.5cm 1cm 6.5cm},clip=True,scale=0.75]{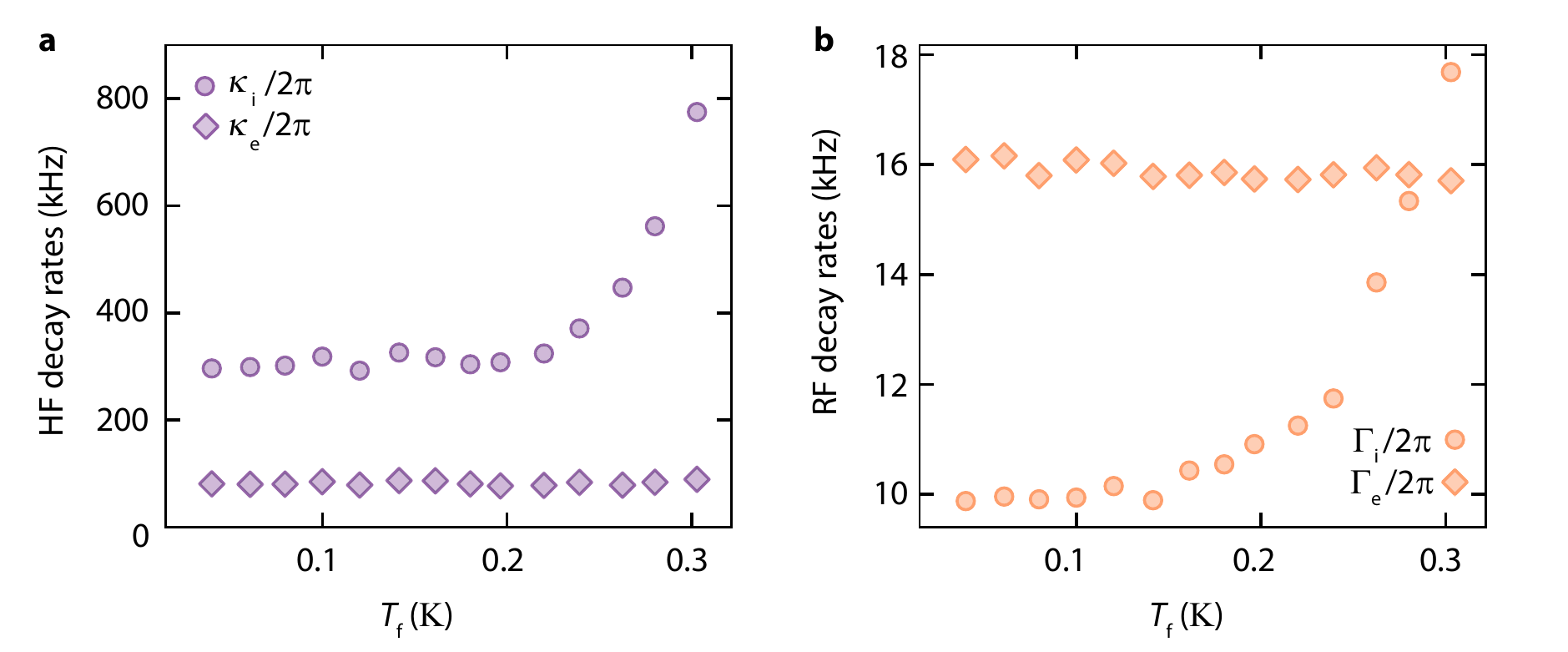}}
	\caption{\textsf{\textbf{Linewidths vs temperature.} Decay rates of the microwave mode $\kappa$, $\kappa_\mathrm{e}$ in panel \textbf{a} and of the radio-frequency mode $\Gamma_\mathrm{i}$ and $\Gamma_\mathrm{e}$ in panel \textbf{b} depending on fridge temperature $T_\mathrm{f}$.}}
	\label{fig:decays_Ts}
\end{figure}

Afterwards, we set a pump tone of constant power to the red sideband of the HF cavity and detect both, the photon-pressure induced transparency (PPIT) in a reflection measurement of the HF cavity and the upconverted RF thermal noise power spectrum.
We adjust the pump power for each temperature to keep the cooperativity $\mathcal{C}\leq 1$ in order to avoid being too close to the normal-mode splitting regime but still obtain a good signal.
From the PPIT data, we determine the actual cooperativity.
Using the resonance frequencies and the linewidths, we determine the thermal photon occupation number in the RF mode relative to the added noise photons from the detected thermal noise curve for each temperature as described in Supplementary Note~6.
We repeat the output noise detection for two distinct experimental conditions which correspond to two different RF mode temperatures.
The feedline of the RF circuit is connected to a cryogenic radio-frequency amplifier and as we do not have a radio-frequency isolator or circulator, the noise emitted by this amplifier will reach the RF input with only small attenuation.
Hence, we can increase or decrease the RF input noise by switching this amplifier on or off, respectively, cf. Supplementary Fig~\ref{fig:ampON}\textbf{b}.
As first step in the data analysis and temperature calibration, we use the "amplifier off" dataset, i.e., the "cold" RF mode to estimate the number of added photons $n_\mathrm{add}'$.
We experimentally observe that the effective RF mode occupation increases almost linearly for the larger fridge temperatures and saturates at a constant value for low fridge temperatures, cf. Supplementary Fig.~11\textbf{c} or main paper Fig.~3\textbf{b}.
This behaviour indicates that for large fridge temperatures the RF bath follows the base temperature, but that for low temperatures the RF bath temperature is not completely equilibrating to the fridge.
To capture this observation phenomenologically, we assume the effective bath temperature to be given by $T_\mathrm{RF} = \sqrt{T_\mathrm{f}^2 + T_\mathrm{r}^2}$ with the fridge temperature $T_\mathrm{f}$ and the residual mode temperature $T_\mathrm{r}$.
The resulting RF mode occupation is then given by
\begin{equation}
n^\mathrm{RF} =  \left(e^{\frac{\hbar\omega_0}{k_\mathrm{B}T_\mathrm{RF}}}-1\right)^{-1}.
\end{equation}
This function will lead to a curve very similar to the observations, i.e., to a linear increase of $n^\mathrm{RF}$ for large base temperatures and to a gradual saturation at a residual occupation at low temperatures at $T_\mathrm{RF} \approx T_\mathrm{r}\gg T_\mathrm{f}$.
We use the normalized occupation data extracted from fits to the experimental power spectral densities $\tilde{n} = n_\mathrm{th}^\mathrm{RF}(T_\mathrm{f})/\left(n_\mathrm{add}' + \frac{1}{2} \right)$ using Eq.~(\ref{eq:PSD}) now and fit these extracted and normalized data points with the function $\alpha n^\mathrm{RF}(T_\mathrm{f})$ with $\alpha = \left( n_\mathrm{add}' + 1/2 \right)^{-1}$ using $T_\mathrm{r}$ and $\alpha$ as fit parameters.
We obtain $T_\mathrm{r} \approx 141\,$mK and $n_\mathrm{add}' = \frac{1}{\alpha} - \frac{1}{2} \approx 10.6$.
However, our results depend slightly on the chosen function for the effective temperature.
We can get a similar behaviour for any function $T_\mathrm{RF} = \left( T_\mathrm{f}^k + T_\mathrm{r}^k \right)^{\frac{1}{k}} $ with $k$ being approximately in the range $1.5 \leq k \leq 4$ and which leads to $n_\mathrm{add}'$ values in the range $9 \leq n_\mathrm{add}' \leq 12$.
To take this spread into account, which arises from using only a phenomenological model function, we will use an average value of $n_\mathrm{add}' \approx 11 \pm 2$.
This value for $n_\mathrm{add}'$ is consistent with an HF HEMT noise temperature of $\sim2\,$K (datasheet) and $2.7\,$dB attenuation between sample and amplifier.
The results we obtain with $n_\mathrm{add}' = 11$ for the RF mode occupation in both RF amplifier configurations are shown in Supplementary Fig.~\ref{fig:ampON}\textbf{c}.
For the RF amplifier switched off, we obtain a residual occupation of $7\pm1$ photons at the fridge base temperature, which stays nearly constant until the fridge reaches about $150\,$mK.
Note, that this residual value of $\sim 7$ photons is considerably larger than what would be theoretically expected from the Bose distribution for a complete thermalization to the base temperature, which would be $n_\mathrm{Bose}^\mathrm{RF}(20\,\rm mK) \approx 0.5$, cf. also black line in Supplementary Fig.~11\textbf{c}.
We will discuss possible reasons for this deviation below.
From around $100-150\,$mK upwards, the RF occupation starts to increase and approaches the Bose occupation shown as black line.
For the configuration with the amplifier switched on, the residual occupation is about a factor of three larger due to the increased noise coming along the input/output line from the amplifier and coupling into the circuit.
Still, a slight increase of the total occupation with increasing fridge temperature is visible, which can be attributed to an increase of the internal RF mode bath temperature.
In both amplifier states, the residual occupation of the RF mode is considerably larger than what is expected from the Bose factor and for full thermalization.
The observation that the RF amplifier state (on vs off) has such a large impact, however, lets us believe that the main contribution to the deviation from the Bose occupation is the noise propagating along the RF feedline from the unisolated amplifier to the sample.
The amplifier is mounted in between the $800\,$mK plate and the $3\,$K plate, and except for a directional coupler, a $1\,$dB attenuator and cable losses is not isolated further from the device, cf. also Supplementary Fig.~3.
Although for a detailed and concise analysis of the origin and the mechanism behind this residual occupation, a precise knowledge of the individual bath temperatures and their dependence on the fridge temperature would be necessary, we will try to shed some light onto the internal and external contribution to the RF mode occupation with an extension of the phenomenological model used already above.
To model the thermal occupation taking into account contributions from internal and external baths, we use
\begin{equation}
n_\mathrm{th}^\mathrm{RF} = \frac{\Gamma_\mathrm{i}}{\Gamma_0}n_\mathrm{i}^\mathrm{RF} + \frac{\Gamma_\mathrm{e}}{\Gamma_0}n_\mathrm{e}^\mathrm{RF}
\end{equation}
where the internal and external occupations are given by 
\begin{equation}
n_\mathrm{i/e}^\mathrm{RF} =  \left(e^{\frac{\hbar\omega_0}{k_\mathrm{B}T_\mathrm{i/e}}}-1\right)^{-1}.
\end{equation}
For the effective internal and external temperatures $T_\mathrm{i}$ and $T_\mathrm{e}$, respectively, we take into account possible devations from the fridge temperature by a residual temperature $T_\mathrm{r, i/e}$, which e.g. considers the RF amplifier noise arriving at the sample RF input.
To phenomenologiclly model a gradual adjustment of the effective bath temperatures to the fridge temperature, we use $T_\mathrm{i/e} = \sqrt{T_\mathrm{f}^2 + T_\mathrm{r, i/e}^2}$.
In addition, we fit the temperature dependence of $\Gamma_\mathrm{i}$ and take it into account in the calculation of $n_\mathrm{th}^\mathrm{RF}$.
For the case of RF amplifier off, we observe an increased RF mode linewidth $\Gamma_0 \approx 2\pi\cdot 40\,$kHz, which we attribute to two-level systems being saturated with the amplifier on.
With all these factors considered, we obtain the lines shown in Fig.~\ref{fig:ampON}\textbf{c} from a simultaneous fit of the ON and OFF datasets, giving good qualitative agreement with the data.
As fit parameters, we used $T_\mathrm{r, i}, T_\mathrm{r, e}^\mathrm{on}$ and $T_\mathrm{r, e}^\mathrm{off}$.
Hence, the parameter set for both curves is identical, except for the residual temperature of the external bath $T_\mathrm{r, e}$, which is modified by the amplifier power state, and for $\Gamma_0$, which differs between the amplifier ON and OFF states.
The extracted temperatures are $T_\mathrm{r, i} = 31\,$mK, $T_\mathrm{r, e}^\mathrm{on} = 743\,$mK and $T_\mathrm{r, e}^\mathrm{off} = 285\,$mK.
Furthermore, the internal and external linewidths at base temperature were fixed to $\Gamma_\mathrm{e} = 2\pi\cdot 16\,$kHz, $\Gamma_\mathrm{i}^\mathrm{on} = 2\pi\cdot 10\,$kHz, $\Gamma_\mathrm{i}^\mathrm{off} = 2\pi\cdot 24\,$kHz.
Within this model, we find our intuitive explanation confirmed, i.e., that the noise propagating from the hot RF amplifier to the device is the main origin for the large residual occupation at base temperature.
We also note, however, that the results of this fitting depend significantly on the exact model used for the effective bath temperatures and that more refined experiments would be required to fully and reliably determine the mechanisms and noise processes of the RF mode.

\section*{Supplementary Note 8: Cooling the RF mode with increased thermal occupancy}

We repeat the cooling experiment discussed in the main paper Fig.~3 also for the RF amplifier switched on, leaving the RF mode occupied with about 20.5 thermal photons at fridge base temperature.
The resulting spectra of the sideband-cooling in this state are shown in Supplementary Fig.~\ref{fig:ampON} and besides a larger amplitude due to the increased occupancy look nearly identical to the spectra shown in the main paper.
The corresponding cooled RF photons are plotted in \textbf{e} together with the data for the amplifier switched off.

\begin{figure}[h]
	\centerline {\includegraphics[trim={1.0cm 9.5cm 1.0cm 0.0cm},clip=True, width=0.95\textwidth]{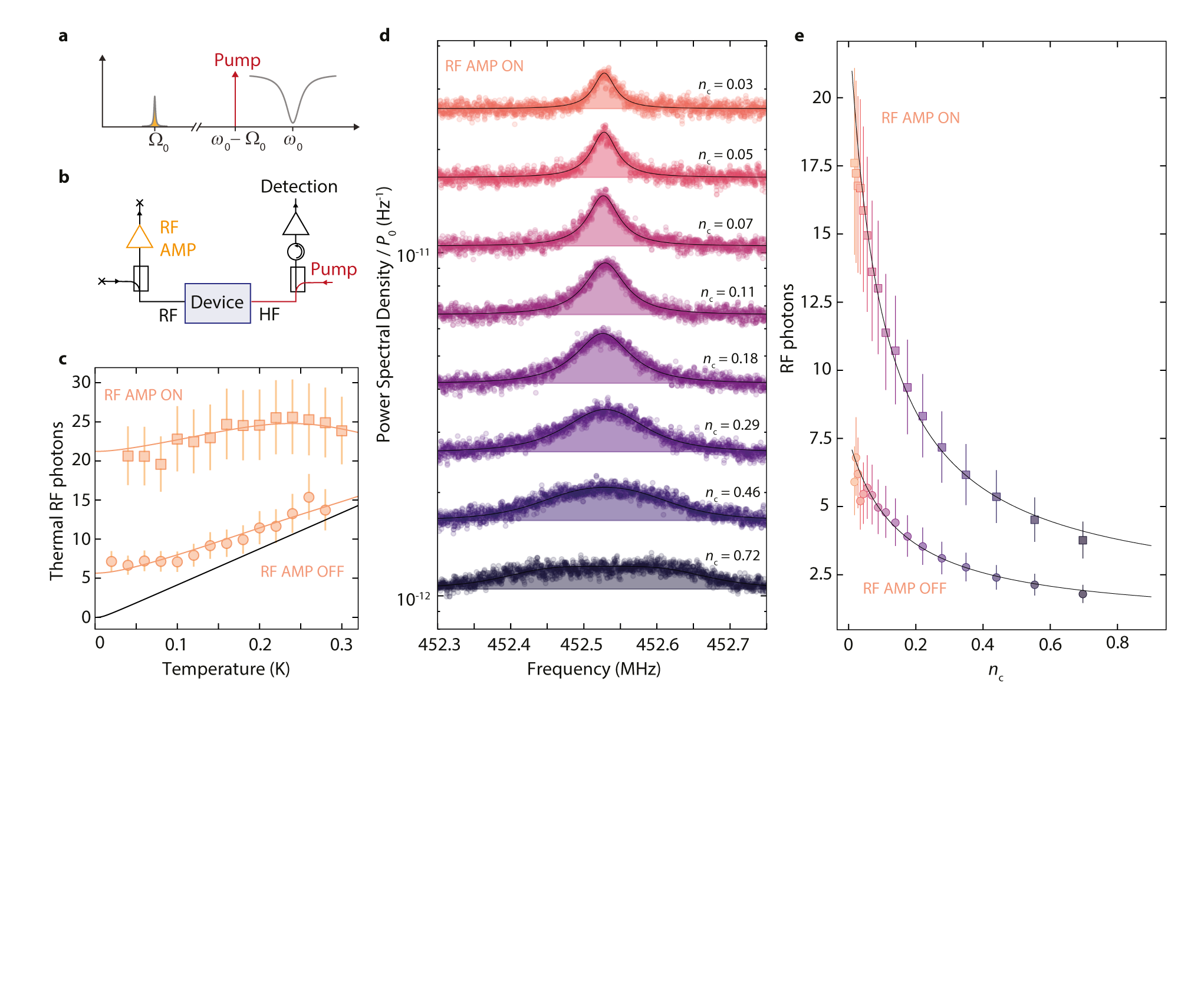}}
	\caption{\textsf{\textbf{Photon-pressure sideband-cooling of a hot and a hotter RF resonator.} \textbf{a} For the observation of upconverted thermal noise of the RF resonator, a pump tone is set to the red sideband of the high-frequency mode $\omega_\mathrm{p} = \omega_0-\Omega_0$ and the cavity output field around $\omega = \omega_0$ is detected with a signal analyzer. The RF input/output side of the device is connected to a cryogenic radio-frequency amplifier, which is used for the reflection characterization of the RF mode, cf. panel \textbf{b}. The state of this RF amplifier can be used to control the thermal occupation of the RF mode. When it is switched ON, its output noise increases the thermal occupation of the RF mode as shown in panel \textbf{c}, where the thermal photon number vs fridge temperature is plotted for both cases, RF amplifier switched ON and RF amplifier switched OFF. Symbols are data, black line is the Bose factor, orange lines are models for the thermal occupation and discussed in the text. From the thermal calibration, we determine the thermal occupation of the RF mode at fridge base temperature to be $n_\mathrm{RF}^\mathrm{on} \sim 20.5$ and $n_\mathrm{RF}^\mathrm{off} \sim 7$. \textbf{d} shows the measured high-frequency output power spectral density for increasing red-sideband pump power, normalized to the on-chip pump power $P_0$ for RF amplifier ON. Frequency axis is given with respect to the constant pump frequency. Circles are data, lines and shaded areas are fits. With increasing pump strength, i.e., increasing intracavity photon number $n_\mathrm{c}$, the RF resonance gets broadened by photon-pressure damping, and its total thermal noise power gets reduced, which corresponds to cooling of the mode. In \textbf{e}, the thermal RF mode occupation is shown as symbols vs pump photon number $n_\mathrm{c}$ for both cases, RF amplifier ON (squares) and RF amplifier OFF (circles). Error bars for the amplifier ON (OFF) data correspond to uncertainties of $\pm 2$ HF photons of added noise in the detection chain and $\pm 1\,$kHz ($\pm 2\,$kHz) in bare RF linewidth $\Gamma_0^{\textrm{on}} = 26\,$kHz ($\Gamma_0^{\textrm{off}} = 40\,$kHz).}}
	\label{fig:ampON}
\end{figure}

\section*{Supplementary Note 9: Theory of normal-mode thermometry}
\subsection*{The high-frequency response function with normal mode susceptibilities}
From the equations of motion, we obtained the response of the system around the HF mode as
\begin{equation}
S_{11}^\mathrm{HF} = 1-\kappa_\mathrm{e}\frac{\chi_\mathrm{c}}{1+g^2\chi_c\chi_0}
\end{equation}
under the assumption of pumping around the red sideband and the sideband-resolved regime. 
The resonance condition $\left(\chi_\mathrm{c}^\mathrm{eff}\right)^{-1} = 0$ for the effective HF cavity susceptibility
\begin{equation}
\chi_\mathrm{c}^\mathrm{eff} = \frac{\chi_\mathrm{c}}{1+g^2\chi_\mathrm{c}\chi_0}
\end{equation}
provided us with the complex solutions of the effective susceptibility
\begin{equation}
\tilde{\omega}_\pm = \omega_0 + \frac{\delta}{2} + i\frac{\kappa + \Gamma_0}{4} \pm \sqrt{g^2 - \left(\frac{\kappa - \Gamma_0 +2i\delta}{4}\right)^2}
\end{equation}
where $\delta$ is the pump detuning from the red HF cavity sideband.
Now, we define the normal mode susceptibilities
\begin{equation}
\chi_+ = \frac{1}{\frac{\kappa_+}{2} + i\Delta_+}, ~~~ \chi_- = \frac{1}{\frac{\kappa_-}{2} + i\Delta_-}
\end{equation}
where $\Delta_\pm = \omega - \omega_\pm$ and
\begin{equation}
\omega_\pm = \mathrm{Re}[\tilde{\omega_\pm}], ~~~ \kappa_\pm = \mathrm{Im}[\tilde{\omega_\pm}]
\end{equation}
are the real and imaginary parts, respectively, of the complex solutions.
With these, we can rewrite the HF response function as
\begin{equation}
S_{11}^\mathrm{HF} = 1-\kappa_\mathrm{e, +}\chi_+ -\kappa_\mathrm{e, -}\chi_-
\end{equation}
which is exact with the (complex and frequency-dependent) external linewidths
\begin{equation}
\kappa_\mathrm{e, \pm} = \mp\frac{i\kappa_\mathrm{e}}{2\chi_0\sqrt{g^2 - \left(\frac{\kappa - \Gamma_0 + 2i\delta}{4}\right)^2}}.
\end{equation}
For the regime of considerable coupling $g \gg \kappa/2, \Gamma_0/2$ and possibly large detunings $\Delta \lesssim g$, we approximate this by
\begin{equation}
\kappa_\mathrm{e, \pm} \approx \frac{\kappa_\mathrm{e}}{2}\left(1\pm \frac{\delta}{\sqrt{\delta^2 + 4g^2}}\right).
\end{equation}
\subsection*{Normal-mode thermometry}
Using the approximated normal-mode representation of the HF cavity in the strong-coupling regime, we get for the output field power spectral density in units of quanta
\begin{equation}
\frac{S_\mathrm{nms}}{\hbar\omega} = \frac{1}{2} + n_\mathrm{add}' + 4\frac{\kappa_\mathrm{e,+} \kappa_\mathrm{i,+}}{\kappa_+^2 + 4\Delta_+^2}\left(n_\mathrm{i, +} - n_\mathrm{e, +} \right) + 4\frac{\kappa_\mathrm{e,-}\kappa_\mathrm{i,-}}{\kappa_-^2 + 4\Delta_-^2}\left(n_\mathrm{i, -} - n_\mathrm{e, -} \right)
\end{equation}
where $\kappa_\mathrm{i, \pm} = \kappa_\pm - \kappa_\mathrm{e, \pm}$ and $n_\mathrm{e, +}, n_\mathrm{e, -}$ and $n_\mathrm{i, +}, n_\mathrm{i, -}$ are the effective external and internal bath occupations of the normal modes, respectively.
If the external baths are given again by the fridge temperature, we get $n_\mathrm{e, \pm} \ll n_\mathrm{i, \pm} \approx \frac{\kappa_\pm}{\kappa_\mathrm{i}, \pm} n_\pm$.
From the condition $S_\mathrm{nms} = S$, we can then follow
\begin{equation}
\kappa_\mathrm{e,+}n_+ + \kappa_\mathrm{e,-}n_- = \kappa_\mathrm{e} n_\mathrm{cool}^\mathrm{HF}
\end{equation}
where $n_\mathrm{cool}^\mathrm{HF}$ is the thermal occupation (imbalance) of the HF mode while cooling the RF mode with a red sideband tone. 
As $n_\mathrm{cool}^\mathrm{HF} \gg n_\mathrm{e}^\mathrm{HF}$, the imbalance occupation corresponds in good approximation to the total occupation. 
So this way we can calculate the effective HF cavity occupation which is given by
\begin{equation}
n_\mathrm{cool}^\mathrm{HF} = \frac{\kappa}{\kappa + \Gamma_0}\frac{4g^2 + \Gamma_0(\kappa + \Gamma_0)\left[1+\frac{4\delta^2}{(\kappa + \Gamma_0)^2}\right]}{4g^2 + \kappa\Gamma_0\left[1+\frac{4\delta^2}{(\kappa + \Gamma_0)^2}\right]}n_\mathrm{th}^\mathrm{HF} + \frac{\Gamma_0}{\kappa + \Gamma_0}\frac{4g^2}{4g^2 + \kappa\Gamma_0\left[1+\frac{4\delta^2}{(\kappa + \Gamma_0)^2}\right]}n_\mathrm{th}^\mathrm{RF}.
\end{equation}
when red-sideband driving.
The total number of noise photons in the hybridized mode regime is then given by
\begin{equation}
n_\mathrm{tot} = n_\mathrm{cool}^\mathrm{HF} + n_\mathrm{cool}^\mathrm{RF}.
\end{equation}

\section*{Supplementary Note 10: Determination of system parameters for normal-mode thermometry}

\begin{figure}[h]
	\centerline {\includegraphics[trim={0cm 7cm 0cm 0cm},clip=True,scale=0.6]{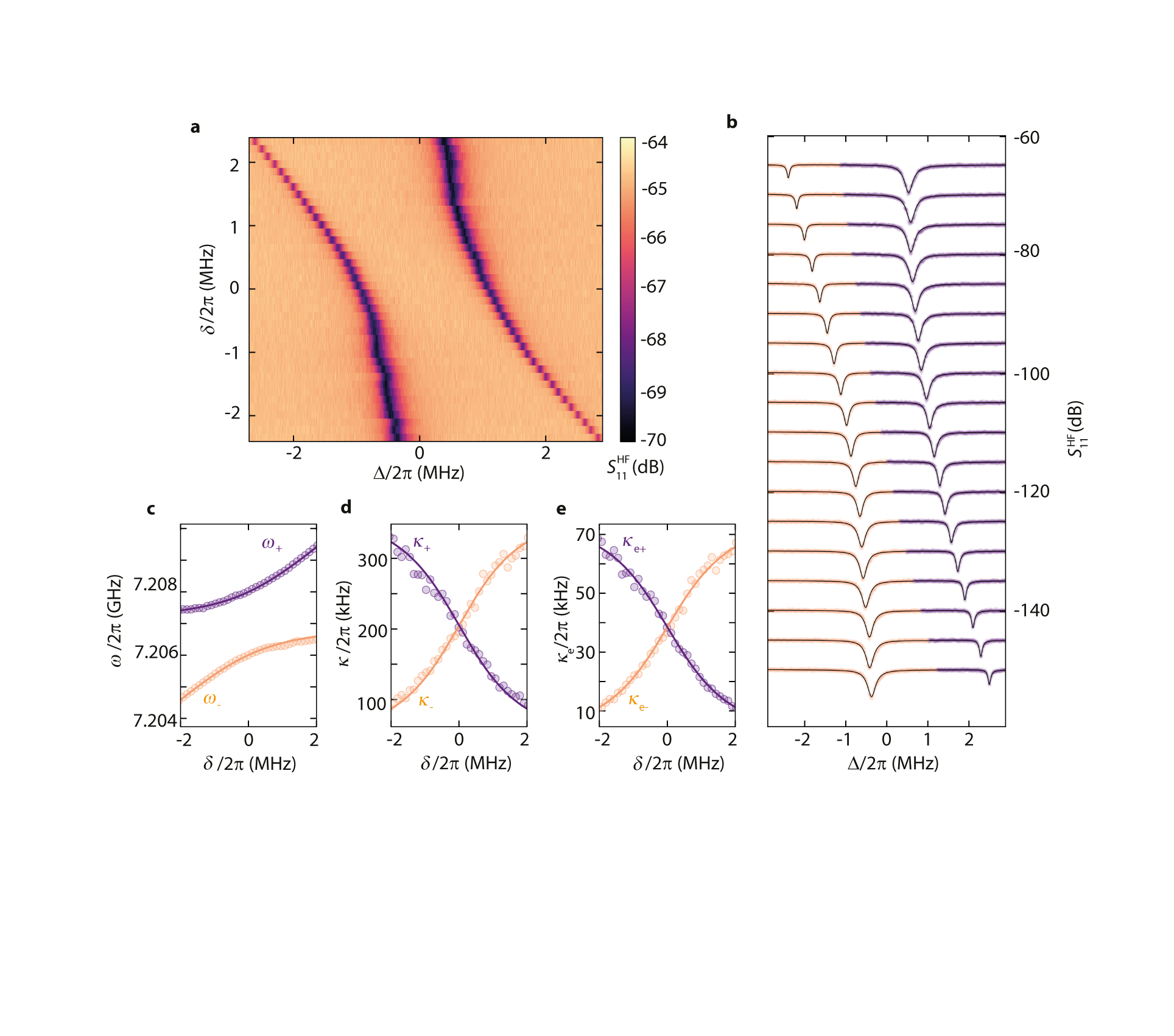}}
	\caption{\textsf{\textbf{HF response in the strong-coupling regime and normal-mode fit parameters.} \textbf{a} Color-coded HF reflection response in the strong-coupling regime. The pump tone is swept through the HF mode red sideband with $\omega_\mathrm{p} = \omega_0 - \Omega_0 + \delta$. The response frequency is given relative to the bare HF cavity mode $\Delta = \omega - \omega_0$. \textbf{b} Linecuts of \textbf{a}, showing the individually fitted parts of the response as orange and purple. The top curve is plotted as measured (unshifted), subsequent curves are manually downshifted by $-5\,$dB each for clarity. Shown is every second linescan of \textbf{a}. Fit curves are plotted as black lines. From the fit curves shown in \textbf{b}, we get the three relevant normal-mode parameters resonance frequency $\omega_\pm$, total decay rate $\kappa_\pm$ and external decay rate $\kappa_\mathrm{e,\pm}$, plotted in \textbf{c}, \textbf{d}, and \textbf{e}, respectively. }}
	\label{fig:NMS}
\end{figure}

For the analysis of the normal-mode thermal spectra, it is essential to know the individual mode parameters, which we obtain from a characterization of the HF reflection response in the strong-coupling regime.
Supplementary Fig.~\ref{fig:NMS}\textbf{a} shows $S_{11}^\mathrm{HF}$ for approximately $n_\mathrm{c} = 100$ pump intracavity photons.
This data was measured simultaneously with the power spectral densities (PSD) shown in main paper Fig.~4, but during the measurement of the PSDs the network analyzer was switched off.
Both datasets were acquired while iteratively sweeping a pump tone through the red sideband with $\omega_\mathrm{p} = \omega_0 - \Omega_0 + \delta$.
As the pump tone approaches exactly the red sideband frequency, we observe photon-pressure induced hybridization between the modes with a resonant splitting of $g/\pi\sim 2.1\,$MHz, about one order of magnitude larger than the normal-mode linewidths $(\kappa + \Gamma_0)/2 = 2\pi\cdot 200\,$kHz and a factor of three larger than the RF thermal decoherence rate $\Gamma_0 n_\mathrm{th}^\mathrm{RF} \approx 2\pi\cdot300\,$kHz.
For an analysis of the normal modes, we denote the lower-frequency mode with the superscript $-$ and the higher-frequency mode with $+$, i.e., the resonance frequencies are identified as $\omega_-$ and $\omega_+$ and the total and external linewidths as $\kappa_-$, $\kappa_+$ and $\kappa_\mathrm{e,-}$, $\kappa_\mathrm{e,+}$, respectively.
The theoretical description of these parameters was given in Supplementary Note~9.
To determine these parameters from the experimental data, each linescan of Supplementary Fig.~\ref{fig:NMS}\textbf{a} was split into two parts, each of them containing one of the two resonances.
This splitting is shown in \textbf{b} by using two different colors, orange for the range of $\omega_-$ and purple for the range of $\omega_+$.
Both sub-responses were fitted individually using Eq.~(\ref{eq:ResponsefuncHF}).
The returned fit parameters $\omega_\pm$, $\kappa_\pm$ and $\kappa_\mathrm{e\pm}$ are plotted in Supplementary Figs.~\ref{fig:NMS}\textbf{c}, \textbf{d}, and \textbf{e}, respectively.
The theoretical dependences as derived in Supplementary Note~9 are plotted as lines and show excellent agreement with the data.
When the two modes are completely hybridized ($\delta = 0$), we find for both a total decay rate $\kappa_+ = \kappa_- = 2\pi\cdot 200\,$kHz, which is what we expect from the theoretical expression for this case $\kappa_\pm = (\kappa + \Gamma_0)/2$.
Also the external decay rates, usually defined by the coupling capacitance to the feedline, are half the value of the bare HF mode $\kappa_\mathrm{e,+} = \kappa_\mathrm{e,-} = 2\pi\cdot40\,$kHz.

\begin{thebibliography}{45}
	
	\expandafter\ifx\csname natexlab\endcsname\relax\def\natexlab#1{#1}\fi
	\expandafter\ifx\csname bibnamefont\endcsname\relax
	\def\bibnamefont#1{#1}\fi
	\expandafter\ifx\csname bibfnamefont\endcsname\relax
	\def\bibfnamefont#1{#1}\fi
	\expandafter\ifx\csname citenamefont\endcsname\relax
	\def\citenamefont#1{#1}\fi
	\expandafter\ifx\csname url\endcsname\relax
	\def\url#1{\texttt{#1}}\fi
	\expandafter\ifx\csname urlprefix\endcsname\relax\def\urlprefix{URL }\fi
	\providecommand{\bibinfo}[2]{#2}
	\providecommand{\eprint}[2][]{\url{#2}}
	
	
	
	\subsection*{References}
	
	\bibitem[{\citenamefont{Aspelmeyer et~al.}(2014)\citenamefont{Aspelmeyer, Kippenberg, Marquardt}}]{Aspelmeyer14}
	\bibinfo{author}{\bibfnamefont{M.}~\bibnamefont{Aspelmeyer}},
	\bibinfo{author}{\bibfnamefont{T.~J.}~\bibnamefont{Kippenberg}},
	\bibinfo{author}{\bibfnamefont{F.}~\bibnamefont{Marquardt}},
	\bibinfo{title}{\bibfnamefont{Cavity optomechanics}}.
	\textit{\bibinfo{journal}{Reviews of Modern Physics}} \textbf{\bibinfo{volume}{86}},
	\bibinfo{pages}{1391} (\bibinfo{year}{2014}).
	
	\bibitem[{\citenamefont{Teufel et~al.}(2009)\citenamefont{Teufel, Donner, Castellanos-Beltran, Harlow, Lehnert}}]{Teufel09}
	\bibinfo{author}{\bibfnamefont{J.~D.}~\bibnamefont{Teufel}},
	\bibinfo{author}{\bibfnamefont{T.}~\bibnamefont{Donner}},
	\bibinfo{author}{\bibfnamefont{M.~A.}~\bibnamefont{Castellanos-Beltran}},
	\bibinfo{author}{\bibfnamefont{J.~W.}~\bibnamefont{Harlow}},
	\bibinfo{author}{\bibfnamefont{K.~W.}~\bibnamefont{Lehnert}},
	\bibinfo{title}{\bibfnamefont{Nanomechanical motion measured with an imprecision below that at the standard quantum limit}}.
	\textit{\bibinfo{journal}{Nature Nanotechnology}} \textbf{\bibinfo{volume}{4}},
	\bibinfo{pages}{820-823} (\bibinfo{year}{2009}).
	
	\bibitem[{\citenamefont{Wollman et~al.}(2015)\citenamefont{Wollman, Lei, Weinstein, Suh, Kronwald, Marquardt, Clerk, Schwab}}]{Wollman15}
	\bibinfo{author}{\bibfnamefont{E.~E.}~\bibnamefont{Wollman}},
	\bibinfo{author}{\bibfnamefont{C.~U.}~\bibnamefont{Lei}},
	\bibinfo{author}{\bibfnamefont{A.~J.}~\bibnamefont{Weinstein}},
	\bibinfo{author}{\bibfnamefont{J.}~\bibnamefont{Suh}},
	\bibinfo{author}{\bibfnamefont{A.}~\bibnamefont{Kronwald}},
	\bibinfo{author}{\bibfnamefont{F.}~\bibnamefont{Marquardt}},
	\bibinfo{author}{\bibfnamefont{A.~A.}~\bibnamefont{Clerk}},
	\bibinfo{author}{\bibfnamefont{K.~C.}~\bibnamefont{Schwab}},
	\bibinfo{title}{\bibfnamefont{Quantum squeezing of motion in a mechanical resonator}}.
	\textit{\bibinfo{journal}{Science}} \textbf{\bibinfo{volume}{349}},
	\bibinfo{pages}{952-955} (\bibinfo{year}{2015}).
	
	\bibitem[{\citenamefont{Xu et~al.}(2016)\citenamefont{Xu, Mason, Jiang, Harris}}]{Xu16}
	\bibinfo{author}{\bibfnamefont{H.}~\bibnamefont{Xu}},
	\bibinfo{author}{\bibfnamefont{D.}~\bibnamefont{Mason}},
	\bibinfo{author}{\bibfnamefont{Luyao}~\bibnamefont{Jiang}},
	\bibinfo{author}{\bibfnamefont{J.~G.~E.}~\bibnamefont{Harris}},
	\bibinfo{title}{\bibfnamefont{Topological energy transfer in an optomechanical system with exceptional points}}.
	\textit{\bibinfo{journal}{Nature}} \textbf{\bibinfo{volume}{537}},
	\bibinfo{pages}{80-83} (\bibinfo{year}{2016}).
	
	\bibitem[{\citenamefont{Reed et~al.}(2017)\citenamefont{Reed, Mayer, Teufel, Burkhart, Pfaff, Reagor, Sletten, Ma, Schoelkopf, Knill, Lehnert}}]{Reed17}
	\bibinfo{author}{\bibfnamefont{A.~P.}~\bibnamefont{Reed}},
	\bibinfo{author}{\bibfnamefont{K.~H.}~\bibnamefont{Mayer}},
	\bibinfo{author}{\bibfnamefont{J.~D.}~\bibnamefont{Teufel}},
	\bibinfo{author}{\bibfnamefont{L.~D.}~\bibnamefont{Burkhart}},
	\bibinfo{author}{\bibfnamefont{W.}~\bibnamefont{Pfaff}},
	\bibinfo{author}{\bibfnamefont{M.}~\bibnamefont{Reagor}},
	\bibinfo{author}{\bibfnamefont{L.}~\bibnamefont{Sletten}},
	\bibinfo{author}{\bibfnamefont{X.}~\bibnamefont{Ma}},
	\bibinfo{author}{\bibfnamefont{R.~J.}~\bibnamefont{Schoelkopf}},
	\bibinfo{author}{\bibfnamefont{E.}~\bibnamefont{Knill}},
	\bibinfo{author}{\bibfnamefont{K.~W.}~\bibnamefont{Lehnert}},
	\bibinfo{title}{\bibfnamefont{Faithful conversion of propagating quantum information to mechanical motion}}.
	\textit{\bibinfo{journal}{Nature Physics}} \textbf{\bibinfo{volume}{13}},
	\bibinfo{pages}{1163-1167} (\bibinfo{year}{2017}).
	
	\bibitem[{\citenamefont{Riedinger et~al.}(2018)\citenamefont{Riedinger, Wallucks, Marinkovi\'c, L\"oschnauer, Aspelmeyer, Hong, Gr\"oblacher}}]{Riedinger18}
	\bibinfo{author}{\bibfnamefont{R.}~\bibnamefont{Riedinger}},
	\bibinfo{author}{\bibfnamefont{A.}~\bibnamefont{Wallucks}},
	\bibinfo{author}{\bibfnamefont{I.}~\bibnamefont{Marinkovi\'c}},
	\bibinfo{author}{\bibfnamefont{C.}~\bibnamefont{L\"oschnauer}},
	\bibinfo{author}{\bibfnamefont{M.}~\bibnamefont{Aspelmeyer}},
	\bibinfo{author}{\bibfnamefont{S.}~\bibnamefont{Hong}},
	\bibinfo{author}{\bibfnamefont{S.}~\bibnamefont{Gr\"oblacher}},
	\bibinfo{title}{\bibfnamefont{Remote quantum entanglement between two micromechanical oscillators}}.
	\textit{\bibinfo{journal}{Nature}} \textbf{\bibinfo{volume}{556}},
	\bibinfo{pages}{473-477} (\bibinfo{year}{2018}).
	
	\bibitem[{\citenamefont{Ockeloen-Korppi et~al.}(2018)\citenamefont{Ockeloen-Korppi, Damsk\"agg, Pirkkalainen, Asjad, Clerk, Massel, Woolley, Sillanp\"a\"a}}]{OckeloenKorppi18}
	\bibinfo{author}{\bibfnamefont{C.~F.}~\bibnamefont{Ockeloen-Korppi}},
	\bibinfo{author}{\bibfnamefont{E.}~\bibnamefont{Damsk\"agg}},
	\bibinfo{author}{\bibfnamefont{J.-M.}~\bibnamefont{Pirkkalainen}},
	\bibinfo{author}{\bibfnamefont{M.}~\bibnamefont{Asjad}},
	\bibinfo{author}{\bibfnamefont{A.~A.}~\bibnamefont{Clerk}},
	\bibinfo{author}{\bibfnamefont{F.}~\bibnamefont{Massel}},
	\bibinfo{author}{\bibfnamefont{M.~J.}~\bibnamefont{Woolley}},
	\bibinfo{author}{\bibfnamefont{M.~A.}~\bibnamefont{Sillanp\"a\"a}},
	\bibinfo{title}{\bibfnamefont{Stabilized entanglement of massive mechanical oscillators}}.
	\textit{\bibinfo{journal}{Nature}} \textbf{\bibinfo{volume}{556}},
	\bibinfo{pages}{478-482} (\bibinfo{year}{2018}).
	
	\bibitem[{\citenamefont{Xu et~al.}(2019)\citenamefont{Xu, Jiang, Clerk, Harris}}]{Xu19}
	\bibinfo{author}{\bibfnamefont{H.}~\bibnamefont{Xu}},
	\bibinfo{author}{\bibfnamefont{Luyao}~\bibnamefont{Jiang}},
	\bibinfo{author}{\bibfnamefont{A.~A.}~\bibnamefont{Clerk}},
	\bibinfo{author}{\bibfnamefont{J.~G.~E.}~\bibnamefont{Harris}},
	\bibinfo{title}{\bibfnamefont{Nonreciprocal control and cooling of phonon modes in an optomechanical system}}.
	\textit{\bibinfo{journal}{Nature}} \textbf{\bibinfo{volume}{568}},
	\bibinfo{pages}{65-69} (\bibinfo{year}{2019}).
	
	\bibitem[{\citenamefont{Cohadon et~al.}(1999)\citenamefont{Cohadon, Heidmann, Pinard}}]{Cohadon99}
	\bibinfo{author}{\bibfnamefont{P.~F.}~\bibnamefont{Cohadon}},
	\bibinfo{author}{\bibfnamefont{A.}~\bibnamefont{Heidmann}},
	\bibinfo{author}{\bibfnamefont{M.}~\bibnamefont{Pinard}},
	\bibinfo{title}{\bibfnamefont{Cooling of a Mirror by Radiation Pressure}}.
	\textit{\bibinfo{journal}{Physical Review Letters}} \textbf{\bibinfo{volume}{83}},
	\bibinfo{pages}{3174-3177} (\bibinfo{year}{1999}).
	
	\bibitem[{\citenamefont{Braginsky and Vyatchanin}(2002)\citenamefont{Braginsky, Vyatchanin}}]{Braginsky02}
	\bibinfo{author}{\bibfnamefont{V.~B.}~\bibnamefont{Braginsky}},
	\bibinfo{author}{\bibfnamefont{S.~P.}~\bibnamefont{Vyatchanin}},
	\bibinfo{title}{\bibfnamefont{Low quantum noise tranquilizer for Fabry-Perot interferometer}}.
	\textit{\bibinfo{journal}{Physics Letters A}} \textbf{\bibinfo{volume}{293}},
	\bibinfo{pages}{228-234} (\bibinfo{year}{2002}).
	
	\bibitem[{\citenamefont{H\"ohberger Metzger and Karrai}(2004)\citenamefont{H\"ohberger Metzger, Karrai}}]{Metzger04}
	\bibinfo{author}{\bibfnamefont{C.}~\bibnamefont{H\"ohberger~Metzger}},
	\bibinfo{author}{\bibfnamefont{K.}~\bibnamefont{Karrai}},
	\bibinfo{title}{\bibfnamefont{Cavity cooling of a microlever}}.
	\textit{\bibinfo{journal}{Nature}} \textbf{\bibinfo{volume}{432}},
	\bibinfo{pages}{1002-1005} (\bibinfo{year}{2004}).
	
	\bibitem[{\citenamefont{Gigan et~al.}(2006)\citenamefont{Gigan, B\"ohm, Paternostro, Blaser, Langer, Hertzberg, Schwab, B\"auerle, Aspelmeyer, Zeilinger}}]{Gigan06}
	\bibinfo{author}{\bibfnamefont{S.}~\bibnamefont{Gigan}},
	\bibinfo{author}{\bibfnamefont{H.~R.}~\bibnamefont{B\"ohm}},
	\bibinfo{author}{\bibfnamefont{M.}~\bibnamefont{Paternostro}},
	\bibinfo{author}{\bibfnamefont{F.}~\bibnamefont{Blaser}},
	\bibinfo{author}{\bibfnamefont{G.}~\bibnamefont{Langer}},
	\bibinfo{author}{\bibfnamefont{J.~B.}~\bibnamefont{Hertzberg}},
	\bibinfo{author}{\bibfnamefont{K.~C.}~\bibnamefont{Schwab}},
	\bibinfo{author}{\bibfnamefont{D.}~\bibnamefont{B\"auerle}},
	\bibinfo{author}{\bibfnamefont{M.}~\bibnamefont{Aspelmeyer}},
	\bibinfo{author}{\bibfnamefont{A.}~\bibnamefont{Zeilinger}},
	\bibinfo{title}{\bibfnamefont{Self-cooling of a micromirror by radiation pressure}}.
	\textit{\bibinfo{journal}{Nature}} \textbf{\bibinfo{volume}{444}},
	\bibinfo{pages}{67-70} (\bibinfo{year}{2006}).
	
	\bibitem[{\citenamefont{Arcizet et~al.}(2006)\citenamefont{Arcizet, Cohadon, Briant, Pinard, Heidmann}}]{Arcizet06}
	\bibinfo{author}{\bibfnamefont{O.}~\bibnamefont{Arcizet}},
	\bibinfo{author}{\bibfnamefont{P.~F.}~\bibnamefont{Cohadon}},
	\bibinfo{author}{\bibfnamefont{T.}~\bibnamefont{Briant}},
	\bibinfo{author}{\bibfnamefont{M.}~\bibnamefont{Pinard}},
	\bibinfo{author}{\bibfnamefont{A.}~\bibnamefont{Heidmann}},
	\bibinfo{title}{\bibfnamefont{Radiation-pressure cooling and optomechanical instability of a micromirror}}.
	\textit{\bibinfo{journal}{Nature}} \textbf{\bibinfo{volume}{444}},
	\bibinfo{pages}{71-74} (\bibinfo{year}{2006}).
	
	\bibitem[{\citenamefont{Kleckner and Bouwmeester}(2006)\citenamefont{Kleckner, Bouwmeester}}]{Kleckner06}
	\bibinfo{author}{\bibfnamefont{D.}~\bibnamefont{Kleckner}},
	\bibinfo{author}{\bibfnamefont{D.}~\bibnamefont{Bouwmeester}},
	\bibinfo{title}{\bibfnamefont{Sub-kelvin optical cooling of a micromechanical resonator}}.
	\textit{\bibinfo{journal}{Nature}} \textbf{\bibinfo{volume}{444}},
	\bibinfo{pages}{75-78} (\bibinfo{year}{2006}).
	
	\bibitem[{\citenamefont{Schliesser et~al.}(2006)\citenamefont{Schliesser, Del'Haye, Nooshi, Vahala, Kippenberg}}]{Schliesser06}
	\bibinfo{author}{\bibfnamefont{A.}~\bibnamefont{Schliesser}},
	\bibinfo{author}{\bibfnamefont{P.}~\bibnamefont{Del'Haye}},
	\bibinfo{author}{\bibfnamefont{N.}~\bibnamefont{Nooshi}},
	\bibinfo{author}{\bibfnamefont{K.~J.}~\bibnamefont{Vahala}},
	\bibinfo{author}{\bibfnamefont{T.~J.}~\bibnamefont{Kippenberg}},
	\bibinfo{title}{\bibfnamefont{Radiation Pressure Cooling of a Micromechanical Oscillator Using Dynamical Backaction}}.
	\textit{\bibinfo{journal}{Physical Review Letters}} \textbf{\bibinfo{volume}{97}},
	\bibinfo{pages}{243905} (\bibinfo{year}{2006}).
	
	\bibitem[{\citenamefont{Wineland et~al.}(1978)\citenamefont{Wineland, Drullinger, Walls}}]{Wineland78}
	\bibinfo{author}{\bibfnamefont{D.~J.}~\bibnamefont{Wineland}},
	\bibinfo{author}{\bibfnamefont{R.~E.}~\bibnamefont{Drullinger}},
	\bibinfo{author}{\bibfnamefont{F.~L.}~\bibnamefont{Walls}},
	\bibinfo{title}{\bibfnamefont{Radiation-Pressure Cooling of Bound Resonant Absorbers}}.
	\textit{\bibinfo{journal}{Physical Review Letters}} \textbf{\bibinfo{volume}{40}},
	\bibinfo{pages}{1639-1642} (\bibinfo{year}{1978}).
	
	\bibitem[{\citenamefont{Neuhauser et~al.}(1978)\citenamefont{Neuhauser, Hohenstatt, Toschek, Dehmelt}}]{Neuhauser78}
	\bibinfo{author}{\bibfnamefont{W.}~\bibnamefont{Neuhauser}},
	\bibinfo{author}{\bibfnamefont{M.}~\bibnamefont{Hohenstatt}},
	\bibinfo{author}{\bibfnamefont{H.}~\bibnamefont{Dehmelt}},
	\bibinfo{title}{\bibfnamefont{Optical-sideband-cooling of Visible Atom Cloud Confined in Parabolic Well}}.
	\textit{\bibinfo{journal}{Physical Review Letters}} \textbf{\bibinfo{volume}{41}},
	\bibinfo{pages}{233-236} (\bibinfo{year}{1978}).
	
	\bibitem[{\citenamefont{Diedrich et~al.}(1989)\citenamefont{Diedrich, Bergquist, Itano, Wineland}}]{Diedrich89}
	\bibinfo{author}{\bibfnamefont{F.}~\bibnamefont{Diedrich}},
	\bibinfo{author}{\bibfnamefont{J.~C.}~\bibnamefont{Bergquist}},
	\bibinfo{author}{\bibfnamefont{W.~M.}~\bibnamefont{Itano}},
	\bibinfo{author}{\bibfnamefont{D.~J.}~\bibnamefont{Wineland}},
	\bibinfo{title}{\bibfnamefont{Laser Cooling to the Zero-Point Energy of Motion}}.
	\textit{\bibinfo{journal}{Physical Review Letters}} \textbf{\bibinfo{volume}{62}},
	\bibinfo{pages}{403-406} (\bibinfo{year}{1989}).
	
	\bibitem[{\citenamefont{Hamann et~al.}(1998)\citenamefont{Hamann, Haycock, Klose, Pax, Deutsch, Jessen}}]{Hamann98}
	\bibinfo{author}{\bibfnamefont{S.~E.}~\bibnamefont{Hamann}},
	\bibinfo{author}{\bibfnamefont{D.~L.}~\bibnamefont{Haycock}},
	\bibinfo{author}{\bibfnamefont{G.}~\bibnamefont{Klose}},
	\bibinfo{author}{\bibfnamefont{P.~H.}~\bibnamefont{Pax}},
	\bibinfo{author}{\bibfnamefont{I.~H.}~\bibnamefont{Deutsch}},
	\bibinfo{author}{\bibfnamefont{P.~S.}~\bibnamefont{Jessen}},
	\bibinfo{title}{\bibfnamefont{Resolved-Sideband Raman Cooling to the Ground State of an Optical Lattice}}.
	\textit{\bibinfo{journal}{Physical Review Letters}} \textbf{\bibinfo{volume}{80}},
	\bibinfo{pages}{4149-4152} (\bibinfo{year}{1998}).
	
	\bibitem[{\citenamefont{Teufel et~al.}(2011)\citenamefont{Teufel, Donner, Li, Harlow, Allman, Cicak, Sirois, Whittaker, Lehnert, Simmonds}}]{Teufel11}
	\bibinfo{author}{\bibfnamefont{J.~D.}~\bibnamefont{Teufel}},
	\bibinfo{author}{\bibfnamefont{T.}~\bibnamefont{Donner}},
	\bibinfo{author}{\bibfnamefont{D.}~\bibnamefont{Li}},
	\bibinfo{author}{\bibfnamefont{J.~W.}~\bibnamefont{Harlow}},
	\bibinfo{author}{\bibfnamefont{M.~S.}~\bibnamefont{Allman}},
	\bibinfo{author}{\bibfnamefont{K.}~\bibnamefont{Cicak}},
	\bibinfo{author}{\bibfnamefont{A.~J.}~\bibnamefont{Sirois}},
	\bibinfo{author}{\bibfnamefont{J.~D.}~\bibnamefont{Whittaker}},
	\bibinfo{author}{\bibfnamefont{K.~W.}~\bibnamefont{Lehnert}},
	\bibinfo{author}{\bibfnamefont{R.~W.}~\bibnamefont{Simmonds}},
	\bibinfo{title}{\bibfnamefont{Sideband-cooling of micromechanical motion to the quantum ground state}}.
	\textit{\bibinfo{journal}{Nature}} \textbf{\bibinfo{volume}{475}},
	\bibinfo{pages}{359-363} (\bibinfo{year}{2011}).
	
	\bibitem[{\citenamefont{Chan et~al.}(2011)\citenamefont{Chan, Mayer~Alegre, Safavi-Naeini, Hill, Krause, Gr\"oblacher, Aspelmeyer, Painter}}]{Chan11}
	\bibinfo{author}{\bibfnamefont{J.}~\bibnamefont{Chan}},
	\bibinfo{author}{\bibfnamefont{T.~P.}~\bibnamefont{Mayer~Alegre}},
	\bibinfo{author}{\bibfnamefont{A.~H.}~\bibnamefont{Safavi-Naeini}},
	\bibinfo{author}{\bibfnamefont{J.~T.}~\bibnamefont{Hill}},
	\bibinfo{author}{\bibfnamefont{A.}~\bibnamefont{Krause}},
	\bibinfo{author}{\bibfnamefont{S.}~\bibnamefont{Gr\"oblacher}},
	\bibinfo{author}{\bibfnamefont{M.}~\bibnamefont{Aspelmeyer}},
	\bibinfo{author}{\bibfnamefont{O.}~\bibnamefont{Painter}},
	\bibinfo{title}{\bibfnamefont{Laser cooling of a nanomechanical oscillator into its quantum ground state}}.
	\textit{\bibinfo{journal}{Nature}} \textbf{\bibinfo{volume}{478}},
	\bibinfo{pages}{89-92} (\bibinfo{year}{2011}).
	
	\bibitem[{\citenamefont{Wilson-Rae et~al.}(2007)\citenamefont{Wilson-Rae, Nooshi, Zwerger, Kippenberg}}]{WilsonRae07}
	\bibinfo{author}{\bibfnamefont{I.}~\bibnamefont{Wilson-Rae}},
	\bibinfo{author}{\bibfnamefont{N.}~\bibnamefont{Nooshi}},
	\bibinfo{author}{\bibfnamefont{W.}~\bibnamefont{Zwerger}},
	\bibinfo{author}{\bibfnamefont{T.~J.}~\bibnamefont{Kippenberg}},
	\bibinfo{title}{\bibfnamefont{Theory of Ground State Cooling of a Mechanical Oscillator Using Dynamical Backaction}}.
	\textit{\bibinfo{journal}{Physical Review Letters}} \textbf{\bibinfo{volume}{99}},
	\bibinfo{pages}{093901} (\bibinfo{year}{2007}).
	
	\bibitem[{\citenamefont{Marquardt et~al.}(2007)\citenamefont{Marquardt, Chen, Clerk, Girvin}}]{Marquardt07}
	\bibinfo{author}{\bibfnamefont{F.}~\bibnamefont{Marquardt}},
	\bibinfo{author}{\bibfnamefont{J.~P.}~\bibnamefont{Chen}},
	\bibinfo{author}{\bibfnamefont{A.~A.}~\bibnamefont{Clerk}},
	\bibinfo{author}{\bibfnamefont{S.~M.}~\bibnamefont{Girvin}},
	\bibinfo{title}{\bibfnamefont{Quantum Theory of Cavity-Assisted sideband-cooling of Mechanical Motion}}.
	\textit{\bibinfo{journal}{Physical Review Letters}} \textbf{\bibinfo{volume}{99}},
	\bibinfo{pages}{093902} (\bibinfo{year}{2007}).
	
	\bibitem[{\citenamefont{Meekhof et~al.}(1996)\citenamefont{Meekhof, Monroe, King, Itano, Wineland}}]{Meekhof96}
	\bibinfo{author}{\bibfnamefont{D.~M.}~\bibnamefont{Meekhof}},
	\bibinfo{author}{\bibfnamefont{C.}~\bibnamefont{Monroe}},
	\bibinfo{author}{\bibfnamefont{B.~E.}~\bibnamefont{King}},
	\bibinfo{author}{\bibfnamefont{W.~M.}~\bibnamefont{Itano}},
	\bibinfo{author}{\bibfnamefont{D.~J.}~\bibnamefont{Wineland}},
	\bibinfo{title}{\bibfnamefont{Generation of Nonclassical Motional States of a Trapped Atom}}.
	\textit{\bibinfo{journal}{Physical Review Letters}} \textbf{\bibinfo{volume}{76}},
	\bibinfo{pages}{1796-1799} (\bibinfo{year}{1996}).
	
	\bibitem[{\citenamefont{Eichler and Petta}(2018)\citenamefont{Eichler, Petta}}]{Eichler18}
	\bibinfo{author}{\bibfnamefont{C.}~\bibnamefont{Eichler}},
	\bibinfo{author}{\bibfnamefont{J.~R.}~\bibnamefont{Petta}},
	\bibinfo{title}{\bibfnamefont{Realizing a Circuit Analog of an Optomechanical System with Longitudinally Coupled Superconducting Resonators}}.
	\textit{\bibinfo{journal}{Physical Review Letters}} \textbf{\bibinfo{volume}{120}},
	\bibinfo{pages}{227702} (\bibinfo{year}{2018}).
	
	\bibitem[{\citenamefont{Bothner et~al.}(2020)\citenamefont{Bothner, Rodrigues, Steele}}]{Bothner20}
	\bibinfo{author}{\bibfnamefont{D.}~\bibnamefont{Bothner}},
	\bibinfo{author}{\bibfnamefont{I.~C.}~\bibnamefont{Rodrigues}},
	\bibinfo{author}{\bibfnamefont{G.~A.}~\bibnamefont{Steele}},
	\bibinfo{title}{\bibfnamefont{Photon-pressure strong coupling between two superconducting circuits}}.
	\textit{\bibinfo{journal}{Nature Physics}} \textbf{\bibinfo{volume}{17}},
	\bibinfo{pages}{85-91} (\bibinfo{year}{2021}).
	
	\bibitem[{\citenamefont{Johansson et~al.}(2014)\citenamefont{Johansson, Johansson, Nori}}]{Johansson14}
	\bibinfo{author}{\bibfnamefont{J.~R.}~\bibnamefont{Johansson}},
	\bibinfo{author}{\bibfnamefont{G.}~\bibnamefont{Johansson}},
	\bibinfo{author}{\bibfnamefont{F.}~\bibnamefont{Nori}},
	\bibinfo{title}{\bibfnamefont{Optomechanical-like coupling between superconducting resonators}}.
	\textit{\bibinfo{journal}{Physical Review A}} \textbf{\bibinfo{volume}{90}},
	\bibinfo{pages}{053833} (\bibinfo{year}{2014}).
	
	\bibitem[{\citenamefont{Hardal et~al.}(2017)\citenamefont{Hardal, Aslan, Wilson, M\"ustecapl{\i}o\u{g}lu}}]{Hardal17}
	\bibinfo{author}{\bibfnamefont{A.~\"U.~C.}~\bibnamefont{Hardal}},
	\bibinfo{author}{\bibfnamefont{N.}~\bibnamefont{Aslan}},
	\bibinfo{author}{\bibfnamefont{C.~M.}~\bibnamefont{Wilson}},
	\bibinfo{author}{\bibfnamefont{\"O.~E.}~\bibnamefont{M\"ustecapl{\i}o\u{g}lu}},
	\bibinfo{title}{\bibfnamefont{Quantum heat engine with coupled superconducting resonators}}.
	\textit{\bibinfo{journal}{Physical Review E}} \textbf{\bibinfo{volume}{96}},
	\bibinfo{pages}{062120} (\bibinfo{year}{2017}).
	
	\bibitem[{\citenamefont{Weigand and Terhal}(2020)\citenamefont{Weigand, Terhal}}]{Weigand20}
	\bibinfo{author}{\bibfnamefont{D.~J.}~\bibnamefont{Weigand}},
	\bibinfo{author}{\bibfnamefont{B.~M.}~\bibnamefont{Terhal}},
	\bibinfo{title}{\bibfnamefont{Realizing modular quadrature measurements via a tunable photon-pressure coupling in circuit QED}}.
	\textit{\bibinfo{journal}{Physical Review A}} \textbf{\bibinfo{volume}{101}},
	\bibinfo{pages}{053840} (\bibinfo{year}{2020}).
	
	\bibitem[{\citenamefont{Metelmann and Clerk}(2014)\citenamefont{Metelmann, Clerk}}]{Metelmann14}
	\bibinfo{author}{\bibfnamefont{A.}~\bibnamefont{Metelmann}},
	\bibinfo{author}{\bibfnamefont{A.~A.}~\bibnamefont{Clerk}},
	\bibinfo{title}{\bibfnamefont{Quantum-Limited Amplification via Reservoir Engineering}}.
	\textit{\bibinfo{journal}{Physical Review Letters}} \textbf{\bibinfo{volume}{112}},
	\bibinfo{pages}{133904} (\bibinfo{year}{2014}).
	
	\bibitem[{\citenamefont{Massel et~al.}(2011)\citenamefont{Massel, Heikkil\"a, Pirkkalainen, Cho, Saloniemi, Hakonen, Sillanp\"a\"a}}]{Massel11}
	\bibinfo{author}{\bibfnamefont{F.}~\bibnamefont{Massel}},
	\bibinfo{author}{\bibfnamefont{T.~T.}~\bibnamefont{Heikkil\"a}},
	\bibinfo{author}{\bibfnamefont{J.-M.}~\bibnamefont{Pirkkalainen}},
	\bibinfo{author}{\bibfnamefont{S.~U.}~\bibnamefont{Cho}},
	\bibinfo{author}{\bibfnamefont{H.}~\bibnamefont{Saloniemi}},
	\bibinfo{author}{\bibfnamefont{P.~J.}~\bibnamefont{Hakonen}},
	\bibinfo{author}{\bibfnamefont{M.~A.}~\bibnamefont{Sillanp\"a\"a}},
	\bibinfo{title}{\bibfnamefont{Microwave amplification with nanomechanical resonators}}.
	\textit{\bibinfo{journal}{Nature}} \textbf{\bibinfo{volume}{480}},
	\bibinfo{pages}{351-354} (\bibinfo{year}{2011}).
	
	\bibitem[{\citenamefont{Nunnenkamp et~al.}(2014)\citenamefont{Nunnenkamp, Sudhir, Feofanov, Roulet, Kippenberg}}]{Nunnenkamp14}
	\bibinfo{author}{\bibfnamefont{A.}~\bibnamefont{Nunnenkamp}},
	\bibinfo{author}{\bibfnamefont{V.}~\bibnamefont{Sudhir}},
	\bibinfo{author}{\bibfnamefont{A.~K.}~\bibnamefont{Feofanov}},
	\bibinfo{author}{\bibfnamefont{A.}~\bibnamefont{Roulet}},
	\bibinfo{author}{\bibfnamefont{T.~J.}~\bibnamefont{Kippenberg}},
	\bibinfo{title}{\bibfnamefont{Quantum-Limited Amplification and Parametric Instability in the Reversed Dissipation Regime of Cavity Optomechanics}}.
	\textit{\bibinfo{journal}{Physical Review Letters}} \textbf{\bibinfo{volume}{113}},
	\bibinfo{pages}{023604} (\bibinfo{year}{2014}).
	
	\bibitem[{\citenamefont{Ockeloen-Korppi et~al.}(2016)\citenamefont{Ockeloen-Korppi, Damsk\"agg, Prkkalainen, Heikkil\"a, Massel, Sillanp\"a\"a}}]{OckeloenKorppi16}
	\bibinfo{author}{\bibfnamefont{C.~F.}~\bibnamefont{Ockeloen-Korppi}},
	\bibinfo{author}{\bibfnamefont{E.}~\bibnamefont{Damsk\"agg}},
	\bibinfo{author}{\bibfnamefont{J.-M.}~\bibnamefont{Pirkkalainen}},
	\bibinfo{author}{\bibfnamefont{T.~T.}~\bibnamefont{Heikkil\"a}},
	\bibinfo{author}{\bibfnamefont{F.}~\bibnamefont{Massel}},
	\bibinfo{author}{\bibfnamefont{M.~A.}~\bibnamefont{Sillanp\"a\"a}},
	\bibinfo{title}{\bibfnamefont{Low-Noise Amplification and Frequency Conversion with a Multiport Microwave Optomechanical Device}}.
	\textit{\bibinfo{journal}{Physical Review X}} \textbf{\bibinfo{volume}{6}},
	\bibinfo{pages}{041024} (\bibinfo{year}{2016}).
	
	\bibitem[{\citenamefont{Bothner et~al.}(2020)\citenamefont{Bothner, Yanai, Iniguez-Rabago, Yuan, Blanter, Steele}}]{Bothner20a}
	\bibinfo{author}{\bibfnamefont{D.}~\bibnamefont{Bothner}},
	\bibinfo{author}{\bibfnamefont{S.}~\bibnamefont{Yanai}},
	\bibinfo{author}{\bibfnamefont{A.}~\bibnamefont{Iniguez-Rabago}},
	\bibinfo{author}{\bibfnamefont{M.}~\bibnamefont{Yuan}},
	\bibinfo{author}{\bibfnamefont{Ya.~M.}~\bibnamefont{Blanter}},
	\bibinfo{author}{\bibfnamefont{G.~A.}~\bibnamefont{Steele}},
	\bibinfo{title}{\bibfnamefont{Cavity electromechanics with parametric mechanical driving}}.
	\textit{\bibinfo{journal}{Nature Communications}} \textbf{\bibinfo{volume}{11}},
	\bibinfo{pages}{1589} (\bibinfo{year}{2020}).
	
	\bibitem[{\citenamefont{Malz et~al.}(2018)\citenamefont{Malz, T\'oth, Bernier, Feofanov, Kippenberg, Nunnenkamp}}]{Malz18}
	\bibinfo{author}{\bibfnamefont{D.}~\bibnamefont{Malz}},
	\bibinfo{author}{\bibfnamefont{L.~D.}~\bibnamefont{T\'oth}},
	\bibinfo{author}{\bibfnamefont{N.~R.}~\bibnamefont{Bernier}},
	\bibinfo{author}{\bibfnamefont{A.~F.}~\bibnamefont{Feofanov}},
	\bibinfo{author}{\bibfnamefont{T.~J.}~\bibnamefont{Kippenberg}},
	\bibinfo{author}{\bibfnamefont{A.}~\bibnamefont{Nunnenkamp}},
	\bibinfo{title}{\bibfnamefont{Quantum-Limited Directional Amplifiers with Optomechanics}}.
	\textit{\bibinfo{journal}{Physical Review Letters}} \textbf{\bibinfo{volume}{120}},
	\bibinfo{pages}{023601} (\bibinfo{year}{2018}).
	
	\bibitem[{\citenamefont{Bernier et~al.}(2017)\citenamefont{Bernier, T\'oth, Koottandavida, Ioannou, Malz, Nunnenkamp, Feofanov, Kippenberg}}]{Bernier17}
	\bibinfo{author}{\bibfnamefont{N.~R.}~\bibnamefont{Bernier}},
	\bibinfo{author}{\bibfnamefont{L.~D.}~\bibnamefont{Toth}},
	\bibinfo{author}{\bibfnamefont{A.}~\bibnamefont{Koottandavida}},
	\bibinfo{author}{\bibfnamefont{M.~A.}~\bibnamefont{Ioannou}},
	\bibinfo{author}{\bibfnamefont{D.}~\bibnamefont{Malz}},
	\bibinfo{author}{\bibfnamefont{A.}~\bibnamefont{Nunnenkamp}},
	\bibinfo{author}{\bibfnamefont{A.~K.}~\bibnamefont{Feofanov}},
	\bibinfo{author}{\bibfnamefont{T.~J.}~\bibnamefont{Kippenberg}},
	\bibinfo{title}{\bibfnamefont{Nonreciprocal reconfigurable microwave optomechanical circuit}}.
	\textit{\bibinfo{journal}{Nature Communications}} \textbf{\bibinfo{volume}{8}},
	\bibinfo{pages}{604} (\bibinfo{year}{2017}).
	
	\bibitem[{\citenamefont{Barzanjeh et~al.}(2017)\citenamefont{Barzanjeh, Wulf, Peruzzo, Kalaee, Dieterle, Painter, Fink}}]{Barzanjeh17}
	\bibinfo{author}{\bibfnamefont{S.}~\bibnamefont{Barzanjeh}},
	\bibinfo{author}{\bibfnamefont{M.}~\bibnamefont{Wulf}},
	\bibinfo{author}{\bibfnamefont{M.}~\bibnamefont{Peruzzo}},
	\bibinfo{author}{\bibfnamefont{M.}~\bibnamefont{Kalaee}},
	\bibinfo{author}{\bibfnamefont{P.~B.}~\bibnamefont{Dieterle}},
	\bibinfo{author}{\bibfnamefont{O.}~\bibnamefont{Painter}},
	\bibinfo{author}{\bibfnamefont{J.~M.}~\bibnamefont{Fink}},
	\bibinfo{title}{\bibfnamefont{Mechanical on-chip microwave circulator}}.
	\textit{\bibinfo{journal}{Nature Communications}} \textbf{\bibinfo{volume}{8}},
	\bibinfo{pages}{953} (\bibinfo{year}{2017}).
	
	\bibitem[{\citenamefont{Safavi-Naeini et~al.}(20113)\citenamefont{Safavi-Naeini, Mayer Alegre, Chan, Eichenfield, Winger, Lin, Hill, Chang, Painter}}]{SafaviNaeini11}
	\bibinfo{author}{\bibfnamefont{A.~H.}~\bibnamefont{Safavi-Naeini}},
	\bibinfo{author}{\bibfnamefont{T.~P.}~\bibnamefont{Mayer Alegre}},
	\bibinfo{author}{\bibfnamefont{J.}~\bibnamefont{Chan}},
	\bibinfo{author}{\bibfnamefont{M.}~\bibnamefont{Eichenfield}},
	\bibinfo{author}{\bibfnamefont{M.}~\bibnamefont{Winger}},
	\bibinfo{author}{\bibfnamefont{Q.}~\bibnamefont{Lin}},
	\bibinfo{author}{\bibfnamefont{J.~T.}~\bibnamefont{Hill}},
	\bibinfo{author}{\bibfnamefont{D.~E.}~\bibnamefont{Chang}},
	\bibinfo{author}{\bibfnamefont{O.}~\bibnamefont{Painter}},
	\bibinfo{title}{\bibfnamefont{Electromagnetically induced transparency and slow light with optomechanics}}.
	\textit{\bibinfo{journal}{Nature}} \textbf{\bibinfo{volume}{472}},
	\bibinfo{pages}{69-73} (\bibinfo{year}{2011}).
	
	\bibitem[{\citenamefont{Zhou et~al.}(2013)\citenamefont{Zhou, Hocke, Schliesser, Marx, Huebl, Gross, Kippenberg}}]{Zhou13}
	\bibinfo{author}{\bibfnamefont{X.}~\bibnamefont{Zhou}},
	\bibinfo{author}{\bibfnamefont{F.}~\bibnamefont{Hocke}},
	\bibinfo{author}{\bibfnamefont{A.}~\bibnamefont{Schliesser}},
	\bibinfo{author}{\bibfnamefont{A.}~\bibnamefont{Marx}},
	\bibinfo{author}{\bibfnamefont{H.}~\bibnamefont{Huebl}},
	\bibinfo{author}{\bibfnamefont{R.}~\bibnamefont{Gross}},
	\bibinfo{author}{\bibfnamefont{T.~J.}~\bibnamefont{Kippenberg}},
	\bibinfo{title}{\bibfnamefont{Slowing, advancing and switching of microwave signls using circuit nanoelectromechanics}}.
	\textit{\bibinfo{journal}{Nature Physics}} \textbf{\bibinfo{volume}{9}},
	\bibinfo{pages}{179-184} (\bibinfo{year}{2013}).
	
	\bibitem[{\citenamefont{Fang et~al.}(2017)\citenamefont{Fang, Luo, Metelmann, Matheny, Marquardt, Clerk, Painter}}]{Fang17}
	\bibinfo{author}{\bibfnamefont{K.}~\bibnamefont{Fang}},
	\bibinfo{author}{\bibfnamefont{J.}~\bibnamefont{Luo}},
	\bibinfo{author}{\bibfnamefont{A.}~\bibnamefont{Metelmann}},
	\bibinfo{author}{\bibfnamefont{M.~H.}~\bibnamefont{Matheny}},
	\bibinfo{author}{\bibfnamefont{F.}~\bibnamefont{Marquardt}},
	\bibinfo{author}{\bibfnamefont{A.~A.}~\bibnamefont{Clerk}},
	\bibinfo{author}{\bibfnamefont{O.}~\bibnamefont{Painter}},
	\bibinfo{title}{\bibfnamefont{Generalized non-reciprocity in an optomechanical circuit via synthetic magnetism and reservoir engineering}}.
	\textit{\bibinfo{journal}{Nature Physics}} \textbf{\bibinfo{volume}{13}},
	\bibinfo{pages}{465-471} (\bibinfo{year}{2017}).
	
	\bibitem[{\citenamefont{Toth et~al.}(2017)\citenamefont{Toth, Bernier, Nunnenkamp, Feofanov, Kippenberg}}]{Toth17}
	\bibinfo{author}{\bibfnamefont{L.~D.}~\bibnamefont{T\'oth}},
	\bibinfo{author}{\bibfnamefont{N.~R.}~\bibnamefont{Bernier}},
	\bibinfo{author}{\bibfnamefont{A.}~\bibnamefont{Nunnenkamp}},
	\bibinfo{author}{\bibfnamefont{A.~K.}~\bibnamefont{Feofanov}},
	\bibinfo{author}{\bibfnamefont{T.~J.}~\bibnamefont{Kippenberg}},
	\bibinfo{title}{\bibfnamefont{A dissipative quantum reservoir for microwave light using a mechanical oscillator}}.
	\textit{\bibinfo{journal}{Nature Physics}} \textbf{\bibinfo{volume}{13}},
	\bibinfo{pages}{787-792} (\bibinfo{year}{2017}).
	
	\bibitem[{\citenamefont{Arute et~al.}(2019)\citenamefont{Arute, Arya, Babbush, Bacon, Bardin, Barends, Biswas, Boixo, Brandao, Buell, Burkett, Chen, Chen, Chiaro, Collins, Courtney, Dunsworth, Farhi, Foxen, Fowler, Gidney, Giustina, Graff, Guerin, Habegger, Harrigan, Hartmann, Ho, Hoffmann, Huang, Humble, Isakov, Jeffrey, Jiang, Kafri, Kechedzhi, Kelly, Klimov, Knysh, Korotkov, Kostritsa, Landhuis, Lindmark, Lucero, Lyakh, Mandr\`a, McClean, McEwen, Megrant, Mi, Michielsen, Mohseni, Mutus, Naaman, Neeley, Neill, Niu, Ostby, Petukhov, Platt, Quintana, Rieffel, Roushan, Rubin, Sank, Satzinger, Smelyanskiy, Sung, Trevithick, Vainsencher, Villalonga, White, Yao, Yeh, Zalcman, Neven, Martinis}}]{Arute19}
	\bibinfo{author}{\bibfnamefont{F.}~\bibnamefont{Arute}},
	\bibinfo{author}{\bibfnamefont{K.}~\bibnamefont{Arya}},
	\bibinfo{author}{\bibfnamefont{R.}~\bibnamefont{Babbush}},
	\bibinfo{author}{\bibfnamefont{D.}~\bibnamefont{Bacon}},
	\bibinfo{author}{\bibfnamefont{J.~C.}~\bibnamefont{Bardin}},
	\bibinfo{author}{\bibfnamefont{R.}~\bibnamefont{Barends}},
	\bibinfo{author}{\bibfnamefont{R.}~\bibnamefont{Biswas}},
	\bibinfo{author}{\bibfnamefont{S.}~\bibnamefont{Boixo}},
	\bibinfo{author}{\bibfnamefont{F.~G.~S.~L.}~\bibnamefont{Brandao}},
	\bibinfo{author}{\bibfnamefont{D.~A.}~\bibnamefont{Buell}},
	\bibinfo{author}{\bibfnamefont{B.}~\bibnamefont{Burkett}},
	\bibinfo{author}{\bibfnamefont{Y.}~\bibnamefont{Chen}},
	\bibinfo{author}{\bibfnamefont{Z.}~\bibnamefont{Chen}},
	\bibinfo{author}{\bibfnamefont{B.}~\bibnamefont{Chiaro}},
	\bibinfo{author}{\bibfnamefont{R.}~\bibnamefont{Collins}},
	\bibinfo{author}{\bibfnamefont{W.}~\bibnamefont{Courtney}},
	\bibinfo{author}{\bibfnamefont{A.}~\bibnamefont{Dunsworth}},
	\bibinfo{author}{\bibfnamefont{E.}~\bibnamefont{Farhi}},
	\bibinfo{author}{\bibfnamefont{B.}~\bibnamefont{Foxen}},
	\bibinfo{author}{\bibfnamefont{A.}~\bibnamefont{Fowler}},
	\bibinfo{author}{\bibfnamefont{C.}~\bibnamefont{Gidney}},
	\bibinfo{author}{\bibfnamefont{M.}~\bibnamefont{Giustina}},
	\bibinfo{author}{\bibfnamefont{R.}~\bibnamefont{Graff}},
	\bibinfo{author}{\bibfnamefont{K.}~\bibnamefont{Guerin}},
	\bibinfo{author}{\bibfnamefont{S.}~\bibnamefont{Habegger}},
	\bibinfo{author}{\bibfnamefont{M.~P.}~\bibnamefont{Harrigan}},
	\bibinfo{author}{\bibfnamefont{M.~J.}~\bibnamefont{Hartmann}},
	\bibinfo{author}{\bibfnamefont{A.}~\bibnamefont{Ho}},
	\bibinfo{author}{\bibfnamefont{M.}~\bibnamefont{Hoffmann}},
	\bibinfo{author}{\bibfnamefont{T.}~\bibnamefont{Huang}},
	\bibinfo{author}{\bibfnamefont{T.~S.}~\bibnamefont{Humble}},
	\bibinfo{author}{\bibfnamefont{S.~V.}~\bibnamefont{Isakov}},
	\bibinfo{author}{\bibfnamefont{I.}~\bibnamefont{Jeffrey}},
	\bibinfo{author}{\bibfnamefont{Z.}~\bibnamefont{Jiang}},
	\bibinfo{author}{\bibfnamefont{D.}~\bibnamefont{Kafri}},
	\bibinfo{author}{\bibfnamefont{K.}~\bibnamefont{Kechedzhi}},
	\bibinfo{author}{\bibfnamefont{J.}~\bibnamefont{Kelly}},
	\bibinfo{author}{\bibfnamefont{P.~V.}~\bibnamefont{Klimov}},
	\bibinfo{author}{\bibfnamefont{S.}~\bibnamefont{Knysh}},
	\bibinfo{author}{\bibfnamefont{A.}~\bibnamefont{Korotkov}},
	\bibinfo{author}{\bibfnamefont{F.}~\bibnamefont{Kostritsa}},
	\bibinfo{author}{\bibfnamefont{D.}~\bibnamefont{Landhuis}},
	\bibinfo{author}{\bibfnamefont{M.}~\bibnamefont{Lindmark}},
	\bibinfo{author}{\bibfnamefont{E.}~\bibnamefont{Lucero}},
	\bibinfo{author}{\bibfnamefont{D.}~\bibnamefont{Lyakh}},
	\bibinfo{author}{\bibfnamefont{S.}~\bibnamefont{Mandr\`a}},
	\bibinfo{author}{\bibfnamefont{J.~R.}~\bibnamefont{McClean}},
	\bibinfo{author}{\bibfnamefont{M.}~\bibnamefont{McEwen}},
	\bibinfo{author}{\bibfnamefont{A.}~\bibnamefont{Megrant}},
	\bibinfo{author}{\bibfnamefont{X.}~\bibnamefont{Mi}},
	\bibinfo{author}{\bibfnamefont{K.}~\bibnamefont{Michielsen}},
	\bibinfo{author}{\bibfnamefont{M.}~\bibnamefont{Mohseni}},
	\bibinfo{author}{\bibfnamefont{J.}~\bibnamefont{Mutus}},
	\bibinfo{author}{\bibfnamefont{O.}~\bibnamefont{Naaman}},
	\bibinfo{author}{\bibfnamefont{M.}~\bibnamefont{Neeley}},
	\bibinfo{author}{\bibfnamefont{C.}~\bibnamefont{Neill}},
	\bibinfo{author}{\bibfnamefont{M.~Y.}~\bibnamefont{Niu}},
	\bibinfo{author}{\bibfnamefont{E.}~\bibnamefont{Ostby}},
	\bibinfo{author}{\bibfnamefont{A.}~\bibnamefont{Petukhov}},
	\bibinfo{author}{\bibfnamefont{J.~C.}~\bibnamefont{Platt}},
	\bibinfo{author}{\bibfnamefont{C.}~\bibnamefont{Quintana}},
	\bibinfo{author}{\bibfnamefont{E.~J.}~\bibnamefont{Rieffel}},
	\bibinfo{author}{\bibfnamefont{P.}~\bibnamefont{Roushan}},
	\bibinfo{author}{\bibfnamefont{N.~C.}~\bibnamefont{Rubin}},
	\bibinfo{author}{\bibfnamefont{D.}~\bibnamefont{Sank}},
	\bibinfo{author}{\bibfnamefont{K.~J.}~\bibnamefont{Satzinger}},
	\bibinfo{author}{\bibfnamefont{V.}~\bibnamefont{Smelyanskiy}},
	\bibinfo{author}{\bibfnamefont{K.~J.}~\bibnamefont{Sung}},
	\bibinfo{author}{\bibfnamefont{M.~D.}~\bibnamefont{Trevithick}},
	\bibinfo{author}{\bibfnamefont{A.}~\bibnamefont{Vainsencher}},
	\bibinfo{author}{\bibfnamefont{B.}~\bibnamefont{Villalonga}},
	\bibinfo{author}{\bibfnamefont{T.}~\bibnamefont{White}},
	\bibinfo{author}{\bibfnamefont{Z.~J.}~\bibnamefont{Yao}},
	\bibinfo{author}{\bibfnamefont{P.}~\bibnamefont{Yeh}},
	\bibinfo{author}{\bibfnamefont{A.}~\bibnamefont{Zalcman}},
	\bibinfo{author}{\bibfnamefont{H.}~\bibnamefont{Neven}},
	\bibinfo{author}{\bibfnamefont{J.~M.}~\bibnamefont{Martinis}},
	\bibinfo{title}{\bibfnamefont{Quantum supremacy using a programmable superconducting processor}}.
	\textit{\bibinfo{journal}{Nature}} \textbf{\bibinfo{volume}{574}},
	\bibinfo{pages}{505-510} (\bibinfo{year}{2019}).
	
	\bibitem[{\citenamefont{Backes et~al.}(2021)\citenamefont{Backes, Palken, Kenany, Brubaker, Cahn, Droster, Hilton, Ghosh, Jackson, Lamoreaux, Leder, Lehnert, Lewis, Malnou, Maruyama, Rapidis, Simanovskaia, Singh, Speller, Urdinaran, Vale, Assendelft, Bibber, Wang}}]{Backes21}
	\bibinfo{author}{\bibfnamefont{K.~M.}~\bibnamefont{Backes}},
	\bibinfo{author}{\bibfnamefont{D.~A.}~\bibnamefont{Palken}},
	\bibinfo{author}{\bibfnamefont{S.}~\bibnamefont{Al Kenany}},
	\bibinfo{author}{\bibfnamefont{M.~M.}~\bibnamefont{Brubaker}},
	\bibinfo{author}{\bibfnamefont{S.~B.}~\bibnamefont{Cahn}},
	\bibinfo{author}{\bibfnamefont{A.}~\bibnamefont{Droster}},
	\bibinfo{author}{\bibfnamefont{G.~C.}~\bibnamefont{Hilton}},
	\bibinfo{author}{\bibfnamefont{S.}~\bibnamefont{Ghosh}},
	\bibinfo{author}{\bibfnamefont{H.}~\bibnamefont{Jackson}},
	\bibinfo{author}{\bibfnamefont{S.~K.}~\bibnamefont{Lamoreaux}},
	\bibinfo{author}{\bibfnamefont{A.~F.}~\bibnamefont{Leder}},
	\bibinfo{author}{\bibfnamefont{K.~W.}~\bibnamefont{Lehnert}},
	\bibinfo{author}{\bibfnamefont{S.~M.}~\bibnamefont{Lewis}},
	\bibinfo{author}{\bibfnamefont{M.}~\bibnamefont{Malnou}},
	\bibinfo{author}{\bibfnamefont{R.~H.}~\bibnamefont{Maruyama}},
	\bibinfo{author}{\bibfnamefont{N.~M.}~\bibnamefont{Rapidis}},
	\bibinfo{author}{\bibfnamefont{M.}~\bibnamefont{Simanovskaia}},
	\bibinfo{author}{\bibfnamefont{S.}~\bibnamefont{Singh}},
	\bibinfo{author}{\bibfnamefont{D.~H.}~\bibnamefont{Speller}},
	\bibinfo{author}{\bibfnamefont{I.}~\bibnamefont{Urdinaran}},
	\bibinfo{author}{\bibfnamefont{L.~R.}~\bibnamefont{Vale}},
	\bibinfo{author}{\bibfnamefont{E.~C.}~\bibnamefont{van Assendelft}},
	\bibinfo{author}{\bibfnamefont{K.}~\bibnamefont{van Bibber}},
	\bibinfo{author}{\bibfnamefont{H.}~\bibnamefont{Wang}},
	\bibinfo{title}{\bibfnamefont{A quantum enhanced search for dark matter axions}}.
	\textit{\bibinfo{journal}{Nature}} \textbf{\bibinfo{volume}{590}},
	\bibinfo{pages}{238-242} (\bibinfo{year}{2021}).
	
	\bibitem[{\citenamefont{Chaudhuri et~al.}(2019)\citenamefont{Chaudhuri}}]{Chaudhuri19}
	\bibinfo{author}{\bibfnamefont{S.}~\bibnamefont{Chaudhuri}},
	\bibinfo{title}{\bibfnamefont{The Dark Matter Radio: A Quantum-enhanced Search for QCD Axion Dark Matter}}.
	\bibinfo{pages}{PhD thesis, Stanford University} (\bibinfo{year}{2019}).
	
	\bibitem[{\citenamefont{Verhagen et~al.}(2012)\citenamefont{Verhagen, Del\'eglise, Weis, Schliesser, Kippenberg}}]{Verhagen12}
	\bibinfo{author}{\bibfnamefont{E.}~\bibnamefont{Verhagen}},
	\bibinfo{author}{\bibfnamefont{S.}~\bibnamefont{Del\'eglise}},
	\bibinfo{author}{\bibfnamefont{S.}~\bibnamefont{Weis}},
	\bibinfo{author}{\bibfnamefont{A.}~\bibnamefont{Schliesser}},
	\bibinfo{author}{\bibfnamefont{T.~J.}~\bibnamefont{Kippenberg}},
	\bibinfo{title}{\bibfnamefont{Quantum-coherent coupling of a mechanical oscillator to an optical cavity mode}}.
	\textit{\bibinfo{journal}{Nature}} \textbf{\bibinfo{volume}{482}},
	\bibinfo{pages}{63-67} (\bibinfo{year}{2012}).
	
	\bibitem[{\citenamefont{Levenson-Falk et~al.}(2011)\citenamefont{Levenson-Falk, Vijay, Siddiqi}}]{LevensonFalk11}
	\bibinfo{author}{\bibfnamefont{E.~M.}~\bibnamefont{Levenson-Falk}},
	\bibinfo{author}{\bibfnamefont{R.}~\bibnamefont{Vijay}},
	\bibinfo{author}{\bibfnamefont{I.}~\bibnamefont{Siddiqi}},
	\bibinfo{title}{\bibfnamefont{Nonlinear microwave response of aluminum weak-link Josephson oscillators}}.
	\textit{\bibinfo{journal}{Applied Physics Letters}} \textbf{\bibinfo{volume}{98}},
	\bibinfo{pages}{123115} (\bibinfo{year}{2011})
	
	\bibitem[{\citenamefont{Kennedy et~al.}(2019)\citenamefont{Kennedy, Burnett, Fenton, Constantino, Warburton, Morton, Dupont-Ferrier}}]{Kennedy19}
	\bibinfo{author}{\bibfnamefont{O.~W.}~\bibnamefont{Kennedy}},
	\bibinfo{author}{\bibfnamefont{J.}~\bibnamefont{Burnett}},
	\bibinfo{author}{\bibfnamefont{J.~C.}~\bibnamefont{Fenton}},
	\bibinfo{author}{\bibfnamefont{N.~G.~N.}~\bibnamefont{Constantino}},
	\bibinfo{author}{\bibfnamefont{P.~A.}~\bibnamefont{Warburton}},
	\bibinfo{author}{\bibfnamefont{J.~J.~L.}~\bibnamefont{Morton}},
	\bibinfo{author}{\bibfnamefont{E.}~\bibnamefont{Dupont-Ferrier}},
	\bibinfo{title}{\bibfnamefont{Tunable Nb Superconducting Resonator Based on a Constriction Nano-SQUID Fabricated with a Ne Focused Ion Beam}}.
	\textit{\bibinfo{journal}{Physical Review Applied}} \textbf{\bibinfo{volume}{11}},
	\bibinfo{pages}{014006} (\bibinfo{year}{2019}).
	
	\bibitem[{\citenamefont{Rodrigues et~al.}(2019)\citenamefont{Rodrigues, Bothner, Steele}}]{Rodrigues19}
	\bibinfo{author}{\bibfnamefont{I.~C.}~\bibnamefont{Rodrigues}},
	\bibinfo{author}{\bibfnamefont{D.}~\bibnamefont{Bothner}},
	\bibinfo{author}{\bibfnamefont{G.~A.}~\bibnamefont{Steele}},
	\bibinfo{title}{\bibfnamefont{Coupling microwave photons to a mechanical resonator using quantum interference}}.
	\textit{\bibinfo{journal}{Nature Communications}} \textbf{\bibinfo{volume}{10}},
	\bibinfo{pages}{5359} (\bibinfo{year}{2019}).
	
	\bibitem[{\citenamefont{Vijay et~al.}(2009)\citenamefont{Vijay, Sau, Cohen, Siddiqi}}]{Vijay09}
	\bibinfo{author}{\bibfnamefont{R.}~\bibnamefont{Vijay}},
	\bibinfo{author}{\bibfnamefont{J.~D.}~\bibnamefont{Sau}},
	\bibinfo{author}{\bibfnamefont{M.~L.}~\bibnamefont{Cohen}},
	\bibinfo{author}{\bibfnamefont{I.}~\bibnamefont{Siddiqi}}.
	\bibinfo{title}{\bibfnamefont{Optimizing Anharmonicity in Nanoscale Weak Link Josephson Junction Oscillators}}.
	\textit{\bibinfo{journal}{Physical Review Letters}} \textbf{\bibinfo{volume}{103}},
	\bibinfo{pages}{087003} (\bibinfo{year}{2009}).
	
	\bibitem[{\citenamefont{Vijay et~al.}(2010)\citenamefont{Vijay, Levenson-Falk, Slichter, Siddiqi}}]{Vijay10}
	\bibinfo{author}{\bibfnamefont{R.}~\bibnamefont{Vijay}},
	\bibinfo{author}{\bibfnamefont{E.~M.}~\bibnamefont{Levenson-Falk}},
	\bibinfo{author}{\bibfnamefont{D.~H.}~\bibnamefont{Slichter}},
	\bibinfo{author}{\bibfnamefont{I.}~\bibnamefont{Siddiqi}}.
	\bibinfo{title}{\bibfnamefont{Approaching ideal weak link behaviour with three dimensional aluminum nanobridges}}.
	\textit{\bibinfo{journal}{Applied Physics Letters}} \textbf{\bibinfo{volume}{96}},
	\bibinfo{pages}{223112} (\bibinfo{year}{2010}).
	
	\bibitem[{\citenamefont{Igreja and Dias}(2004)\citenamefont{Igreja, Dias}}]{Igreja04}
	\bibinfo{author}{\bibfnamefont{R.}~\bibnamefont{Igreja}}
	\bibinfo{author}{\bibfnamefont{C.~J.}~\bibnamefont{Dias}},
	\bibinfo{title}{\bibfnamefont{Analytical evaluation of the interdigital electrodes capacitance for a multi-layered structure}}.
	\textit{\bibinfo{journal}{Sensors and Actuators A}} \textbf{\bibinfo{volume}{112}},
	\bibinfo{pages}{291-301} (\bibinfo{year}{2004})
	
	
	\bibitem[{\citenamefont{Gely and Steele}(2019)\citenamefont{Gely, Steele}}]{Gely19a}
	\bibinfo{author}{\bibfnamefont{M.~F.}~\bibnamefont{Gely}},
	\bibinfo{author}{\bibfnamefont{G.~A.}~\bibnamefont{Steele}},
	\bibinfo{title}{\bibfnamefont{QuCAT: Quantum Circuit Analyzer Tool in Python}}.
	\textit{\bibinfo{journal}{New Journal of Physics}} \textbf{\bibinfo{volume}{22}},
	\bibinfo{pages}{013025} (\bibinfo{year}{2020}).
	
	\bibitem[{\citenamefont{Nunnenkamp et~al.}(2011)\citenamefont{Nunnenkamp, B\o rkje, Girvin}}]{Nunnenkamp11}
	\bibinfo{author}{\bibfnamefont{A.}~\bibnamefont{Nunnenkamp}},
	\bibinfo{author}{\bibfnamefont{K.}~\bibnamefont{B\o rkje}},
	\bibinfo{author}{\bibfnamefont{S.~M.}~\bibnamefont{Girvin}},
	\bibinfo{title}{\bibfnamefont{Single-Photon Optomechanics}}.
	\textit{\bibinfo{journal}{Physical Review Letters}} \textbf{\bibinfo{volume}{107}},
	\bibinfo{pages}{063602} (\bibinfo{year}{2011}).
	
	\bibitem[{\citenamefont{Rabl}(2011)\citenamefont{Rabl}}]{Rabl11}
	\bibinfo{author}{\bibfnamefont{P.}~\bibnamefont{Rabl}},
	\bibinfo{title}{\bibfnamefont{Photon Blockade Effect in Optomechanical Systems}}.
	\textit{\bibinfo{journal}{Physical Review Letters}} \textbf{\bibinfo{volume}{107}},
	\bibinfo{pages}{063601} (\bibinfo{year}{2011}).
	
	\bibitem[{\citenamefont{Weinstein et~al.}(2014)\citenamefont{Weinstein, Lei, Wollman, Suh, etelmann, Clerk, Schwab}}]{Weinstein14}
	\bibinfo{author}{\bibfnamefont{A.~J.}~\bibnamefont{Weinstein}},
	\bibinfo{author}{\bibfnamefont{C.~U.}~\bibnamefont{Lei}},
	\bibinfo{author}{\bibfnamefont{E.~E.}~\bibnamefont{Wollman}},
	\bibinfo{author}{\bibfnamefont{J.}~\bibnamefont{Suh}},
	\bibinfo{author}{\bibfnamefont{A.}~\bibnamefont{Metelmann}},
	\bibinfo{author}{\bibfnamefont{A.~A.}~\bibnamefont{Clerk}},
	\bibinfo{author}{\bibfnamefont{K.~C.}~\bibnamefont{Schwab}}.
	\bibinfo{title}{\bibfnamefont{Observation and Interpretation of Motional Sideband Asymmetry in a Quantum Electromechanical Device}}.
	\textit{\bibinfo{journal}{Physical Review X}} \textbf{\bibinfo{volume}{4}},
	\bibinfo{pages}{041003} (\bibinfo{year}{2014}).
	
	\bibitem[{\citenamefont{Dobrindt et~al.}(2008)\citenamefont{Dobrindt, Wilson-Rae, Kippenberg}}]{Dobrindt08}
	\bibinfo{author}{\bibfnamefont{J.~M.}~\bibnamefont{Dobrindt}},
	\bibinfo{author}{\bibfnamefont{I.}~\bibnamefont{Wilson-Rae}},
	\bibinfo{author}{\bibfnamefont{T.~J.}~\bibnamefont{Kippenberg}},
	\bibinfo{title}{\bibfnamefont{Parametric Normal-Mode Splitting in Cavity Optomechanics}}.
	\textit{\bibinfo{journal}{Physical Review Letters}} \textbf{\bibinfo{volume}{101}},
	\bibinfo{pages}{263602} (\bibinfo{year}{2008}).
	
	\bibitem[{\citenamefont{Teufel et~al.}(2011)\citenamefont{Teufel, Li, Allman, Cicak, Sirois, Whittaker, Simmonds}}]{Teufel11a}
	\bibinfo{author}{\bibfnamefont{J.~D.}~\bibnamefont{Teufel}},
	\bibinfo{author}{\bibfnamefont{Dale}~\bibnamefont{Li}},
	\bibinfo{author}{\bibfnamefont{M.~S.}~\bibnamefont{Allman}},
	\bibinfo{author}{\bibfnamefont{K.}~\bibnamefont{Cicak}},
	\bibinfo{author}{\bibfnamefont{A.~J.}~\bibnamefont{Sirois}},
	\bibinfo{author}{\bibfnamefont{J.~D.}~\bibnamefont{Whittaker}},
	\bibinfo{author}{\bibfnamefont{R.~W.}~\bibnamefont{Simmonds}}.
	\bibinfo{title}{\bibfnamefont{Circuit cavity electromechanics in the strong-coupling regime}}.
	\textit{\bibinfo{journal}{Nature}} \textbf{\bibinfo{volume}{471}},
	\bibinfo{pages}{204-208} (\bibinfo{year}{2011}).
	
	
	\bibitem[{\citenamefont{Xu et~al.}(2020)\citenamefont{Xu, Han, Zou, Fu, Xu, Zhong, Jiang, Tang}}]{Xu20a}
	\bibinfo{author}{\bibfnamefont{M.}~\bibnamefont{Xu}},
	\bibinfo{author}{\bibfnamefont{X.}~\bibnamefont{Han}},
	\bibinfo{author}{\bibfnamefont{C.-L.}~\bibnamefont{Zou}},
	\bibinfo{author}{\bibfnamefont{W.}~\bibnamefont{Fu}},
	\bibinfo{author}{\bibfnamefont{Y.}~\bibnamefont{Xu}},
	\bibinfo{author}{\bibfnamefont{C.}~\bibnamefont{Zhong}},
	\bibinfo{author}{\bibfnamefont{L.}~\bibnamefont{Jiang}},
	\bibinfo{author}{\bibfnamefont{H.~X.}~\bibnamefont{Tang}},
	\bibinfo{title}{\bibfnamefont{Radiative cooling of a superconducting resonator}}.
	\textit{\bibinfo{journal}{Physical Review Letters}} \textbf{\bibinfo{volume}{124}},
	\bibinfo{pages}{033602} (\bibinfo{year}{2020}).
	
	\bibitem[{\citenamefont{Albanese et~al.}(2020)\citenamefont{Albanese, Probst, Ranjan, Zollitsch, Pechal, Wallraff, Morton, Vion, Esteve, Flurin, Bertet}}]{Albanese20}
	\bibinfo{author}{\bibfnamefont{B.}~\bibnamefont{Albanese}},
	\bibinfo{author}{\bibfnamefont{S.}~\bibnamefont{Probst}},
	\bibinfo{author}{\bibfnamefont{V.}~\bibnamefont{Ranjan}},
	\bibinfo{author}{\bibfnamefont{C.}~\bibnamefont{Zollitsch}},
	\bibinfo{author}{\bibfnamefont{M.}~\bibnamefont{Pechal}},
	\bibinfo{author}{\bibfnamefont{A.}~\bibnamefont{Wallraff}},
	\bibinfo{author}{\bibfnamefont{J.~J.~L.}~\bibnamefont{Morton}},
	\bibinfo{author}{\bibfnamefont{D.}~\bibnamefont{Vion}},
	\bibinfo{author}{\bibfnamefont{D.}~\bibnamefont{Esteve}},
	\bibinfo{author}{\bibfnamefont{E.}~\bibnamefont{Flurin}},
	\bibinfo{author}{\bibfnamefont{P.}~\bibnamefont{Bertet}},
	\bibinfo{title}{\bibfnamefont{Radiative cooling of a spin ensemble}}.
	\textit{\bibinfo{journal}{Nature Physics}} \textbf{\bibinfo{volume}{16}},
	\bibinfo{pages}{751-755} (\bibinfo{year}{2020}).
	
	\bibitem[{\citenamefont{Yuan et~al.}(2015)\citenamefont{Yuan, Singh, Blanter, Steele}}]{Yuan15}
	\bibinfo{author}{\bibfnamefont{M.}~\bibnamefont{Yuan}},
	\bibinfo{author}{\bibfnamefont{V.}~\bibnamefont{Singh}},
	\bibinfo{author}{\bibfnamefont{Ya.~M.}~\bibnamefont{Blanter}},
	\bibinfo{author}{\bibfnamefont{G.~A.}~\bibnamefont{Steele}},
	\bibinfo{title}{\bibfnamefont{Large cooperativity and microkelvin cooling with a three-dimensional optomechanical cavity}}.
	\textit{\bibinfo{journal}{Nature Communications}} \textbf{\bibinfo{volume}{6}},
	\bibinfo{pages}{8491} (\bibinfo{year}{2015}).
	
	\bibitem[{\citenamefont{Valenzuela et~al.}(2006)\citenamefont{Valenzuela, Oliver, Berns, Berggren, Levitov, Orlando}}]{Valenzuela06}
	\bibinfo{author}{\bibfnamefont{S.~O.}~\bibnamefont{Valenzuela}},
	\bibinfo{author}{\bibfnamefont{W.~D.}~\bibnamefont{Oliver}},
	\bibinfo{author}{\bibfnamefont{D.~M.}~\bibnamefont{Berns}},
	\bibinfo{author}{\bibfnamefont{K.~K.}~\bibnamefont{Berggren}},
	\bibinfo{author}{\bibfnamefont{L.~S.}~\bibnamefont{Levitov}},
	\bibinfo{author}{\bibfnamefont{T.~P.}~\bibnamefont{Orlando}},
	\bibinfo{title}{\bibfnamefont{Microwave-Induced Cooling of a Superconducting Qubit}}.
	\textit{\bibinfo{journal}{Science}} \textbf{\bibinfo{volume}{314}},
	\bibinfo{pages}{1589-1592} (\bibinfo{year}{2006}).
	
	\bibitem[{\citenamefont{Gely et~al.}(2019)\citenamefont{Gely, Kounalakis, Dickel, Dalle, Vatr\'e, Baker, Jenkins, Steele}}]{Gely19}
	\bibinfo{author}{\bibfnamefont{M.~F.}~\bibnamefont{Gely}},
	\bibinfo{author}{\bibfnamefont{M.}~\bibnamefont{Kounalakis}},
	\bibinfo{author}{\bibfnamefont{C.}~\bibnamefont{Dickel}},
	\bibinfo{author}{\bibfnamefont{J.}~\bibnamefont{Dalle}},
	\bibinfo{author}{\bibfnamefont{R.}~\bibnamefont{Vatr\'e}},
	\bibinfo{author}{\bibfnamefont{B.}~\bibnamefont{Baker}},
	\bibinfo{author}{\bibfnamefont{M.~D.}~\bibnamefont{Jenkins}},
	\bibinfo{author}{\bibfnamefont{G.~A.}~\bibnamefont{Steele}},
	\bibinfo{title}{\bibfnamefont{Observation and stabilization of photonic Fock states in a hot radio-frequency resonator}}.
	\textit{\bibinfo{journal}{Science}} \textbf{\bibinfo{volume}{363}},
	\bibinfo{pages}{1072-1075} (\bibinfo{year}{2019}).
	
\end{thebibliography}
\end{document}